\documentclass[fleqn]{aa}  

\usepackage{graphicx}
\usepackage{xcolor}
\usepackage{hyperref}
\usepackage{txfonts}
\usepackage{calligra}
\usepackage{calrsfs}
\usepackage[T1]{fontenc}
\usepackage[mathcal]{eucal}
\newcommand{\vectX}{\bf {\it X}}
\newcommand{\vectTheta}{\bf {\it \Theta}}
\newcommand{\matrGamma}{\bf \Gamma}

\begin{document}

   \title{{\it Godzilla}, a monster lurks in the Sunburst galaxy.}

   \titlerunning{Godzilla}
  \authorrunning{Diego et al.}

   \author{J.M. Diego
         \inst{1}\fnmsep\thanks{jdiego@ifca.unican.es}
          \and
        M. Pascale\inst{2}
          \and
       B.~J.~Kavanagh\inst{1}
        \and
        P. Kelly\inst{3}
          \and
        L. Dai\inst{4}
          \and
        B. Frye\inst{5}
          \and
        T. Broadhurst\inst{6,7,8}
          }

   \institute{Instituto de F\'isica de Cantabria (CSIC-UC). Avda. Los Castros s/n. 39005 Santander, Spain
         \and
         Department of Astronomy, University of California, 501 Campbell Hall \#3411, Berkeley, CA 94720, USA
         \and
         School of Physics and Astronomy, University of Minnesota, 116 Church Street SE, Minneapolis, MN 55455, USA
         \and
         Department of Physics, University of California, 366 Physics North MC 7300, Berkeley, CA 94720, US
         \and
         Department of Astronomy/Steward Observatory, University of Arizona, 933 N Cherry Ave., Tucson, AZ 85721, USA
         \and
         Department of Theoretical Physics, University of the Basque Country UPV-EHU, 48040 Bilbao, Spain
         \and
         Donostia International Physics Center (DIPC), 20018 Donostia, The Basque Country
         \and
         IKERBASQUE, Basque Foundation for Science, Alameda Urquijo, 36-5 48008 Bilbao, Spain
             }


 \abstract{
     We model the strong lensing effect in the galaxy cluster PSZ1 G311.65-18.48 (z=0.443) with an improved version of the hybrid method WSLAP+. We extend the number of constraints by including the position of critical points, which are combined with the classic positional constraints of the lensed galaxies.  We pay special attention to a transient candidate source (Tr) previously discovered in the giant Sunburst arc (z=2.37). Our lens model predicts Tr to be within a fraction of an arcsecond from the critical curve, having a larger magnification factor than previously found, but still not large enough to explain the observed flux and lack of counterimages. 
     Possible candidate counterimages are discussed that would lower the magnification required to explain Tr, but extreme magnification factors ($\mu>1000$) are still required, even in that case. The presence of a small mass perturber with a mass comparable to a dwarf galaxy ($M\sim 10^8 \,{\rm M}_{\odot}$) near the position of Tr is needed  in order to explain the required  magnification and morphology of the lensed galaxy. We discuss how the existence of this perturber could potentially be used to constrain models of dark matter. The large apparent brightness and unresolved nature of the magnified object implies a combination of extreme magnification and a very luminous and compact source ($r<0.3$ pc). Possible candidates are discussed, including an hyperluminous star or an accretion disc around an intermediate-mass black hole (IMBH). Based on spectral information, we argue that a luminous blue variable (LBV) star caught during an outburst is the most likely candidate. Owing to the extreme magnification and luminosity of this source we dub it {\it Godzilla}.    
   }

   \maketitle
%

\section{Introduction}
PSZ1 G311.65-18.48  is a massive cluster at $z_l=0.443$ which was first identified thanks to its strong Sunyaev-Zeldovich signature in Planck data \citep{Planck2014}. Optical follow up of this cluster from the ESO revealed a strongly lensed bright galaxy at redshift $z_s=2.3702$ \citep{Dahle2016}. The lensed galaxy formed  a giant arc with nearly circular symmetry, suggesting a roundish morphology (in projection) for the cluster mass. Space HST images and detailed spectroscopy from the ground have allowed for the identification of a powerful ionizing source at the giant arc. More specifically, \cite{RiveraThorsen2017} found evidence of escaping ionizing Lyman $\alpha$ photons from the source at $z_s=2.3702$. Because of this fact, the giant arc was dubbed the Sunburst arc. The hypothesis of escaping ionizing photons was later confirmed in \cite{RiveraThorsen2019} and \cite{Vanzella2020B}. Ionizing emission was identified in an unprecedentedly large number of multiple images (12) of the same unresolved feature, or knot. This knot was later constrained to have a very small size, most likely a compact star cluster (effective radius of $\approx 8$ pc) with an estimated mass of $10^7 {\rm M}_{\odot}$ \citep{Vanzella2021}. \\

A peculiar object, classified as a transient candidate (referred to as Tr hereafter), is identified in the Sunburst arc in \cite{Vanzella2020}, with a magnitude $F814W\approx 22$ and at the same redshift as the Sunburst arc. In that work, the authors report a minimum magnification of 20 for Tr, but it could be higher due to the proximity of Tr to the critical curve. The authors argue that the magnification can not be larger than 100 based on its location in a region where no critical curves are expected nearby. However this argument can be revisited since it is based on the distance to the estimated position of critical points (or symmetry points). A constraint in the position of such symmetry points is not enough to constrain the minimum distance to the critical curve, or the presence of a small mass perturber, that could bring the critical curve closer to the position of Tr. If the magnification is less than 100, the intrinsic luminosity at Tr must be comparable to that of a supernova (SN), or absolute magnitude $M_V\approx -19$. In this case, the SN should be observed at other locations along the Sunburst arc, with a time delay due to the combination of geometric plus Shapiro delays. On cluster scales, these time delays can range from weeks to years, but in configurations resembling perfect Einstein rings, the time delays are expected to be relatively small (less than a year). The counterimages of the alleged SN have not been identified, despite Tr being present in observations spanning $\approx 7$ years, raising questions about the true nature of Tr. 
In an alternative scenario, where the magnification can be significantly larger than 100, it would be possible to consider relatively faint sources to explain Tr, such as bright luminous stars at high redshift. Very large magnification factors for Tr would also naturally explain the apparent lack of counterimages, since only one of them (Tr) would be detectable, thanks to its very large magnification factor. 
Examples of stars at extreme magnification factors have already been observed, with Icarus being the first star at cosmological distances observed through this mechanism \citep{Kelly2018}. Other recent examples can be also found in the literature, for example \cite{Chen2019,Kaurov2019}.

\cite{Vanzella2020} discusses how Tr has stayed bright for almost one year in the rest frame of the source. Visual inspection of recent HST images taken in June 2021 reveal that Tr is still bright at that time, which, combined with earlier observations from March 2014 that already show the presence of Tr (specifically in the z-band image taken with NTT/EFOCS2, as shown in Figure 2 of \cite{Dahle2016}), extend the bright phase to at least $\approx 2$ years in the rest frame. This is in tension with the explanation of Tr as an SN during peak emission.

In follow-up work, \cite{Vanzella2021} delenses the Sunburst arc and constrains its intrinsic size to $\approx 3$ kpc$^2$. Tr is mentioned again in this work, but without discussing its nature. In the same work, two possible faint counterimages of Tr are mentioned, 5.7c and 5.7d. 
However, as we will discuss later, 5.7c and 5.7d in \cite{Vanzella2021} are not valid counterimages of Tr based on their location within the giant arc.  
Hence, Tr remains without clear identifiable counterimages in the literature. 
The lack of bright counterimages for Tr is puzzling, but offers important clues that help constrain the possible nature of this source. Given the time scale over which the object has been observed ($\approx 2$ years in the rest frame), and the time delay between pairs of images appearing near  critical curves (typically much less than a year), a source that is bright enough, like an SN, should be multiply lensed and observed at other locations around the arc. The fact that it is only observed once has been suggested as evidence in support of the transient hypothesis in \cite{Vanzella2020} (for instance a supernova). In the present work we discuss an alternative scenario in which Tr is not a transient, but a relatively persistent but fainter and minute source (a hyperluminous star or similar compact object like an accretion disc) which is being magnified by extreme factors ($\mu > 1000$), such that it can be detected in HST in only one location, without producing detectable counterimages.

Any possible interpretation of Tr needs to be supported by lens models. A lens model is presented by the same team in \cite{Pignataro2021}. In that work, and taking advantage of integral field spectroscopy in a large portion of the cluster with the MUSE instrument, the authors identify 5 lensed galaxies, with 81 identifiable knots that can be used as lensing constraints. Most of these knots belong to the giant Sunburst arc, or system 5.  System 1 in that work is not used as a lens constraint since one of the counterimages falls outside the footprint of the observation with MUSE, and is not confirmed spectroscopically. Hence the lens model in  \cite{Pignataro2021} is derived with 4 out of the 5 reported systems, all of them confirmed spectroscopically, and with system 5 contributing with most of the lensing constraints. In section \ref{sec_data} we discuss brefly the lensing constraints.  The lens model in \cite{Pignataro2021} is based on the public code Lenstool \cite{Kneib1996,Jullo2007,Jullo2009}, a parametric model that places small scale halos in the member galaxies and then one or several large scale halos to represent the cluster halo. Parametric models are powerful and often reliable but, like any other method, are not free of systematic effects. One of the limitations of parametric models is that the halos (small and large) are assumed to have some form of symmetry. Although this may be a good approximation in many scenarios, it does not capture non-symmetric mass distributions due to tidal forces, which are prevalent in galaxy cluster environments. For the particular case of the Sunburst, detailed mass modelling is required in order to interpret features, such as Tr, that fall near critical curves.
The critical curve of the model in \cite{Pignataro2021} reproduces well the symmetry points of the Sunburst. This however comes at a cost. In order to reproduce the observed features, some galaxies need to be optimized independently. In particular, \cite{Pignataro2021} had to include a barred galaxy with a relatively large mass and extreme ellipticity (named 1298 in that work) in order to reproduce some features around the position of Tr. Whether the mass and ellipticity are good approximations to the real mass and ellipticity of this particular galaxy is an open question, but it could also be a manifestation of the limited flexibility of the parametric model. 

Additional lens models can explore the degeneracies in the lens model and provide alternative solutions for the critical curve that winds around the Sunburst arc. In addition, if Tr is indeed a transient event, time delays are a useful prediction from a lens model. Time delay between counterimages can  provide an explanation for the lack of counterimages. They are also helpful in the preparation of observing campaigns, should a counterimage be predicted to appear. This was the case for SN Refsdal, where lens models accurately predicted its reappearance and facilitated its observation with HST \cite{Diego2016,Kelly2016}. Unfortunately \cite{Pignataro2021} provides no information about the time delay associated with Tr. 
In this work we present a new lens model together with its predicted magnification and time delays at the expected positions of the counterimages of Tr, and discuss its implications in relation to the possible nature of Tr. 



We complement the earlier comprehensive study of this very interesting object by providing the first hybrid lens model of PSZ1 G311.65-18.48. We base our lensing analysis on the detailed data compilation in \cite{Pignataro2021}. We use an improved version of our hybrid code WSLAP+ that takes advantage of a new type of constraint, namely the position of critical points, or points where critical curves are known to be passing through. 
These points can be identified from the data following symmetry arguments, since near critical curves identifiable knots are expected to be nearly equidistant to the critical curve. The high-resolution of HST combined with the spectroscopic information provided by MUSE allowed \cite{Pignataro2021} to identify a wealth of features in the Sunburst arc that we exploit here to pinpoint the position of critical points. PSZ1 G311.65-18.48 is the first cluster where this new improvement is tested with WSLAP+.\\

The paper is organized as follows. In section \ref{sec_data} we describe the lensing constraints, the majority of which are directly adopted from \cite{Pignataro2021}, though we also discuss the new type of constraints added to WSLAP+. 
The lens models derived using the lensing constraints and our hybrid lens reconstruction algorithm WSLAP+ are presented in section \ref{sec_results}. 
In section \ref{sec_TrCounter} we discuss Tr, and possible counterimages of Tr, which are later used to constrain the magnification of Tr. Time delays and magnifications derived from the lens model at Tr, and the alleged counterimage positions, are also discussed in this section. 
Section~\ref{Sect_MuAtTr} provides an alternative estimation of the magnification of Tr based on photometric measurements and flux ratios at the candidate counterimage positions. Uncertainties from the lens model on the magnification of Tr are discussed in section~\ref{sec_Transient}. In this section we discuss also how a small scale perturber is needed in order to interpret the observations. 
The mass and location of the perturber is constrained in section~\ref{sec_perturber} and the implications of the existence of such a perturber on  different dark matter models is discussed in section~\ref{Sec_DM}. 
Section~\ref{sec_Constraints} reviews the different constraints on Tr from the lens models and the observations. 
The true nature of Tr is discussed in section~\ref{sec_Godzilla}, where we introduce a possible LBV star, which we dub \textit{Godzilla}. 
We discuss our results in section~\ref{sec_discussion}, and summarize in section~\ref{sec_conclusions}.

For the sake of flow and clarity, and to avoid distractions from the main focus of the paper, we move the more technical and tedious calculations to the appendices. This includes the description of our improved lensing reconstruction method WSLAP+, predictions from the lens model including possible new lensed systems, and modelling the PSF. 
We adopt a standard flat cosmological model with $\Omega_m=0.3$ and $h=0.7$. At the redshift of the lens, and for this cosmological model, one arcsecond corresponds to 5.7 kpc, while at the redshift of the Sunburst one arcsecond corresponds to 8.16 kpc. The distance modulus to the Sunburst arc is 46.45.

\section{Gravitational lensing constraints}\label{sec_data}
We derive our lens model using an improved version of the hybrid method WSLAP+. This algorithm has been described extensively in the past \citep{Diego2005,Diego2007,Diego2016}. We include a description of the technical aspects of this algorithm in Appendix~\ref{Appendix1}. This appendix describes also the new improvement implemented in the code, which now incorporates as new constraints the estimated position of critical curves at a given redshift. In this section we describe the new critical curve constraints, together with the standard strong lensing constraints used to derive the lens model.  \\

\begin{figure} 
   \includegraphics[width=9cm]{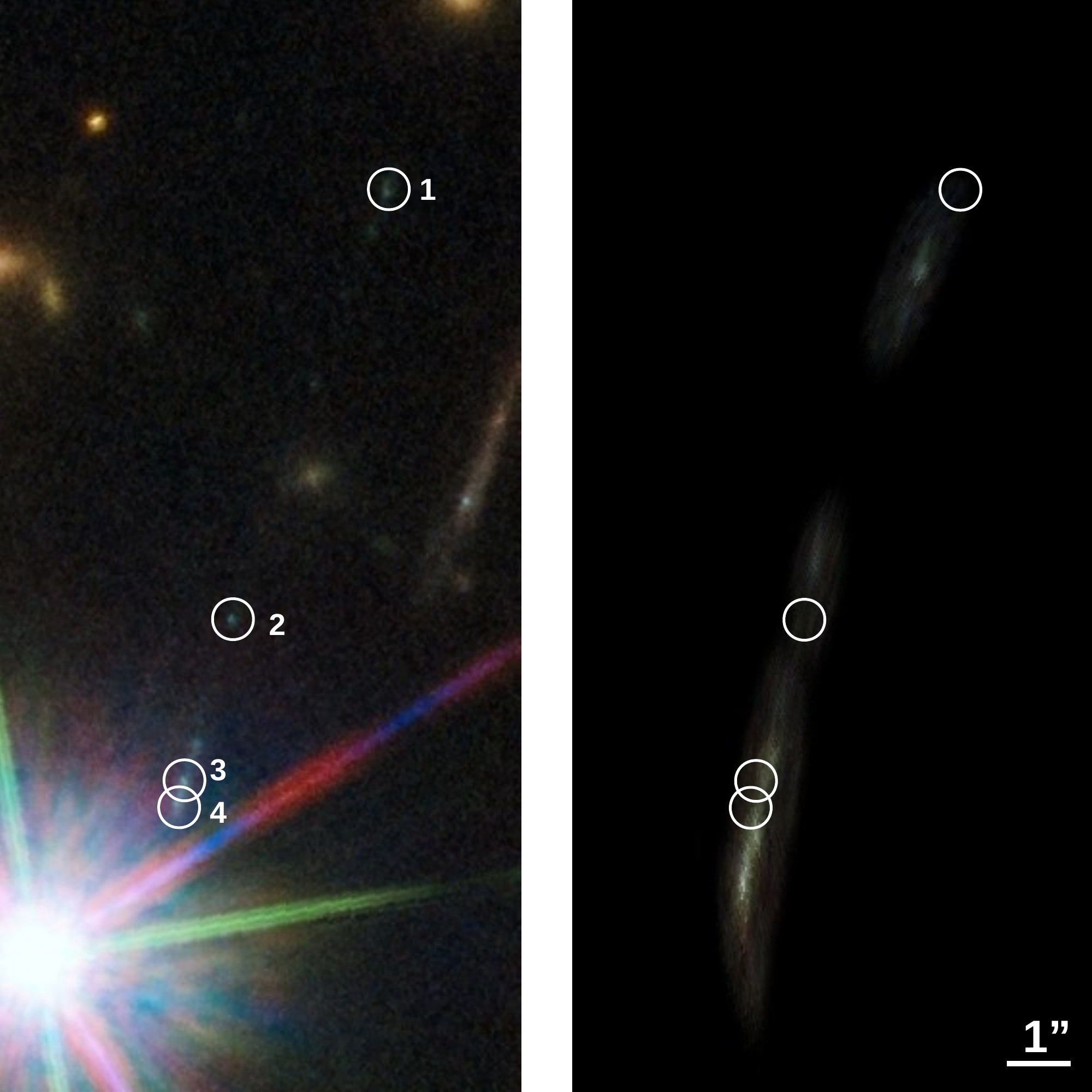}
      \caption{Prediction for system 1 based on the confirmed counterimage in \cite{Pignataro2021}. The left and right panels show the data and predictions respectively. Both morphology and parity are well reproduced by the lens model. The circles in the left panel mark the positions of the 4 knots identified in \cite{Pignataro2021}. These circles are reproduced in the same position in the right panel, in order to better appreciate the offset between the predicted and the observed positions. Offsets of order 1" are typical in free-form reconstructions, especially in regions of the lens plane where constraints are scarce. Note how the tangential magnification between knots 1 and 2 is smaller for the predicted images, while it is larger for knots 3 and 4. 
              }
         \label{Fig_Sys1}
\end{figure}

In order to constrain the lens model, we use the multiple image identification from \cite{Pignataro2021}, which are robust identifications of multiply lensed images thanks to spectroscopic confirmation with MUSE. System 1 at $z=3.505$, which was not used in \cite{Pignataro2021} is used in our model reconstruction. This system was excluded from the analysis of \cite{Pignataro2021} because one image was confirmed spectroscopically by MUSE (the counterimage falls outside the footprint of MUSE observations). In addition, \cite{Pignataro2021} reports that when the non-confirmed counterimage is included in the lens model, a third counterimage for system 3 is predicted but not observed. Earlier models derived with WSLAP+ prior to the publication of the spectroscopic confirmation, predicted system 1 with a redshift larger than 3 (as later confirmed by MUSE data). Our lens model predicts also the correct morphology and position of the unconfirmed arc as shown in Fig.~\ref{Fig_Sys1}. 
Excluding this system from our model poses a similar dilemma to including it in the model of \cite{Pignataro2021}, since at redshift $z=3.505$ a counterimage for image 1b is expected around the position of the unconfirmed image 1a. The color, geometry, flux, and position of image 1a is consistent with the prediction from our earlier model. Given the greater flexibility of WSLAP+, and the solid evidence in favor of the unconfirmed candidate 1a in system 1, we include this system (1a and 1b) in our reconstruction.  The total number of strong lensing knots is then 81, which translates into 162 constraints (see equations in the system of linear equations \ref{eq_lens_system}, where each knot results in two linear equations, one for the $x$ position and one for the $y$ position).

In addition to the positions of the arcs from the 5 systems identified in \cite{Pignataro2021}, we use the high resolution HST images to identify seven points which can be associated with critical points, i.e, a critical curve must pass through these points. This identification is done based on symmetry arguments, since near a critical curve a pair of lensed images must be almost equidistant to the critical curve. Hence, critical points are expected to lay close to the middle point between pairs of images. In table \ref{tab_1} we list the positions of the critical points identified in the HST images, and the angles $\phi$ derived from the direction of the lensed arcs. All critical points are identified using the giant arcs of system 5. Two critical points that can be easily identified in the data between images 5.1d, 5.1e and 5.1f are not included because they are overlapping a member galaxy, or perhaps a background galaxy, as discussed in \cite{Pignataro2021}. The unknown contribution to $\kappa$ from this galaxy makes these two critical points unreliable in our new set of constraints, and hence are not included in our analysis. In total we identify seven reliable critical points that are listed in  table \ref{tab_1}, and shown in Figure \ref{Fig_Data} as black crosses.  The 7 new constraints translate into 14 new linear equations in system \ref{eq_lens_system} (7 for constraints of the type given by equation \ref{eq_New1} and 7 for the type given by equation \ref{eq_New2}).  
Hence, combined with the total number of linear equations in system \ref{eq_lens_system}, the total number of constraints is 176, which is comparable to the number of grid points (177) used to describe the mass distribution. 

\begin{table}
  \begin{minipage}{90mm}
    \caption{Position and angles of the seven critical points used in this work. All points correspond to critical points at the redshift of system 5. The angle is measured counterclockwise from West. Hence negative angles go in clockwise direction from West.}
 \label{tab_1}
 \begin{center}
 \begin{tabular}{|c|ccc|}   
 \hline
 ID   &      RA        &      DEC      & $\phi(^\circ)$   \\
 \hline
 1   &  15:50 06.658   & -78:10:57.63  &  -13.0  \\
 2   &  15:50 06.061   & -78:10:58.22  &  -25.0  \\
 3   &  15:50 05.316   & -78:10:58.85  &  -10.0  \\ 
 4   &  15:50 01.653   & -78:11:06.79  &  -43.0  \\ 
 5   &  15:50 00.121   & -78:11:11.82  &  -58.0  \\
 6   &  15:49 59.862   & -78:11:13.09  &  -60.0  \\
 7   &  15:49 58.723   & -78:11:22.75  &  -78.0  \\
 \hline
\end{tabular}
\end{center}
\end{minipage}
\end{table}
 
 \section{Lens Models}\label{sec_results}
   \begin{figure*}
   \centering
   \includegraphics[width=18.0cm]{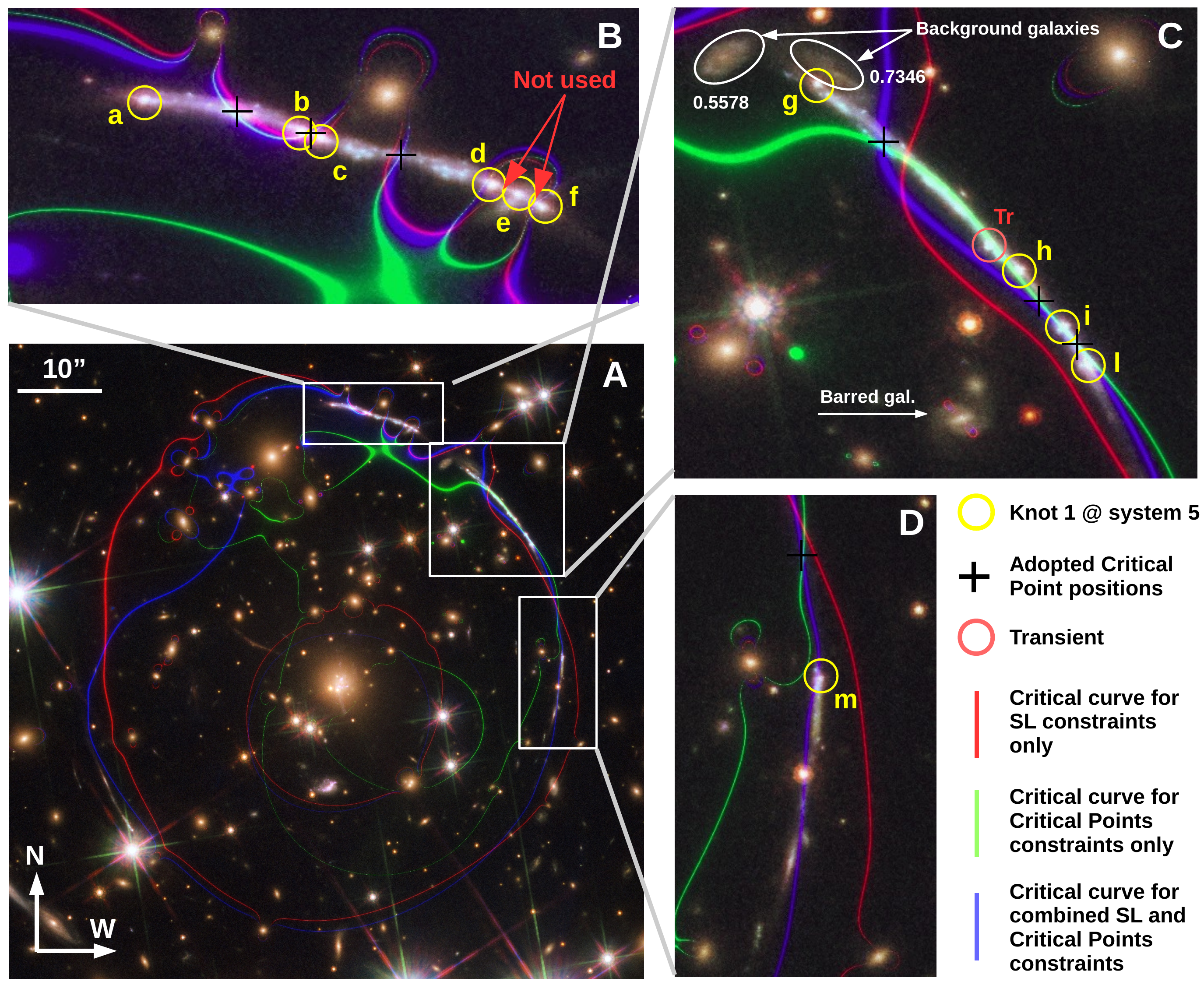}
      \caption{The figure shows the critical curves at the redshift of the Sunburst arc for three different models derived using different combinations of constraints. The red critical curve shows the case where only strong lensing (SL) arc positions are used as constraints. The green critical curve corresponds to the case where only the adopted position of the critical points are used as constraints. Finally the blue critical curve is for the model where both arc positions and critical point positions are used as constraints.  Panel A shows the entire cluster region while panels B, C, and D show zoomed regions around selected areas including key lensing features like the observed position of knot 1 in system 5 (yellow circles). The letters next to each circle follow the labelling scheme for knot 1 in \cite{Pignataro2021}. The inferred position of the critical points used in this work are marked with black crosses. Two background galaxies at redshifts 0.5578 and 0.7346 are marked with white ellipses in panel C. The red arrows mark two critical points not used in our analysis due to the proximity of a lensing galaxy which can bias the values of $\kappa$. The white arrow marks the barred galaxy modelled independently in \cite{Pignataro2021}. Finally, the red circle marks the position of Tr in \cite{Vanzella2020}. The distance between Tr and the blue curve is 0.55".
              }
         \label{Fig_Data}
   \end{figure*}

The minimization process begins by deriving a first model obtained after setting the grid to a regular distribution of 16x16 grid points (i.e all grid points are equally spaced). For this first solution we use only the classic strong lensing constraints (i.e the positions of the knots). This first solution is used to derive a dynamical grid which traces the surface mass density, assigning smaller grid points (i.e smaller FWHM for the Gaussians) to regions of higher mass density.  This process is iterated 3 times, after which both the surface mass density and grid have converged to a stable configuration.

In figure \ref{Fig_Grid} we show the grid obtained after these 3 iterations. This grid is obtained through a Monte-Carlo process where the previous solution for the mass distribution is used to place grid points that follow that distribution. This procedure results in grid points being more concentrated around regions of higher surface mass density. The FWHM of each Gaussian associated to each grid point is inversely proportional to the surface mass density. The number of grid points in this example is 175, comparable to the number of equations in the system of linear equations \ref{eq_lens_system}. 

\cite{Pignataro2021} finds two background sources at redshifts 0.5578 and 0.7346 very close to the Sunburst arc (see figure \ref{Fig_Data}). Due to this proximity, it is expected that these galaxies have some non-negligible effect in the morphology of that giant arc.  In order to account for these galaxies, we simply add two small Gaussian cells (of FWHM 1.8") at their position, and allow WSLAP+ to determine their masses. These additional grid points are marked with a red circle in figure \ref{Fig_Grid} and bring the total number of grid points to 177. Note that the real position of these galaxies is unknown since they are also being lensed by the cluster. However, since the photons from the Sunburst arc and from these galaxies follow very similar geodesics, the path of the Sunburst photons passes close to these galaxies in their respective redshift planes. Placing a perturbing mass at the observed position of these galaxies can mimic the lensing distortion from these galaxies, but caution must be taken to not interpret their masses as the real mass of those galaxies. \\ 

Since critical point constraints have not been used before in WSLAP+, it is interesting to explore the role of the new constraints. Hence we derive three different solutions: i) a solution where only strong lensing (or SL) constraints are used (below we refer to this solution as model $M_1$); ii) a solution where only critical point constraints are used; and iii) a solution that uses both SL and critical point constraints (below we refer to this solution as model $M_2$).  

Figure \ref{Fig_Data} shows the critical curves from the three solutions. In red we show the critical curve for the case where only the 81 strong lensing constraints are used (our model $M_1$). This curve suggests a very round structure for the cluster, similar to the one found by \cite{Pignataro2021}. Close inspection of this model shows that it fails at reproducing with accuracy the position of the critical points (marked with black crosses in the zoomed regions). These points can be easily identified using the multiple knots of system 5 (the Sunburst arc). The position of knot 1 is marked with yellow circles in Figure \ref{Fig_Data} and labelled a--m. This knot corresponds to the compact stellar cluster discussed in \cite{Vanzella2021}. The accuracy on the prediction of the critical points improves, as expected, in the green critical curve, that uses only as constraints the positions of the seven critical points listed in table \ref{tab_1}. In this case, the critical curve passes through all critical points. However, this model fails at reproducing the morphology of the arcs, and differs significantly from the previous model specially in the East section of the cluster, where critical point constraints are not used. The model that combines both the 81 strong lensing constraints and the 7 critical point constraints predicts the critical curve in blue (our model $M_2$). This model captures the desired features of the two previous models. On one hand it is able to reproduce the configuration of all 5 systems while simultaneously reproducing the critical points, although a small offset is observed in the critical point between knots h and i (see panel C in figure \ref{Fig_Data}). We discuss this offset in more detail in section~\ref{sec_Transient} below.

Based on model $M_1$ we showed earlier in figure \ref{Fig_Sys1} the predicted morphology of the candidate 1 of \cite{Pignataro2021} that was not used in their analysis. Our lens model predicts the correct morphology at the correct location for the spectroscopic redshift of its confirmed counterimage. The left panel in the figure shows the real data, while the right panel shows the prediction. The white circles  in both panels are placed at exactly the same coordinates, in order to better appreciate the relative error between the predicted and observed positions. \cite{Pignataro2021} excluded this image from their analysis based on the fact that it falls outside the footprint of MUSE (hence it can not be confirmed spectroscopically), but also with the argument that adding this counterimage results in a lens model that predicts a third (unobserved) counterimage for system 4. Our lens model does not predict that third counterimage, while making a fair prediction of system 1, hence adding great confidence in this system (although still pending its spectroscopic confirmation). 

In terms of integrated mass, the total integrated mass within a radius of 40" from the BCG is ${\rm M}(<40") = 2.47\times10^{14} {\rm M}_{\odot}$ for model $M_1$ and ${\rm M}(<40") = 2.54\times10^{14} {\rm M}_{\odot}$ for model $M_2$. Beyond this radius there are no lensing constraints and the model can not be properly constrained.  In \cite{Pignataro2021}, the authors quote a mass of $\sim 2\times10^{14} {\rm M}_{\odot}$ within $\sim 200$ kpc. For this radius, we find ${\rm M}(< 200 {\rm kpc}) = 2.20\times10^{14} {\rm M}_{\odot}$ and ${\rm M}(< 200 {\rm kpc}) = 2.23\times10^{14} {\rm M}_{\odot}$ for models $M_1$ and $M_2$ respectively.

\section{Candidate counterimages of Tr and time delays}\label{sec_TrCounter}
The source Tr, first discussed in \cite{Vanzella2020}, has no obvious counterimages. This apparent lack of counterimages was one of the reasons in \cite{Vanzella2020} to classify Tr as a stellar transient object (the stellar part due to its unique spectral features, similar to those of stars like Eta Carinae). As mentioned earlier, \cite{Vanzella2021} suggests two possible counterimages for Tr, and labelled 5.7c and 5.7d in that work. However, the position of these knots in the giant arcs is inconsistent with the expected position of the counterimages of Tr, which should appear between knots 1 and 2 \citep[see][]{Vanzella2021}.
In this section we present possible counterimage candidates for Tr. Based on the observed position of Tr with respect to knots 1 and 2, and the lens model prediction, any counterimage of Tr must be also between these knots 1 and 2 in the giant arcs, and in general closer to brighter knot 1 than to the fainter knot 2 (as shown by the model prediction in figure~\ref{Fig_TrPosPredicion} in the appendices). 

\begin{figure} 
   \includegraphics[width=9cm]{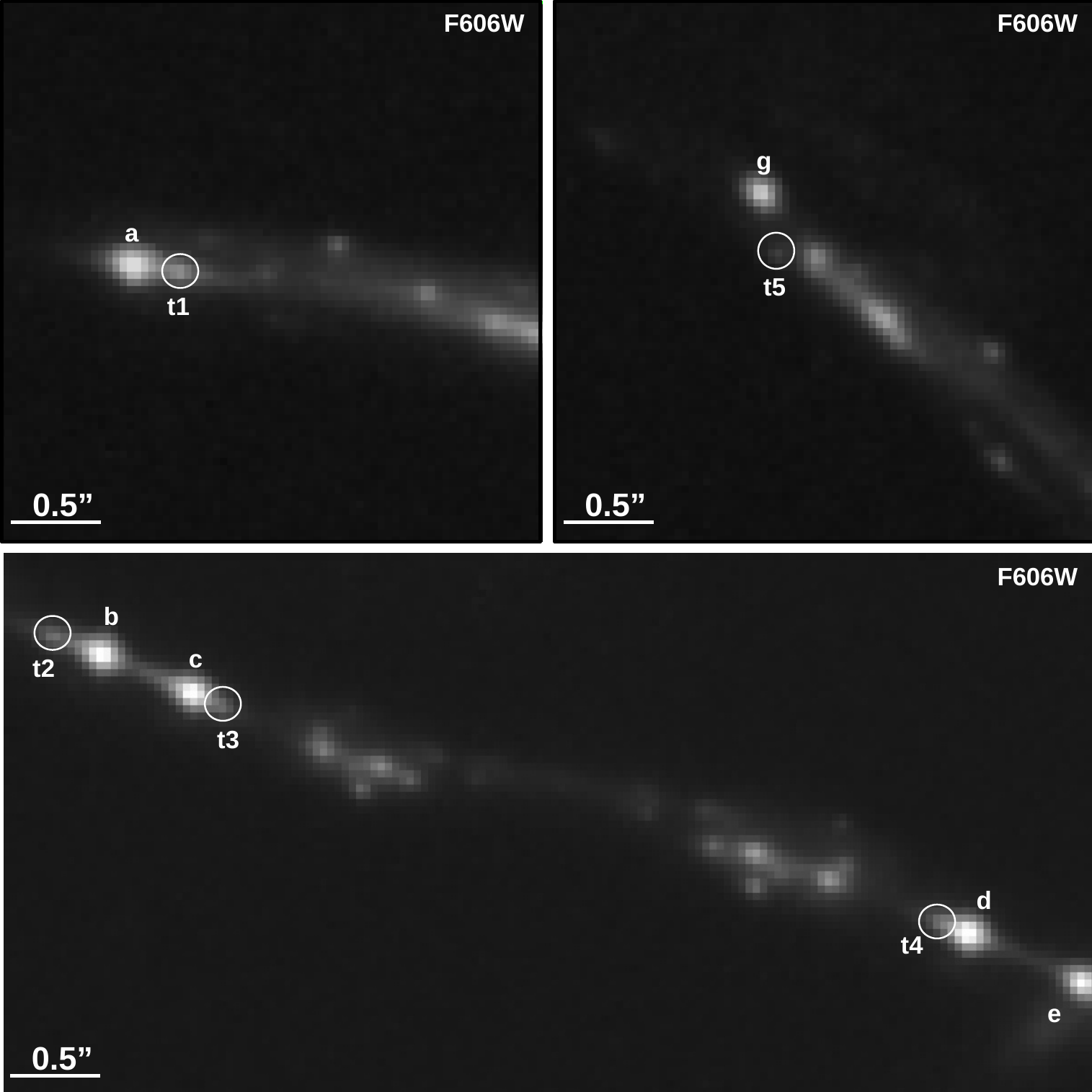}
      \caption{Suggested candidates for the counterimages of Tr based on our lens model. Labels "a" through "g" are used to mark the position of the bright LyC knot 1, as in figure~\ref{Fig_Data}. All images, except t5 are partially blended with knot 1.  
              }
         \label{Fig_Tr2Tr3}
\end{figure}

\begin{table*}
 \centering
  \begin{minipage}{180mm}
    \caption{Columns 1 and 2 show the RA, DEC of Tr and its five counterimage candidates.  Column 3 lists the magnitudes in the filter F606W (or in the filter F390W and in the Vega system in parenthesis). 
    Columns 4 ad 5 show the magnification and time delays predicted by lens model $M_1$, while for model $M_2$ are shown in columns 6 and 7. Time delays are expressed in days. Magnification ratios with respect to Tr are shown in columns 8 and 9 for models $M_1$ and $M_2$ respectively.  Column 10 shows the magnification ratio with respect to Tr estimated from the ratio of distances between the knots 1 and 2 bracketing the positions of Tr and t1--t5. The last column shows the flux ratio between the observed fluxes and the flux at Tr. }
 \label{tab_3}
 \begin{center}
 \begin{tabular}{|c|cc|c|cc|cc|cc|cc|}   
 \hline
     & \small{1} & \small{2} & \small{3} & \small{4} & \small{5} & \small{6} & \small{7} & \small{8} & \small{9} & \small{10} & \small{11}\\
  Id &  RA     &      DEC     &   M$_{\rm AB}$F606W (F390W) & $\mu_1$  & $\Delta T_1$ & $\mu_2$ & $\Delta T_2$ & $\mu_1/\mu_1^{{\rm Tr}}$ & $\mu_2/\mu_2^{{\rm Tr}}$ & $\mu_d/\mu_d^{{\rm Tr}}$  & $f/f^{\rm Tr}$ \\
\hline
Tr & 15 50 00.66 & -78 11 09.96 &  22.05 $\pm$ 0.06  (22.17) &      48    &        0     &   177   &       0      &      1                   &              1           &      1        & 1             \\
t1 & 15 50 07.33 & -78 10 57.29 &  24.40 $\pm$ 0.10  (24.59) &       46    &     -155     &   874   &     112      &     0.96                 &           4.94           &    0.92       & 0.115         \\
t2 & 15 50 06.24 & -78 10 58.00 &  24.54 $\pm$ 0.10  (24.54) &      160    &     -145     &  1064   &     230      &     3.33                 &           6.01           &    0.68       & 0.101         \\
t3 & 15 50 05.93 & -78 10 58.38 &  24.58 $\pm$ 0.49  (25.33) &      195    &     -179     &   434   &     197      &     4.06                 &           2.45           &    0.44       & 0.097         \\
t4 & 15 50 04.65 & -78 10 59.57 &  24.44 $\pm$ 0.59  (24.48) &       25    &     -126     &   453   &     176      &     0.52                 &           2.56           &    0.50       & 0.111         \\
t5 & 15 50 02.19 & -78 11 05.28 &  26.21 $\pm$ 0.19  (26.20) &       19    &      -59     &   105   &    -172      &     0.40                 &           0.59           &    0.18       & 0.022         \\
\hline
\end{tabular}
 \end{center}
\end{minipage}
\end{table*}

We find five candidate counterimages that meet these requirements and show them in Fig~\ref{Fig_Tr2Tr3}. For simplicity we name these candidates t1--t5, as indicated in the figure. We note that knots t1 through t4 correspond to the knots 5.4a through 5.4d in \cite{Pignataro2021}, which also identify these knots as being multiply lensed images of the same object, but does not link them to Tr. Knot t5 is not used in \cite{Pignataro2021}, and is not associated with the family of knots 5.4 in that work. Among these images, only t5 is not partially blended with the bright LyC knot 1. A lens model prediction for t5 is shown in appendix \ref{Appendix4}. A lens model prediction for the other candidates is not as reliable since they are significantly farther away from Tr than t5 (t1 is for instance 24" away from Tr vs 6.6" for t5), but like t5, they should fall between knots 1 and 2 of system 5. We do not identify additional candidates close to the remaining 7 LyC knots. This can be explained if the source responsible for Tr lies beyond the corresponding caustic for these knots (like in the case of knots e,f,i,l in Fig.~\ref{Fig_Data}), or because the images have smaller magnification factors resulting in a fully blended image with the LyC knot 1 (like in the case of knots m and n), or simply because they are too faint to be detected (i.e small magnification factors, as predicted also for images m and n). 
Close inspection of t5 shows a larger than expected distance to the bright LyC knot g, casting some doubt on the association of this knot with Tr. Nevertheless, we keep t5 in our list of possible candidates, noting that if it turns out to be a false positive it would imply the magnification derived below for Tr is even larger than when considering t5 a viable counterimage of Tr. 

Adopting these candidates as possible counterimages of Tr, we can constrain the magnification of Tr based on flux ratio arguments. All five candidates are more than 2 magnitudes fainter than Tr. Smooth lens models predict t5 (which is only 6.6" away from Tr) to have very similar flux, but t5 is observed more than 4 magnitudes fainter than Tr. 

Based on the 5 positions t1--t5, we compute the magnification and time delays in those positions. The values of the magnification and time delay  are listed in table \ref{tab_3} for the two lens models. For convenience we list the time delays (in days) relative to the time of arrival of photons at Tr. Column 3 lists the magnitudes observed at these positions. Due to the partial blending of t1--t5 with nearby knots, for these positions we fit a point source convolved by our PSF model to the nearby bright knot (see appendix \ref{Appendix5}), and subtract it before fitting our PSF model to the positions t1--t5.  Columns 4 and 5 show the magnification and time delay predicted by the model $M_1$, which corresponds to the model that uses only the strong lensing constraints (i.e the position of the lensed knots from \cite{Pignataro2021}).
Column 6 and 7 list the values predicted by the model $M_2$, or lens model derived with both types of constraints, the strong lensing and critical point positions.

From table~\ref{tab_3}, comparing the observed magnitudes (listed in the third column), with the predicted magnifications it is obvious that there is a contradiction between the observation and the models, since the latter do not predict Tr as the brightest image, while the observations clearly indicate that Tr is the brightest image. This tension hints at a missing ingredient in the lens model.  
The discrepancy is also evident when looking at the magnification ratios between the predicted magnifications at positions t1--t5 and at the position Tr (columns 8 and 9). Lens model $M_1$ predicts that the brightest image should be t3, while model $M_2$ predicts that the brightest image should be t2. 

We can compare the predicted magnification ratios with the values obtained directly from the data. We infer this ratio in two alternative ways. 
First we compute the separation between the pair of knots 1 and 2 bracketing Tr at the five positions t1--t5. This separation correlates with the underlying magnification so it can be used as a proxy for the magnification at each location. This is a purely geometric estimator and gives an idea of the average magnification between knots 1 and 2 in the six positions listed in table \ref{tab_3}. This estimator is unaffected by microlensing since it does not rely on measured fluxes. In addition, this estimator does not rely on the need for t1--t5 to be real counterimages of Tr, since it does not use the fluxes at these positions.  The magnification ratio derived this way is listed in column 10 of this table, $\mu_d/\mu_d^{{\rm Tr}}$. A more direct approach is by estimating the flux at each of these six positions and then computing the flux ratio. This gives us a direct measurement of the flux ratio in a model-independent way, but is affected by deblending of some of the images t1--t5, and obviously it would be a biased estimator if t1--t5 are not real counterimages of Tr. The result of this second approach is shown in the last column of table \ref{tab_3}, $f/f^{\rm Tr}$.

By construction, both $\mu_d/\mu_d^{{\rm Tr}}$ and $f/f^{\rm Tr}$ must be 1 at Tr.


If we focus on the geometric estimator of the magnification ratio listed in column 10, $\mu_d/\mu_d^{{\rm Tr}}$, that does not require t1--t5 to be counterimages of Tr, additional counterimages of Tr should be easily identifiable at the expected positions t1 through t4 in table~\ref{tab_3}, and even in t5 with magnitude $\approx 24$. 
If we focus on the flux ratios given in the last column, we observe a clear tension between these flux ratios and the ones inferred from the geometric estimator. The flux at Tr is much larger than expected from the simple geometric estimator. 
The magnification at Tr must be $\approx$ 10 times larger than the magnification at any of the positions t1--t5.  Any plausible lens model has to boost the magnification locally around Tr, in order to significantly increase the magnification at that location with respect to the simple ratio provided by the geometric estimator. This simple test suggests the need for a small scale perturber around Tr that can deliver the required local boost in magnification, without affecting the geometry of the arc.

In table~\ref{tab_3} we provide also time delay estimates at the positions t1--t5. These time delays are relative to the time of arrival of EM radiation at Tr and are expressed in days. Model $M_1$ predicts that the image at Tr arrives last, but the maximum time delay is less than half a year. Since Tr has been observed for $\approx 7$ years, time delays cannot be the explanation for the lack of counterimages. The same logic applies to model $M_2$, with the only difference that Tr is the second image to arrive and the maximum time delay is about two months longer than in model $M_1$. 

 Given the fact that time delays cannot explain the lack of counterimages, and the arguments given above about the anomalous flux at Tr, we can only conclude that a  small scale distortion in lens potential is responsible for the anomalously large flux of Tr. Before we come back to this point in section~\ref{sec_perturber}, in the next section we provide an alternative estimation for the magnification of Tr, based on the assumption that t1--t5 are counterimages, and the lens model predictions.

\section{Inferred magnification of Tr from t1--t5}\label{Sect_MuAtTr}
We measure the fluxes in each of the positions t1--t5 and Tr by fitting the PSF model described in section \ref{Appendix5} to the different positions. For the case of t1--t5, we first fit the flux of the nearby bright knot 1, and subtract it from the data before estimating the flux at t1--t5. The derived magnitudes are listed in table \ref{tab_3}. 

Based on the observed flux of Tr, and the fluxes of the 5 candidate counterimages, we can estimate the relative magnification between Tr, and the 5 counterimage candidates. 
Then, based on the lens model predicted magnifications, we can infer the delensed flux of the source (from t1--t5). Finally, from the ratio of the observed flux at Tr and the delensed flux estimates, we can infer the magnification needed to explain the observed flux at Tr. 

\begin{figure} 
   \includegraphics[width=9cm]{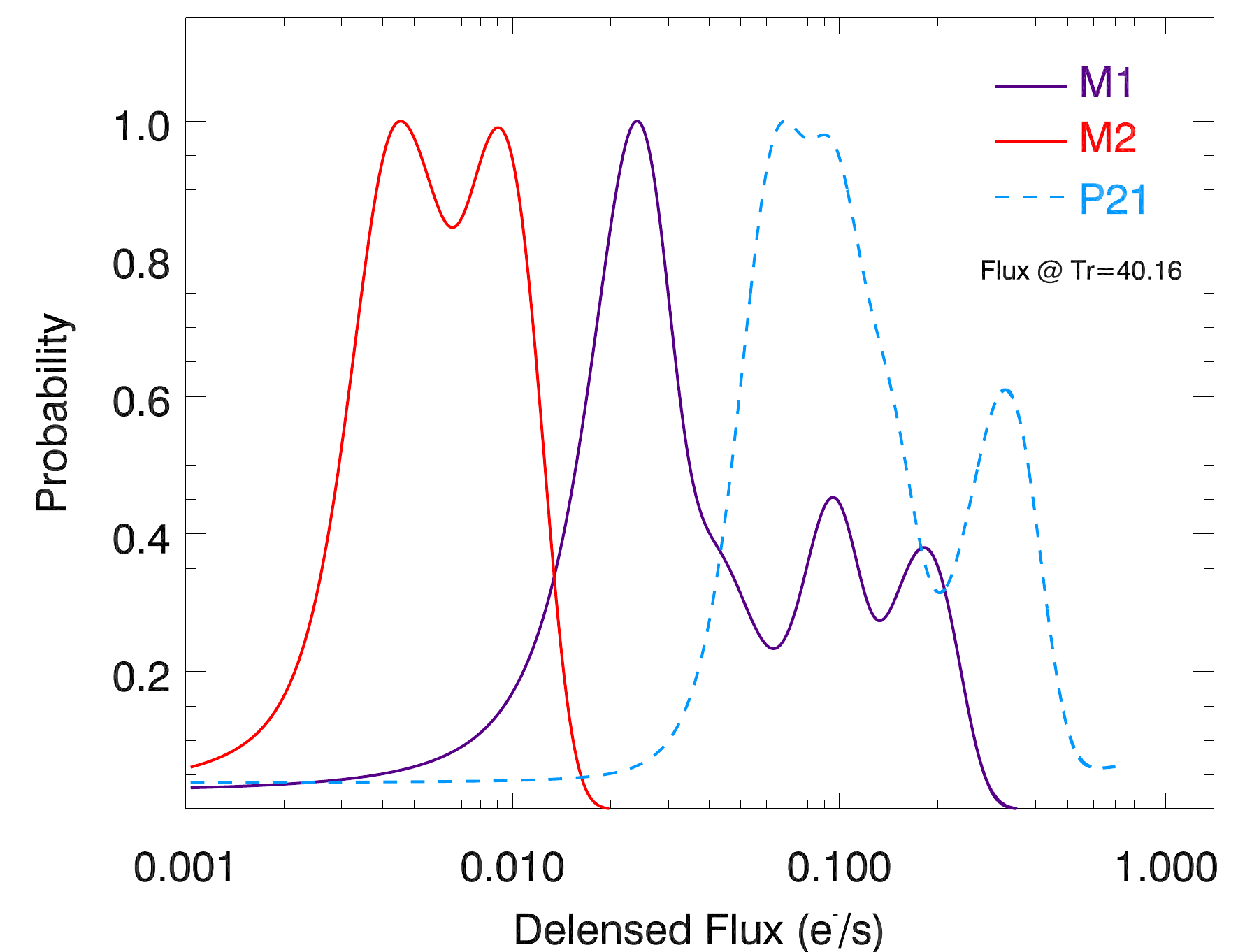}
      \caption{Inferred delensed flux of Tr in the source plane, in internal HST units based on the measured flux at the t1--t5 positions. In 
              these units, the observed flux at Tr is 40.16 in the HST F606W band. If we adopt as the delensed flux the maximum of the probabilities, this implies a magnification factor at Tr of $\approx 1700$ for model $M_1$ and $\approx$ 4000--7000 for model $M_2$. Using the magnification in \cite{Pignataro2021} we infer a magnification of $\approx 600$ at Tr. }
         \label{Fig_DelensedFlux}
\end{figure}

As mentioned earlier, this method will give us only a lower limit for the magnification of Tr, since it is based on the assumption that t1--t5 are real counterimages of Tr. If this assumption is proven to be wrong with future data (for instance by identifying spectral features at t1--t5 that are different than those observed in Tr), then the counterimages remain undetected and must have even smaller fluxes, implying an ever larger magnification for Tr. 
Based on the observed fluxes in positions t1--t5, and the magnification predicted by the two lens models, we can infer the delensed flux. Since we have 5 different estimations of the delensed flux, we combine them assigning a Gaussian distribution to each estimation, and add the Gaussian distributions (i.e we adopt the sum rule of probabilities; $P(A+B)=P(A)+P(B)$ since the delensed flux must be unique). The probability of the delensed flux is then given by;
\begin{equation}
P(f) = \sum_i exp\left(\frac{(f-f_i/\mu_i)^2}{2\sigma_i^2}\right)
\end{equation}
where the index $i$ runs from 1 to 5, $f_i$ is the observed flux for each one of the 5 candidate counterimages. For $\sigma_i$ we take three times the error in the measured flux divided by $\mu_i$. The magnification $\mu_i$ in the above equation is the corresponding value at each position, as predicted by each lensing model. The previous equation can also be interpreted as a weighted median value for the delensed flux. The combined probability for the delensed flux is shown in Fig.~\ref{Fig_DelensedFlux}. The blue curve shows the probability for model $M_1$, while the red curve shows the corresponding probability for model $M_2$. The predicted delensed fluxes in model $M_2$ are naturally smaller than in model $M_1$, owing to the larger predicted magnification. Using the values of the flux at the peak of the probabilities and the observed value of the flux at the position of Tr, we can directly estimate the magnification at Tr for both models, finding $\mu\approx1700$ for model $M_1$, and $\mu\approx$ 4000--7000 for model $M_2$. 
Since in \cite{Pignataro2021}, our positions t1--t4 correspond to their positions 5.4a--5.4d, we can repeat the process using the magnification values at those positions to infer what would be the magnification at Tr, assuming 5.4a--5.4d are counterimages of Tr. In this case we find that since the magnification predicted by the model in \cite{Pignataro2021} is in general smaller than the values from our two models, the inferred magnification at Tr is also smaller. In particular we derive a value of $\mu \approx 600$ at Tr for their model. 
The above estimations are derived using the photometric measurements listed in Table~\ref{tab_3} for the filter F606W. If we use instead the values derived from the F390W filter (values in parenthesis), the inferred magnifications are very similar; $\mu\approx 1400$ for $M_1$, $\mu\approx 8000$ for $M_2$ and $\mu\approx 600$ for the model of \cite{Pignataro2021}.

In summary, if the t1--t5 are real counterimages of Tr, we must conclude that the magnification at Tr must be at least $\approx 600$, adopting the most conservative case scenario model from \cite{Pignataro2021}, but could be as high as $\mu\approx 7000$ if we adopt the most optimistic model.  These magnification factors translate into a gain of 7 to 9.6 magnitudes at Tr, which then would have an absolute magnitude between $M_v\approx -17$ and $M_v\approx -14.3$.

It is important to reiterate that using t1--t5 as candidate counterimages of Tr results on lower limits for the magnification of Tr. Should future observations rule out the possibility of any of these candidates being counterimages of Tr, the flux ratio between Tr and any other possible counterimage (and hence the magnification of Tr) will be larger than the one derived in this work under the assumption that any of the 5 candidates is a real counterimage of Tr.  This would only accentuate the need for a small scale perturber to explain the even larger magnification at Tr. 

Before considering a possible small scale perturber to explain the extreme magnification at Tr, we need to contemplate the possibility that small variations in the large scale potential of the cluster could increase the magnification at Tr. Since the critical curves from the lens model are relatively close to Tr, it is possible that a different configuration in the parameters of the reconstruction algorithm naturally explains the needed magnification, for instance by producing a critical curve passing through Tr (a perfectly aligned critical curve would produce a pair of unresolved images with very large magnification factors). We explore this possibility in the next section. 

\section{Uncertainty in the position of the critical curves near Tr}\label{sec_Transient}
Given the proximity of Tr to the critical curve, small changes in the lens model can result in big changes in the magnification. 

\begin{figure} 
   \includegraphics[width=9cm]{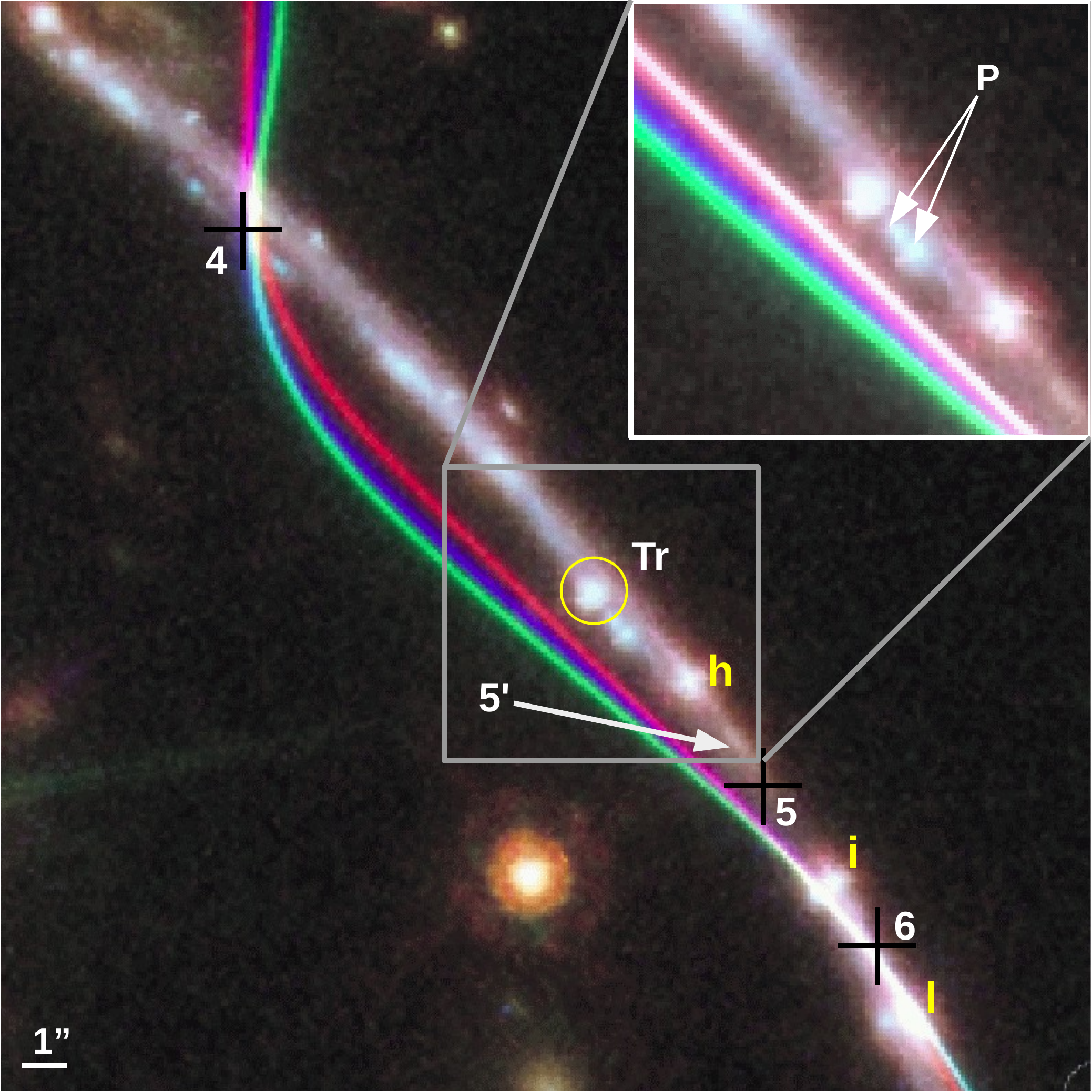}
      \caption{Variability of the critical curves near the position Tr. The blue critical curve corresponds to the same model shown in blue in Figure \ref{Fig_Data}. The red curve is for a model derived with the same configuration but with half the number of iterations. The green curve is for an alternative model for the same number of iterations as the blue curve, but with a different realization of the grid. The crosses mark the position of the critical points used as constraints. Knot number 1 for system 5 is marked with yellow labels (h,i,l). The white arrow marks a possible alternative location (5$^\prime$) for the critical point 5. Using this position 5$^\prime$ as a constraint instead of position 5, brings the critical curve a bit closer to the position Tr (white curve in the inset in the top-right corner).  
              }
         \label{Fig_Tr_vs_CC}
\end{figure}

We explore the uncertainty in the separation between Tr and the critical curve by varying some of the parameters in WSLAP+. Since there is some freedom in the number of iterations and the definition of the grid component, we study how the solution depends on the choices made for these two configurations. We take as a reference the solution obtained when both knot positions and critical point positions are used as constraints. This model corresponds to the blue critical curve in Figure \ref{Fig_Data}. We construct a different random realization of the grid with a similar number of grid points (185 grid points instead of the 177 considered in the reference model). The resulting model predicts a critical curve that is very similar to the one from the reference model, although with small deviations. We show the difference between this model and the reference model near the position Tr in figure \ref{Fig_Tr_vs_CC}, where the reference model is shown as a blue curve and the new model with 185 grid points is shown as a green curve. For this alternative grid, the green curve moves 0.24" away from the position Tr. A different realization (not shown) but with a smaller grid points (146) reverses the shift and puts the critical curve 0.05" closer to Tr.

Next we change the number of iterations. As discussed in earlier work \citep{Diego2005,Diego2015} a solution with a larger number of iterations is not necessarily a better solution. For a larger the number of iterations  the residual in the lens equation gets smaller, but a solution with zero residual will always be biased with respect to the true underlying mass distribution, since the parameterization of the mass distribution (grid or smooth component plus compact component) can never capture all the details of the true mass distribution. Hence, a solution that predicts zero residual with an imperfect parameterization is in general more biased with respect to the true underlying mass distribution than a solution that allows for a small residual. In general good solutions are obtained with WSLAP+ when the distance between the predicted and observed positions is in the range 0.4"--0.7". Smaller separations can be obtained but often at the expense of introducing spurious fluctuations in the mass distribution.  The blue curve in figure \ref{Fig_Tr_vs_CC} is obtained with 500000 iterations. For comparison we show the solution for 250000 iterations as a red curve. We appreciate that the critical curve moves closer to the position of Tr. 

Interestingly, the distance between the critical curves and Tr is comparable to the error between the predicted and observed positions of critical point number 5 ($\approx$ 1", see figure \ref{Fig_Tr_vs_CC}). The parameterization of the lens plane does not have enough flexibility to reproduce this critical point well, but if the shape of the critical curves is correct, they should be displaced approximately 1" towards the arc in order to reproduce the critical point number 5. This shift would put the critical curve on top of Tr, offering a possible explanation for its nature, since then it could be interpreted for instance as a lensed star, such as Icarus \citep[at z=1.49][]{Kelly2018}. 

Close inspection of the arc around critical point 5, suggest that instead of the middle point between knots considered for this critical point, an alternative position for this critical point may be possible. We mark this alternative position (5$^\prime$) with a white arrow in figure \ref{Fig_Tr_vs_CC}. This new position is based on the fact that the bright feature marked with the white arrow is not seen on the other side of critical point 5. Hence, the critical curve is possibly going through this bright feature. We derive a new lens model using critical position 5$^\prime$ instead of 5, and the same configuration as in the red curve (i.e 250000 iterations and grid with 177 points). The resulting model is very similar to the one obtained with critical point 5, and is shown as a white critical curve in the inset of figure \ref{Fig_Tr_vs_CC}). 

Also intriguing is the fact that the parities of knots "i" and "l" (marked in yellow) are not well predicted for any of these models. Knot h is robustly predicted with positive parity. Given the separation between knots "i" and "l", they should have negative and positive parities respectively. This again indicates a lack of flexibility in the lens model around this position that should predict a more arched critical curve between knots "i" and "l".  The model of \cite{Pignataro2021} achieves this by setting a relatively large mass, and large ellipticity to a barred galaxy a few arcseconds south from these knots (see figure \ref{Fig_Data}). Our model does assign a smaller mass (ellipticity is fixed by the distribution of light) to this galaxy which could explain our smoother critical curve around this position. 

Even in the hypothetical case that a smooth lens model produces a critical curve right at the position of Tr, this would pose another problem since immediately a counterimage of knot h in figure~\ref{Fig_Tr_vs_CC} should be expected on the opposite side, and nearly at the same distance from Tr. Figure~\ref{Fig_Tr_vs_CC} clearly shows that this counterimage does not exist and that Tr is not a symmetry point, ruling out directly the possibility of a cluster scale critical curve passing through Tr. Given the impossibility of a smooth model to explain Tr, we consider in the next section a small scale perturbation that provides the answer we are seeking to the conundrum of Tr. 

\section{A perturber to explain Tr}\label{sec_perturber}
As discussed earlier, a critical curve passing through Tr and explaining its large magnification would produce an additional counterimage of knot h in system 5 on the opposite side of Tr, nearly equidistant to it, and similar in flux. Such an additional counterimage is not observed, nor any symmetric features, hence ruling out this possibility. However, one can achieve the extreme magnification needed for Tr, while at the same time avoiding additional counterimages of h, if a small perturber is placed near the position Tr. Proximity to the critical curve guarantees that a small perturber can fall below the detection threshold of HST, while still being able to produce a gravitational lensing effect strong enough to amplify the object at Tr to the needed values (the effective lensing mass of the perturber scales as the macromodel magnification, while the magnification of the cluster does not affect the observed flux from the perturber since the perturber is at the same redshift as the cluster).   \\

We consider the case of a small perturber, which for simplicity we parameterize as circularly symmetric, and with a mass ${\rm M}_E$ inside its Einstein radius. First we study the simpler but pedagogical case where the perturber and the source are perfectly aligned, producing maximum magnification. In this scenario the image forms a perfect Einstein ring, although in reality this is not possible since the shear from the cluster stretches the critical curves from a circular shape into an hour glass shape, as can be appreciated for instance around some of the galaxies near the critical curve in figure \ref{Fig_Data}.  However, this basic approximation will let us do some simple calculations that should be accurate within a factor of a few, and will give us a useful constraint on the minimum mass of the perturber. 
Exploiting the fact that Tr is unresolved, we can then constrain the mass inside the hypothetical Einstein ring, and hence the mass of the perturber. In order to do this we need to take into account the role played by the cluster, since the effective mass of the perturber is amplified by the magnification, $\mu_c$, from the cluster, i.e. $M_\mathrm{eff} \approx \mu_c*M_\mathrm{pert}$.

At the redshift of the cluster and source, the Einstein ring radius scales with the effective mass as 
\begin{equation}
    \theta _e(") = 0.11 \sqrt{ \frac{ M_\mathrm{eff} } { 10^{10} {\rm M}_{\odot} } }\,.
\end{equation}
As shown in appendix \ref{Appendix5}, the maximum diameter of an Einstein ring in order to not be resolved in the HST images is $\approx 30$ mas. 
Hence, adopting a maximum Einstein ring radius of 0.015", we get that the effective mass must be less than $M_\mathrm{eff} < 2\times10^8\, {\rm M}_{\odot}$, which for a conservative limit of $\mu_c \approx 50$ results in $M_\mathrm{pert} < 4\times10^6 \,{\rm M}_{\odot}$ inside its Einstein radius. 

If we consider a perturber with this mass, $M_\mathrm{pert}=4\times10^6\,{\rm M}_{\odot}$, it would form an Einstein ring of $\theta _e(") \approx 0.015"$ which would not be resolved by HST. The magnification is then 
   $\mu= [({\rm R}_e +r_s)^2-({\rm R}_e -r_s)^2]/r_s^2=4{\rm R}_E/r_s$ 
where $r_s$ is the radius of the source, and we have assumed the lensed source forms an Einstein ring with a thickness similar to the thickness of the source. For $r_s=0.01$ parsec, we get $\mu\approx 3400$, a value between the magnifications from models $M_1$ and $M_2$ predicted in section \ref{Sect_MuAtTr}.

\begin{figure} 
   \includegraphics[width=9cm]{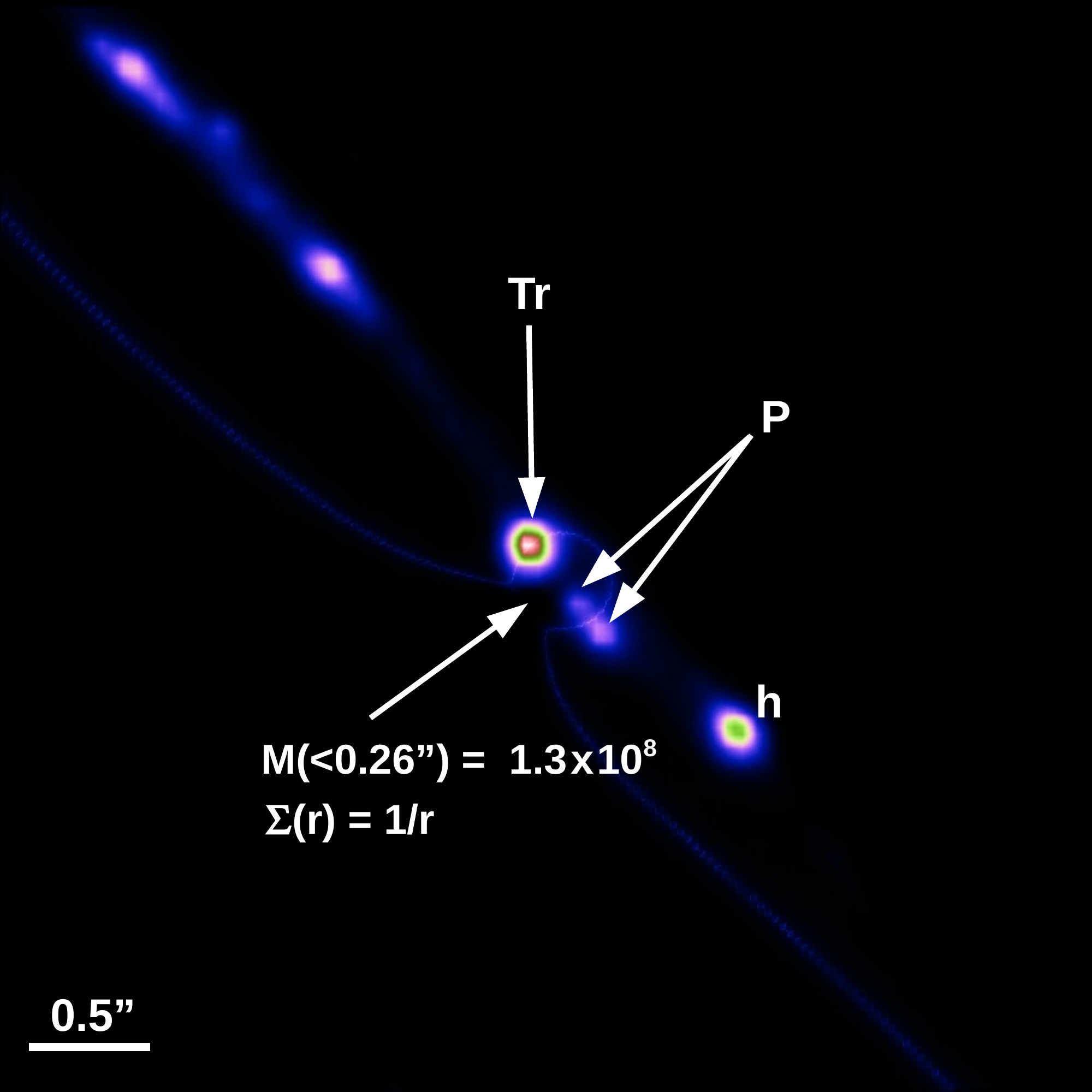}
      \caption{Modification of the critical curve near Tr by a nearby small perturber. This perturber is modeled as a spherical halo with surface mass density profile $\Sigma \propto r^{-1}$ and a small core of 50 parsec. The halo is displaced 0.26" South of Tr. With an enclosed mass of $1.38\times10^8 \,{\rm M}_{\odot}$ within this radius, this perturber is able to simultaneously explain the extreme magnification at Tr and the pair of knots between Tr and knot h of system 5.     
              }
         \label{Fig_Tr_vs_CC2}
\end{figure}

The discussion above is based on the hypothesis that the perturber is perfectly aligned with the position Tr. We have also ignored the fact that the cluster shear breaks the circular symmetry of the Einstein ring around the perturber.  
In the more realistic scenario where Tr and perturber are not perfectly aligned, and the cluster breaks the circular symmetry of the Einstein ring, we would expect a smaller magnification for the same mass of the perturber. 
A perturber that is farther away from Tr, but with a larger mass, can still produce the magnification required to explain Tr, but not in an Einstein ring configuration. In this case, the Einstein ring would be larger than the resolving power of HST, and we would expect to see symmetric features around the perturber's critical curve. 

Interestingly, an additional pair of fainter knots ($\approx$ 1 magnitudes fainter) are found at $\approx 0.35"$ and $\approx 0.5"$ from Tr in the SW direction (this pair is  marked with white arrows and the label P in the inset of figure \ref{Fig_Tr_vs_CC}). From now on we refer to this pair of knots as the "P knots". These P knots are not discussed in earlier work and like Tr, have no recognizable counterparts anywhere else in the lens plane. Although we can not reject the possibility that it is at different redshift, the P knots overlap perfectly with the sunburst arc, and they follow the geometry of the arc, suggesting that they could also be parts of the same galaxy, and are being strongly lensed. We then consider the possibility that a critical curve from a perturber passes through both Tr and the middle point of the P knots. In this case, the mass of the perturber would be larger than the mass considered above ($M_\mathrm{pert} \approx 4\times10^6\, {\rm M}_{\odot}$). Based on the configuration of the P knots, they can not be counterparts of Tr since the maximum magnification is observed at the position Tr, which would correspond to a double image. Since the P knots are already forming a double image, the third image must be fainter, unlike Tr which is brighter. Also, the P knots are bluer than Tr suggesting a different origin.  Hence we conclude that the P knots are due to a different source than Tr, although close to it in the source plane. 
In order to test the hypothesis that a small perturber can explain both the extreme magnification of Tr and the configuration of the P knots, we place a small perturber with circular symmetry 0.26" south from Tr, and add its potential to the WSLAP+ lens model solution. We adopt a $r^{-1}$ density profile for simplicity and find that a mass of $1.38\times10^8 \,{\rm M}_{\odot}$ enclosed within the radius of 0.26" produces a critical curve (in conjunction with the deflection field from our lens model for the cluster) passing through Tr, as well as in between the P knots (see Figure~\ref{Fig_Tr_vs_CC2}). This small perturber would explain the extreme magnification of Tr as an unresolved double image, as well as the P knots as a double image. The third image from Tr would be somewhere else along the giant arc but with a much smaller magnification and possibly overlapping with other bright features in the arc (hence unobserved). The same reasoning applies to the third image of the P knots. The source responsible for the P knots is by itself interesting 
but without spectroscopic confirmation of the P knots, we can only speculate about its nature. Other configurations for the perturber are possible, like for instance a different slope, position or mass, but a detailed modelling of the possible perturber is beyond the scope of this paper, given the limited number of lensing constraints available.  
The inferred mass for the perturber is comparable to the mass of a dwarf galaxy. These galaxies are faint and would not be detected in the current HST images, so it is very possible that the perturber corresponds simply to a dwarf galaxy in the cluster. Similar dwarfs are long known to inhabit nearby clusters like Virgo or Coma \citep{Sandage1984,Thompson1993}. In these dense environments, and due to their relatively weak gravitational potential, dwarf galaxies suffer from quenching of star formation via ram pressure stripping and galaxy-galaxy interactions,  reducing even further their brightness \citep{Rude2020}. 

The existence of a perturber with a mass $\approx 1.4\times10^8 \,{\rm M}_{\odot}$ in a cluster environment is interesting and can be used to constrain dark matter (or DM) models. Before exploring in more detail the source responsible for Tr, in the next section we briefly discuss the implications  that the existence of small-scale halos, like the one discussed in this section, have for DM studies.

\section{Implications for dark matter models}\label{Sec_DM}

In standard $\Lambda$CDM cosmology, dark matter halos are expected to exist down to Earth-mass scales, $M \sim 10^{-6}\,M_\odot$~\citep{Zybin:1999ic,Hofmann:2001bi,Berezinsky:2007qu}, while the threshold halo mass for galaxy formation is on the order $10^8\,M_\odot$~\cite{Nadler:2018iux}. 
Between these scales may exist a wealth of cold invisible substructure.
However, the properties and interactions of the DM particle may suppress such structure on small scales. For example, warm DM is produced with a substantial velocity, giving it a large free-streaming length and leading to a cut-off in the matter power spectrum~\citep{Hogan:2000bv}. Alternatively, strong interactions between cold DM and baryons~\citep{Nadler:2019zrb,Buen-Abad:2021mvc} 
can dampen the growth of structure on small scales. Strong lensing provides a unique opportunity to search for DM substructure and therefore probe the temperature and interactions of DM~\citep{Coogan:2020yux,Ostdiek:2020cqz,Wagner-Carena:2022mrn}. We now estimate the possible constraints which could be derived if the presence of a DM perturber with mass $M_\mathrm{pert} = 1.38\times10^8\,M_\odot$ is confirmed.

Constraints on warm dark matter (WDM) can be derived in terms of the `half-mode mass' $M_\mathrm{hm}$, the mass at which the transfer function relating the WDM and CDM power spectra drops to a value of 0.5. Power on scales $M < M_\mathrm{hm}$ is substantially suppressed by the effects of DM free-streaming. Using the WDM transfer function from~\cite{2012MNRAS.424..684S} and equating $M_\mathrm{hm} = M_\mathrm{pert}$ leads to a constraint on the WDM mass of $m_\mathrm{WDM} \gtrsim 4.2 \,\mathrm{keV}$. This would be slightly weaker than (but comparable to) competing constraints from the Lyman-$\alpha$ forest, Milky Way satellites and other lensing studies~\citep{Irsic:2017ixq,Gilman:2019nap,Enzi:2020ieg}.


In the context of DM-proton scattering, 
\cite{Nadler:2019zrb} estimate the minimum mass $M_\mathrm{crit}$ of DM halos which can form, for which the initial perturbation is not erased by this scattering process. Requiring $M_\mathrm{crit} < M_\mathrm{pert}$ would exclude DM-proton scattering cross sections greater than $\sigma_p = 1.4 \times 10^{-28}\,\mathrm{cm}^2$ for DM particles of mass $100\,\mathrm{MeV}/c^2$~\citep{Nadler:2019zrb}. This would be competitive with similar constraints coming from observations of Milky Way satellites~\citep{Nadler:2019zrb} and complementary to direct searches for DM in this mass range~\citep{EDELWEISS:2019vjv}.




An alternative to a perturber is to consider a model of DM that naturally produces anomalous flux fluctuations near critical curves, as in wave dark matter (or $\psi$DM)  \citep{Schive2016}. In this model the boson density is fully modulated by self-interference, creating pervasive substructure fluctuations in the distribution of dark matter. It is now understood that lensing is sensitive to this  substructure forming a highly structured critical curve with corrugations on the de Broglie scale, as this characteristic scale is preserved in projection \citep{Chan2020}.
This corrugation of the critical curve is predicted to be most convoluted in the case of clusters as the de Broglie scale is relatively small, generating a dense band of corrugations and "islands" where the magnification diverges, and that extend to arcsecond scales.
In this type of model, high magnifications up to $\mu < 10^4$ can be produced out to a fraction of an arcsecond from the critical curve of the equivalent smooth model, so that a high magnification solution for Tr is plausible in the context of $\psi$DM. 
A more detailed analysis of the constraints imposed on this type of model is beyond the scope of this paper but obtaining constraints on the boson mass via the de Broglie scale is conceivable using Tr, and other well-constrained cluster lenses where the intrinsic luminosity of the lensed source is independently estimated.

One of the limitations that prevents us from imposing tighter constraints on dark matter models is the depth and spatial resolution of the images. Future telescopes like the ELT, and the recently launched JWST, will soon improve the  image quality, allowing more detailed studies of Tr and other similar objects\footnote{A proposal to observe Tr with JWST has been already approved, GO 2555.}. JWST observations, in conjunction with HST archival data will soon be used to fit SED models to the photometric observations, narrowing down the possible candidates for Tr. 
Based on these data and detailed lens models, it will be possible to identify objects like Tr, whose observations demand the presence of small scale perturbers (or more generally small scale perturbations of the dark matter field). The accuracy in the mass estimation of these perturbations, as well as their abundance, will improve dramatically in the coming years. Better data will also allow us to reduce the minimum observable mass through this technique, resulting in competitive constraints on DM models.

\section{Lens model constraints on Tr}\label{sec_Constraints}
The nature of Tr remains unclear. The time delay and magnifications predicted by the lens model clearly indicate that if Tr is being magnified by factors of 100 or less, as suggested in earlier work, other counterimages should have been clearly observed. An object like an SN, with absolute magnitude $\approx -19$, and amplified by a minimum factor of 20 (a conservative limit for the lens model in positions near t1--t5), would have been easily observed in HST images. This is not the case, posing a serious problem for the SN interpretation. In order to identify a possible explanation for Tr, we discuss below the different constraints that can be derived from the lens models, and observations. 

\subsection{Time delays}
Assuming a magnification factor of order $\mu\sim 100$ for Tr (that is, a magnification boost of 5 magnitudes), a bright SN at z=2.37 with absolute magnitude of $\approx -19$ would have an apparent magnitude of $\approx 22.5$ (ignoring k-corrections and extinction). This hypothesis to explain Tr is discussed in \cite{Vanzella2020}. 
However, typical SN remain in the bright phase for a relatively short time of typically 1 month, with SN of type II being able to stay bright for several months. In \cite{Vanzella2020}, it is discussed how based on 2016 spectroscopic observations, and 2019 HST observations, Tr remains a bright source for at least 11.9 months in the rest frame (or 3.3 years in the observer frame, extending to over 5 years if one considers the most recent HST observations from June 2021, where Tr is still clearly visible). Visual inspection of NTT/EFOCS2 images taken in March 2014 and shown in \cite{Dahle2016}, reveal a bright unresolved source at the position of Tr  which would imply the source has been persistent for at least 7.2 years (1.93 years in the rest frame). If the SN interpretation is correct, this would make Tr an SN with an unusually long bright phase, although \cite{Vanzella2020} argues that SNe that interact with the circumstellar material can have long durations. 

The possibility that t1--t5 are counterimages of Tr would ease the tension with the SN interpretation and the time delay constraint, since the expected counterimages would be present in the image. However, this would still require an unusually long duration SN. In addition, given the constraint on the minimum magnification ($\mu > 600$), it would imply a relatively faint SN during the bright phase (absolute magnitude fainter than M$_V$=-17).


Tr is probably not a short lived event, but a long-duration event or even a stable source in terms of its flux (in timescale of decades). The SN interpretation is hence unlikely. 

\subsection{Luminosity of the Tr}
As discussed in section~\ref{Sect_MuAtTr}, if t1--t5 are counterimages of Tr, the magnification at Tr ranges from $\mu \approx 600$ to $\mu \approx 7000$. This directly translates into absolute magnitudes in the range $-14.8 < {\rm M}_V < -17.5$ (ignoring k-corrections or extinction). 

If t1--t5 are not counterimages of Tr, and no counterimages are observed elsewhere in the lens plane, this sets an upper limit on its absolute magnitude. HST images with exposure times of approximately 1 hour (similar to the observations of Tr) can reach magnitudes ${\rm M}_{\rm AB}\approx 27.5$ in $F814W$ \citep[see for instance][]{Coe2019}. Adopting as the minimum magnification for the counterimages of Tr a conservative limit of $\mu_{max}=50$ at positions near t1--t5 in  table \ref{tab_3} \citep[this magnification would be even larger in the model of][]{Pignataro2021}, we can infer that Tr must be fainter than $M_{abs} \approx -14.7$, otherwise it would be observed in the HST images in at least one of the t1--t5 positions (again, ignoring k-corrections or extinction). 

Finally, in \cite{Vanzella2021}, the absolute UV luminosity of t1 (5.4a in that work) is reported as $M_{UV}=-16.51 \pm 0.32$, and it is interpreted as a young stellar cluster with age $\gtrsim 7$ Myr, mass $3.6\times10^6 \,{\rm M}_{\odot}$, and radius less than 5.4 pc (only upper limits are quoted for the size since the images are unresolved). The inferred luminosity in  \cite{Vanzella2021} agrees well with the range given at the beginning of this subsection, and would imply that the magnification at Tr is $\mu \approx 1400$. Of course, this estimate is based on the lens model used in that work, and which corresponds to the model in \cite{Pignataro2021} that predicts a magnification of $56 \pm 26$ for t1. Also, it is based on the estimation of just one of the knots t1--t5. The estimate of the luminosity will vary when using other knots of the same 5.4 family. 

\subsection{Minimum magnification of Tr}
We have seen in section~\ref{Sect_MuAtTr} that if t1--t5 are counterimages of Tr, the minimum magnification predicted by the lens models is $\mu \approx 600$ (from the model in \cite{Pignataro2021}). \\

If t1--t5 are not counterimages of Tr, the minimum magnification must be larger, since it needs to account for a larger flux ratio between Tr and the other counterimages (that should have been visible within the time range covered by observations). If no counterimages of Tr are observed, the minimum magnification must be such that it boosts the flux at Tr at least $\approx 5.5$ magnitudes. That is the magnification must be 160 times stronger at Tr than at any other position in the Sunburst arc.  
Adopting a conservative lower limit of $\mu=20$ at any of the expected locations of the counterimages (these positions must be near t1--t5, or between knots 1 and 2 in system 5), this translates into a minimum magnification for Tr of $\mu=3200$. 

\subsection{Maximum magnification of Tr}
Smaller (but intrinsically fainter) sources like individual stars can be magnified to extreme factors owing to the small intrinsic size, since the maximum magnification scales with the inverse of $\sqrt{r}$, where $r$ is the radius of the source. 
At magnification factors of $10^5$,  a superluminous star with absolute magnitude $M_V\approx -9.5$ at $z=2.37$, would appear with a flux similar to the one observed at Tr. These large magnification factors can not be maintained over periods of years since the relative motion between the star and the caustic would eventually reduce the magnification significantly, unless the source is moving in a fine tuned direction parallel to the caustic, which is unlikely. Also, microlenses in the lens plane reduce the maximum magnification, making values above a few tens of thousands very unlikely.\\

\subsection{Maximum magnification in the presence of microlensing}
The ubiquitous presence of microlenses from the cluster intracluster medium results in  microcaustics in the source plane. The number density of microcaustics grows with the magnification from the cluster \citep{Venumadhav2017,Diego2018}. The effective optical depth can be approximated by $\tau_{\rm eff}=\kappa_{\rm ml}\,\mu_c$, where $\kappa_{\rm ml}$ is the coarse-grained surface density of microlenses divided by the critical surface mass density and $\mu_c$ is the cluster magnification factor \citep[see for instance][]{Diego2018}. While the abundance of intracluster stars intervening the line of sight toward Tr is yet to be measured, typical values for $\kappa_{\rm ml}$ found at the cluster Einstein radius range between $10^{-3}$ to $10^{-2}$. Hence, for values of $\mu_c>10^3$ the effective optical depth exceeds unity. In this situation, Tr's motion across a network of micro caustics on the source plane should result in a variable flux. Tr is bright enough in the rest-frame FUV that flux variations at $\sim 5$--$10\,\%$ level should be easily detectable. Non-detection of flux variation would be explained by a value $\tau_{\rm eff}<1$ (i.e. small values for $\kappa_{\rm ml}$, or $\mu_c$, or both). 
If $\tau_{\rm eff} \gg 1$, flux variations can be small. This is known as the "more-is-less" effect, and is analogous to the effect studied in \cite{Dai2021}. In this regime, a high density of micro caustics overlap at any position in the source plane. Crossing one of these microcaustics results in a relatively small change in the total flux, which sums over a large number of micro images.

Random ray deflections due to a population of microlenses ``smooth out'' a sharp macro caustic into a structure of a finite thickness where the {\it persistent} magnification plateaus~\citep{Venumadhav2017}, as if the source has a larger, effective angular size~\citep{2021arXiv210412009D} $\sim \kappa^{1/2}_{\rm ml}\,\theta_{\rm ml} = 0.03$--$0.1\,\mu$as, where $\theta_{\rm ml}\approx 1\,\mu$as is the angular Einstein radius corresponding to the typical main sequence dwarf star $\sim 0.3\,M_\odot$. This translates to a physical scale of $\gtrsim 50$--$170\,$AU for the effective source size. High magnification values that would have to be achieved at source positions closer to a sharp cluster caustic than this ``smooth out'' length scale are not realized in the presence of intracluster microlenses. This limitation on the highest possible magnification is estimated to be a few times $10^4$ for typical strengths of a cluster caustic.

If the source is situated further from the macro caustic than the ``smooth out'' scale, its persistent magnification (i.e. averaged over timescales longer than that of possible microlensing-induced variability) is unaffected by intracluster stars. Nonetheless, flaring events associated with micro caustic crossing can still arise; an estimated peak magnification is $\mu_{\rm pk} \sim (0.4$--$2)\times 10^4\,(R_S/{\rm AU})^{-1/2}$, for $\kappa_{\rm ml}=10^{-3}$--$10^{-2}$ and typical parameters of cluster macro caustic.


\subsection{Maximum size of Tr}\label{sec_size}
The fact that Tr is not resolved is another useful clue. The morphology of the Sunburst arc in the different counterimages shows a similar thickness of $\approx 0.4"$--$0.5"$ or equivalently $\approx$ 2--3 kpc. This size is typical for galaxies at $z>2$, suggesting that in all the counterimages the radial magnification, $\mu_r$, is close to 1. We also recall that \cite{Vanzella2021} constrained the size of the source to be $\approx 3$ kpc$^2$.
Hence most of the magnification can be attributed to the tangential component, $\mu_t$, as expected in giant arcs forming near the Einstein radius of the lens. This is confirmed by our lens model that predicts $\mu_r=1.38$ and  $\mu_r=2.08$ at Tr  for model $M_1$ and model $M_2$ respectively. 
If we adopt the most conservative value of the magnification from the previous subsection (i.e $\mu \approx 600$ from the model in \cite{Pignataro2021}),  the tangential magnification is $\approx 300$, which combined with the fact that Tr is unresolved, constrains the size of Tr to be less than 0.4 parsec. This constraint is obtained after assuming that any separation between two point sources larger than 0.03" is resolved in the HST images in the F606W band, and hence the separation to the critical curve can not be larger than 0.015" (see appendix \ref{Appendix5}).
If instead we assume the minimum magnification derived under the assumption that t1--t5 are not counterimages, i.e., $\mu_{min}=2000$, then the size of the source must be less than $\approx 0.06$ parsec.

\section{Godzilla, an extremely luminous and magnified star at z=2.37.}\label{sec_Godzilla}
The above constraints on persistence in time, luminosity, and size reduces the number of candidates to a small number.  
Transient luminous objects, like classic SN or other short lived events ($<$ 2 years) can be ruled out given the fact that the image has remained bright for $\approx 7$ years and time delays between the expected position of the brightest counterimages must be smaller than 1 year. A bright and compact globular cluster or star forming region could be bright enough to explain the flux of Tr if one assumes a magnification $\mu>600$, but it does not satisfy the constraint on the size, since it would require the flux to be contained in a region smaller than 0.45 parsec. 

Given the constraints on the maximum size, minimum luminosity, and duration of the event, the number of possible candidates is very small. We contemplate two possibilities;  i) an accretion disc around an intermediate mass black hole (IMBH), and ii) a hyperluminous star. \\

\subsection{Intermediate mass black hole}
First we consider a small accretion disc around a black hole. Since the luminosity of the accretion disk grows with the mass of the black hole at its centre, one could in principle have an object that is luminous enough, but contained in a region that is sufficiently small. Accretion discs can be luminous over extended periods of time, easily meeting the constraint on persistence of the source. 

The bolometric luminosity of an accretion disc radiating at the Eddington limit scales with the mass of the BH as \cite{Rybicki1979,MillerColbert2004}
\begin{equation}
    L_{\odot} = 1.3\times10^{38} \left( \frac{M}{\odot}\right) {\rm erg s}^{-1}= 3.38\times10^4\frac{{\rm M}_{BH}}{M_{\odot}}\,. L_{\odot}
\end{equation}
An intermediate mass black hole (IMBH) at $z=2.37$, with mass $\approx 10^5 \,{\rm M}_{\odot}$ would be luminous enough to be observed with magnitude F606W$\approx22$, provided the magnification is $\approx 5\times 10^3$. The part of the energy that is radiated in the UV and optical bands is probably a small fraction of this energy. Some of the energy radiated as X-rays or UV is expected to be reprocessed into the observed optical bands, especially if there is circumstellar material around the source, as suggested by the Bowen fluorescence discussed in \cite{Vanzella2020}. If we assume that 10\% of the bolometric flux is radiated in the UV-optical part of the spectrum, we can compensate for this by increasing the mass of the black hole by a corresponding factor 10. Smaller fractions would translate to correspondingly higher masses. Since the IMBH is simply a working hypothesis, and it is beyond the scope of this paper to constrain with accuracy the possible mass of the IMBH, we simply consider a BH with mass $\approx 10^6 {\rm M}_{\odot}$ in order to explain the observed flux in the UV-optical bands.   

Since accretion discs of IMBH can be significantly larger than a star, we can constrain the maximum mass of the IMBH (and hence the minimum magnification) by using simple scaling relations. 
The half-light radius of an accreting disc, for a given  wavelength $\lambda$,  scales as \citep{Blackburne2011}
\begin{equation}
    r_{1/2}=1.68\times10^{16} {\rm cm}\,\left( \frac{{\rm M}_{\rm BH}}{10^9 {\rm M}_{\odot}} \right)^{2/3}
    \left( \frac{\lambda}{\mu {\rm m}}\right)^{4/3} \,,
\end{equation}
where we have assumed the disc is emitting at the Eddington limit. 
Since the most accurate constraints on the size are given by the HST observations in the UV filters (where Tr remains unresolved), we considered the typical rest frame equivalent wavelength of these filters, which is $\lambda \approx 0.1 \,\mu{\rm m}$. For this wavelength and a mass of $10^6 {\rm M}_{\odot}$, we find $r_{1/2} \approx 1$ AU. Hence the accretion disc of an IMBH with mass $10^6 \,{\rm M}_{\odot}$, in the UV part of the spectrum, would be much smaller than the size constraint found in section \ref{sec_size} for $\mu=5\times10^3$. The mass of the IMBH could be larger by several orders of magnitude and still satisfy the constraint on the maximum size. In this case, since the luminosity would increase by a similar factor, the magnification can be reduced accordingly. As discussed earlier, the magnification at Tr must be at least $\mu > 600$ in order to explain the flux ratio between Tr and the t1--t5 alleged counterimages (and even larger magnification if t1--t5 are not counterimages of Tr), so under the assumptions above, the mass of the possible IMBH should be less than $\sim 10^7 {\rm M}_{\odot}$ in order to fit the observed flux. 

{\it X-ray emission and Ly-$\alpha$:} 
One of the characteristic features of accretion discs is their X-ray emission. Chandra data acquired in 2020 (40 ks, PI M. Bayliss, Obs ID 20442) reveals no source at the position of Tr. Even with no detection, and given the extreme magnification factors considered in this work, we should check if an IMBH with ${\rm M}_{\rm BH} =10^5 \,{\rm M}_{\odot}$ and $\mu=5\times10^3$ is consistent with no detection. Following \cite{Mayers2018}, we estimate that such an IMBH would have an X-ray luminosity of $L_{\rm X[2-10] keV} < 10^{40}$ ergs s$^{-1}$. After magnification this translates into  $L_{\rm X[2-10] keV} < 5\times 10^{43}$ ergs s$^{-1}$. In the same energy range, and for the redshift of Tr, it is found that the deepest observations available with Chandra ($> 2$ Ms exposures) can reach AGN-type objects with luminosity as low as $L_{\rm X[2-10] keV} \approx 3\times 10^{43}$ ergs s$^{-1}$ \cite{Silverman2008}. Hence, we do not expect to see X-ray emission form this source in the much shallower 40 ks observation of this cluster.   \\
Another typical feature in accretion discs is Ly-$\alpha$ emission. Inspection of the spectrum from XShooter reveals no obvious Ly-$\alpha$ emission at the position of Tr. LyC continuum at shorter wavelengths is also undetected in the UV imaging presented in \cite{RiveraThorsen2019}, while it is clearly observed in all 12 positions of the knot 1 of system 5, despite Tr being comparable in brightness in the UV and optical bands as the brightest counterimage of knot 1. It is possible that the Ly-$\alpha$ and LyC emission are heavily absorbed, but that would require a fine-tuned configuration in order to explain the other spectral features shown in \cite{Vanzella2020}.
The lack of  Ly-$\alpha$ and LyC emission weakens the interpretation of an IMBH. As discussed in \cite{Vanzella2020}, these spectral features resemble those of massive LBV stars. In the next section we consider this type of star as the most viable candidate to explain Tr.

\subsection{Godzilla: A hyperluminous star at z=2.37}
Among the most luminous stars in the local universe we find Wolf-Rayet stars like R136a1 with an estimated radius of $\approx 0.2$ AU and  bolometric luminosity $\approx -12.2$ ($M_V \approx -8.2$) \citep{Doran2013,Bestenlehner2020}. At these luminosities, one would need magnification factors $\sim 5\times10^4$, which are tremendously unlikely and very difficult to maintain for more than a few weeks (due to relative motions between the source and the caustic). 
One would naively expect that significantly more massive stars could have much larger luminosities. Unfortunately, the scaling between mass and luminosity discussed in the previous subsection can not be applied to stars above a certain mass. Stars are supported by thermal pressure, and above masses of $\mathcal{O}$(100) ${\rm M}_{\odot}$ they became unstable due to radiation pressure \cite{Figer2005,Zinnecker2007,Crowther2010}. We do not consider here the hypothetical case of very massive stars with mass above several hundred solar masses, although these stars are theoretically possible but short lived \citep{Belkus2007}, nor metal-free Pop III stars. These stars are believed to have existed in the early universe and could be detected through gravitational lensing \citep[][]{Windhorst2018}. However, Godzilla is unlikely a PopIII star given its redshift and presence of metals in its spectrum as shown in \cite[see][]{Vanzella2020}.\\

Hence, at first glance it appears like ordinary stars can not be luminous enough to explain the observed flux at Tr, unless one considers unreasonably high magnification factors. However, there are stars in our local universe that momentarily can increase their luminosity by a substantial amount. \\

Luminous blue variable (LBV) stars like Eta Carinae, also mentioned in \cite{Vanzella2020} are very hot and luminous \citep{Crowther2007,Ramachandran2019}, and show  spectral features similar to the ones observed in Tr. More importantly, during an outburst they can reach the required luminosity. These outbursts can be relatively short, making these stars resemble SN explosions when observed at large distances (SN impostors), but the outbursts can also last decades (like the Great Eruption in Eta Carinae), satisfying the requirement on the duration of the event. This type of star would also explain the Bowen fluorescence observed in \cite{Vanzella2020} as they are often surrounded by circumstellar material from previous outbursts, and other peculiar spectral features like P Cygni profiles from intense stellar winds, clearly observed in the C-IV line in the MUSE spectrum at the position of Tr. 
Currently having an absolute magnitude of $M_V\approx -8.5$, historical records of Eta Carinae show that during the Great Eruption between 1822 and 1864, its luminosity increased by $\approx 5$ magnitudes \cite{Frew2004,Nathan2011}.  Other variable stars like the SN impostor, SN 2002bu have reached even larger luminosities than Eta Carinae \citep[$M_V=-15$][]{Szczygie2012} although it remained in this bright phase during a shorter period of time than  Eta Carinae.  

Assuming our most conservative limit for the magnification factor of $\mu \approx 600$, the source should have an absolute magnitude of $M_{UV} \approx-17.4$ in order to make it consistent with the observed flux. During an outburst phase, no known LBV star has ever been observed in our local universe maintaining this luminosity for two years or more, which would make this star the brightest ever seen. 
At larger magnifications factors of $\mu \approx 2000$, the luminosity reduces to absolute magnitude -15.7, close to but still above the maximum luminosity observed in Eta Carinae during the Great Eruption, but almost as luminous as the SN impostor 2002bu during the 2002 outburst \citep{Szczygie2012}. Even at this large magnification factors, the source responsible for Tr would be a monster star, more luminous than any other star ever observed. It would not be surprising to expect LBV stars more luminous than the ones observed in our local universe after one considers the gain in volume probed at high redshift. Thanks to gravitational lensing, LBV stars can be observed at $z>1$ in a volume $\sim$10 orders of magnitude larger than the volume of the local group, enough to compensate for the very low probability of having $\mu > 1000$. 
If we adopt the maximum magnification inferred from model $M_2$, i.e, $\mu \approx 7000$, the luminosity can be reduced by $\approx 1.3$ magnitudes, bringing the luminosity in line with the luminosity of LBV stars in our local group during an outburst phase, like the Great Eruption in Eta Carinae. \\
Magnifications larger than $\sim 10^4$ would reduce the needed intrinsic luminosity even further but are very unlikely, due to the universal scaling of the lensing probability with magnification ($P(>\mu) \propto \mu^{-2}$). In addition, as discussed earlier, the presence of ubiquitous microlenses disturb the macro-caustic of the cluster+perturber, and effectively reduce the maximum possible magnification of a star close to a cluster caustic \citep{Venumadhav2017,Diego2018}. \\
It seems then very plausible that given the expected abundance of LBV stars at high redshift, and the vast volume accessible through gravitational lensing, that these stars can be detected during their flaring phase. This is especially true for outbursting LBV stars that inhabit  well studied giant arcs, which naturally act as beacons in space of large magnification factors. \\

Based on all the arguments presented so far, we can then establish that the most likely candidate is a very massive and luminous LBV star captured during a long duration and energetic outburst. Because of the combination of extreme luminosity and magnification, from now on we refer to this monster  star as {\it Godzilla}. \\

{\it Probability of observing a Godzilla star:} 

\begin{figure*} 
   \includegraphics[width=18cm]{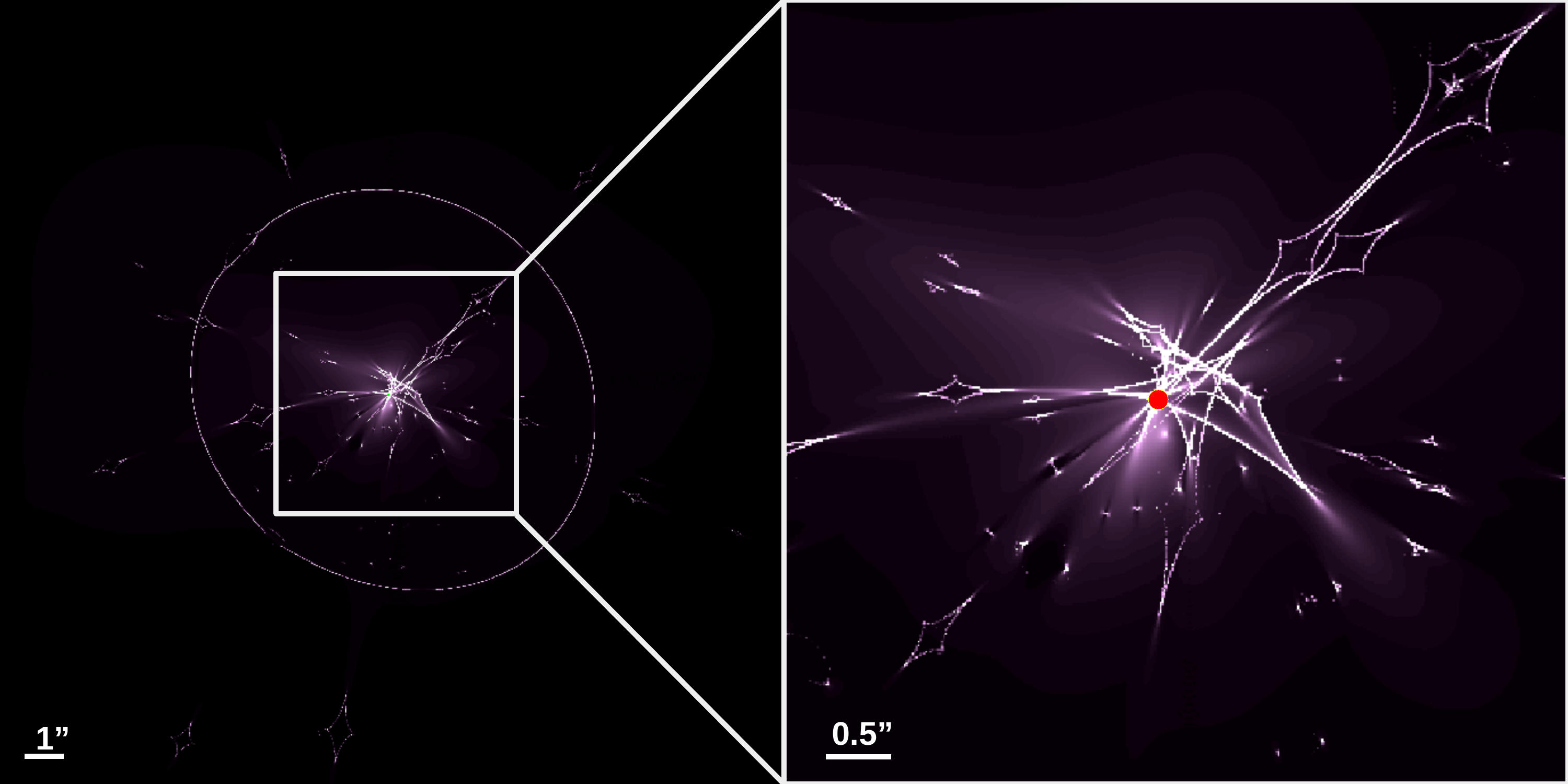}
      \caption{Caustic region at the redshift of Godzilla. The right panel shows the central region, with the predicted position of Godzilla marked with a red dot. The overlapping of several caustics in this region is the reason behind the large multiplicity of knot 1, which appears 12 times. Godzilla must be a fraction of a parsec from one of these caustics.    
              }
         \label{Fig_Godzilla_vs_Caustics}
\end{figure*}
After discussing the possible nature of Godzilla as an outbursting LBV star, we turn our attention to the probability of such an event. With the cluster+perturber model shown in Figure~\ref{Fig_Tr_vs_CC2}, we can estimate the strength of the critical curve at Tr position. We fit the magnification in a direction perpendicular to the critical curve with the universal law $\mu=\mu_o/d$, and find $\mu_o\approx 50"$. We use this result to estimate the distance between the counterimages and the caustic for a given magnification. 
Since the image of Godzilla is unresolved, it must be forming a pair of counterimages, with one counterimage in each side of the critical curve, at a distance of 15 mas at most (see appendix \ref{Appendix5}). From this separation and the law derived before, we infer the magnification is at least $\mu \approx 3330$. If we assume the radial magnification is again $\mu_r \approx 2$, after dividing the 15 mas separation by the tangential magnification we obtain that Godzilla must be at most $\approx 0.073$ pc away from the caustic. This is a very small separation to the caustic. The probability that Godzilla is that close to the cluster+perturber would be equally small.\\ 
However one needs to consider the fact that the galaxy hosting Godzilla is already in an area of large magnification, and that multiple caustics cross that galaxy. The bright knot 1 for instance (close to Godzilla in the source plane), is multiply lensed a record number of 12 times. This implies that multiple overlapping caustics are surrounding Godzilla. The probability of being magnified by a large factor needs to take into account the fact that Godzilla could have been amplified by any of the other nearby caustics. Also, the combined effect of the multiple overlapping caustics has a multiplicative effect in terms of the final amplification factor. This is demonstrated in Figure~\ref{Fig_Godzilla_vs_Caustics}, where we plot the caustic network of the cluster, and the position of Godzilla predicted by our lens model in relation to the caustic network. As can be appreciated, Godzilla is surrounded by several very powerful caustics. Given the limited resolution of the lens model, the caustics can not be well resolved, but the plot clearly shows how multiple caustics overlap near Godzilla's position.  

In order to better estimate the probability of being magnified by an extreme factor of $> 3330$ we rely again on our lens model. We compute the probability of magnification based on the magnification map and find that the probability of magnification scales as the canonical $\mu^{-3}$. In particular we find $dA/d\mu = 10(10^3/\mu)^3$ pc$^2$ (this law offers an excellent fit between $\mu=100$ and $\mu=10000$, which is approximately the maximum magnification that can be measured with enough statistical power given the pixel size in our model of 0.03"). We can integrate this law to infer the area above a given magnification and find; 
\begin{equation} 
    A(>\mu) = 50\left( \frac{10^4}{\mu} \right)^2 \, {pc}^2
\end{equation}
Hence there is an area in the source plane of $\approx$ 450 pc$^2$ with $\mu>3330$. Considering that the area in the host galaxy that contains rich star forming regions is $\approx 500\,\mathrm{pc}\times100\,\mathrm{pc}$ (see the source reconstruction in appendix \ref{Appendix4}), the probability of Godzilla to be in an area of 450 pc$^2$ is $\approx 1\%$. Considering that \cite{Vanzella2021} estimated a very high star formation rate for the Sunburst arc, and that these type of stars must be abundant in the area of $\approx 500\,\mathrm{pc}\times100\,\mathrm{pc}$ considered above, it is not unreasonable to expect one star such as Godzilla to be observed at these magnification factors. 

In a broader sense, we can estimate the abundance of extremely magnified LBV stars (or EMLBV) at cosmological distances. We consider the volume in the redshift shell $1<z<3$, which is $V\approx 900$ Gpc$^3$.
The abundance of LBV can be estimated only in our local volume, but this should be considered as a lower limit since the star formation rate is known to be higher at redshift $z>1$. 
From the recent compilation of \cite{Richardson2018}, there are 40 confirmed LBVs up to the distance of M33 ($d\approx 1$ Mpc). Extrapolating this number to the redshift shell above, we expect a number of at least $10^{13}$ LBV stars in that redshift interval. These stars are not observable through ordinary means, but can be accessed thanks to the boost of gravitational lensing, especially during an outburst phase. 
These outbursts can take place relatively often \citep{Pastorello2010} or can happen with time intervals of centuries. In stars like Eta Carinae, large outbursts have been estimated to take place approximately every 3 centuries during the last millennia \citep{Kiminki2016}. Assuming a conservative average of one outburst per 1000 years for these stars, and a duration for these outbursts of one year  (both in the observer frame), we expect at least  $10^{10}$ stars to be outbursting at any time between redshifts $z=1$ and $z=3$. 
These stars would still have apparent magnitudes $M_g \approx 30$ and remain undetectable without the aid of extreme magnification. If we consider magnification factors larger than $\mu=1000$ (i.e., a boost of 7.5 magnitudes), the probability of magnification $P(\mu>1000) \approx 2\times10^{-9}$ \citep{Diego2019}, resulting in $\approx 20$ EMLBV stars observable at each moment above apparent magnitude $M_g\approx$ 24--25. Some of these stars will be bright enough (like Godzilla) to be detected by upcoming surveys such as Euclid or LSST.  
If we assume a more modest magnification of $\mu>100$, these stars would have apparent magnitudes $M_g\approx$ 27, well within reach of routine observations with JWST. At these magnification we expect two orders of magnitude more, i.e.\ a few thousand magnified outbursting LBV at any given time. Recall that the estimates above are derived under conservative assumptions and that the number of detectable EMLBV stars is probably higher. \\

Recognizing that these EMLBV stars will most likely be found in giant arcs, and that many of these arcs have been already identified and observed, Godzilla is probably the first of a list of EMLBV stars that will soon grow in number. It is very possible that some strongly lensed but unresolved luminous objects, already found in these giant arcs and identified as ultra-compact star forming regions or globular clusters, are instead flaring EMLBV stars like Godzilla. A good example would be the knot t1 (if it is finally confirmed as a counterimage of Godzilla), that has been interpreted as an unresolved young star forming region \citep[see][where t1 corresponds to knot 5.4a in that work]{Vanzella2021}. \\

Future surveys such as Euclid and LSST will detect thousands of new strongly lensed arcs in the coming years. Space telescopes like HST and JWST will be used to observe in greater detail these arcs, opening the door to a large number of stars similar to Godzilla. These observations will unveil new EMLBV stars, but also new caustic crossing stars like Icarus. The main difference between caustic crossing events and EMLBV is timescale. While caustic crossings can last several weeks, EMLBV stars can be observed for years. Also, since outbursting LBV stars are intrinsically more luminous than stars like Icarus, they can appear with magnitudes accessible by large survey telescopes like Euclid and Rubin, opening the door to their identification by these large surveys.  At magnitudes brighter than 23, these stars can be monitored by relatively small ground telescopes, which together with the relatively long timescales of the outbursts, make them ideal targets to study microlensing by stars in the lens, but also compact dark matter candidates like primordial black holes (PBH)~\citep{Green:2020jor}. Given the fact that these stars can be observed only when extreme magnification factors are involved, the overlap of microcaustics from microlenses is expected to be significant. This will result in relatively frequent microcaustic crossings, that will allow us to determine with accuracy the relative motion (direction but also velocity) of the EMLBV star with respect to the web of microcaustics. The presence of a massive PBH with a mass of several tens of solar masses could then be unambiguously established if the path of the EMLBV intersects such a microlens. Well-sampled light curves of EMLBV can then be used to set limits on the abundance of PBHs.


\section{Discussion of the results}
\label{sec_discussion}

Assuming Godzilla is a compact source, the observed flux constrains the source size at given magnification $\mu$. For the discussion below, we assume a fiducial magnification of $\mu=5000$, but present results in its generic form, explicitly including the $\mu$ dependency. From the MUSE measurement of the FUV continuum, the luminosity in the rest-frame wavelength range 1400--2750\,\AA\, is $L_{\rm 1400-2750}=1.2\times 10^{41}\,{\rm erg}\,{\rm s}^{-1}\,\left(\mu/5000\right)^{-1}$ (neglecting dust extinction). This sets a lower bound on the bolometric luminosity at given $\mu$:
\begin{align}
    \label{eq:Llowbd}
    L_{\rm bol} \gtrsim 10^{41}\, {\rm erg}\,{\rm s}^{-1}\,\left(\mu/5000\right)^{-1}.
\end{align}

If the source luminosity is Eddington-limited, a lower bound on the mass $M$ of the central object is
\begin{align}
\label{eq:mass}
    M \gtrsim 1000\,M_\odot\,\left( \mu/5000 \right)^{-1}.
\end{align}
Since the FUV continuum shape resembles that of OB stars, the surface temperature of the continuum source $T_s$ is likely to be similar to those of OB stars. To match the observed flux density $\sim 3\times 10^{-18}\,{\rm erg}^{-1}\,{\rm s}^{-1}\,{\rm cm}^{-2}$\,\AA\,$^{-1}$ at rest-frame $1400$\,\AA\, the source size is
\begin{align}
\label{eq:size}
    R \approx 0.4\,{\rm AU}\, \left( e^{10^5\,{\rm K}/T_s} - 1 \right)^{1/2}\,(\mu/5000)^{-1/2}.
\end{align}
At the fiducial magnification $\mu=5000$, $R\approx 11\,$AU if $T_s=15000\,$K, and $R\approx 2\,$AU if $T_s=30000\,$K.

Combining Eq.~\ref{eq:mass} and Eq.~\ref{eq:size}, we infer the dynamic velocity near the photosphere
\begin{align}
\label{eq:vvir}
    v_{\rm vir} = (G\,M/R)^{1/2} \gtrsim 280\,{\rm km}\,{\rm s}^{-1}\, \left( \frac{\mu}{5000} \right)^{-1/4} \left( \frac{e^{10^5\,{\rm K}/T_s} - 1}{e^{10^5\,{\rm K}/15000\,{\rm K}} - 1} \right)^{-1/4}.
\end{align}
The inferred value decreases with larger $\mu$ or lower $T_s$, albeit with weak power-law dependence. Magnification factors $\mu \gtrsim 10^4$ are physically difficult to realize, and the observed FUV continuum shape suggests $T_s \gtrsim 15000\,$K. Thus the dynamic velocity near the source is significantly higher than the velocity dispersion $\lesssim 100\,{\rm km}\,{\rm s}^{-1}$ of the observed UV emission lines~\citep{Vanzella2020}. This hints at nebular line formation at larger distances from the source (probably powered by photoionization by the continuum source). The significant implication here, particularly constraining for the accretion disk scenario, is that the nebular source size may far exceed the continuum source size --- an extremely high magnification physically possible for the latter may not be realizable for the former. 

The constraints in Eq.~\ref{eq:mass} and Eq.~\ref{eq:vvir} are loosened if the continuum source is super-Eddington. A notable example is an LBV in the outbursting phase when fast ejecta are launched in a non-terminal explosion, interact with dense circumstellar material, and dissipate a tremendous amount of kinetic energy up to $10^{50}\,$erg over a period extending years or even decades. Recent observations have uncovered candidates of long-lasting LBV eruptions in low-$z$ dwarf galaxies. One object, SDSS 1133, reached peak luminosity $M_g=-16$~\citep{Koss2014}  and has an estimated FUV luminosity in the range of $10^{41}$--$10^{42}\,{\rm erg}\,{\rm s}^{-1}$ \citep{Kokubo2021arXiv210107797K}. A similar object PHL 293B persisted for nearly a decade~\citep{Burke2020}, although radiation transfer modeling suggested a moderate $L \approx (2.5-3.5)\times 10^6\,L_\odot$~\citep{Allan2020PHL293B}, which would not be easy to reconcile with Eq.~\ref{eq:Llowbd}. Provided that massive stars must be abundant in the host galaxy~\citep{Vanzella2021}, Godzilla may be a Cosmic Noon example of this outbursting LBV class. This explanation however appears imperfect --- prominent Balmer emission lines, which are widely used as a diagnostic to study the outflowing surroundings of LBVs, are not clearly detected for Godzilla~\citep{Vanzella2020}. Moreover, known LBVs in a prolonged outburst state often show varying fluxes of more than $10\%$ or larger on the timescale of years. 

\subsection{Spectral features:}
\begin{figure} 
   \includegraphics[width=9cm]{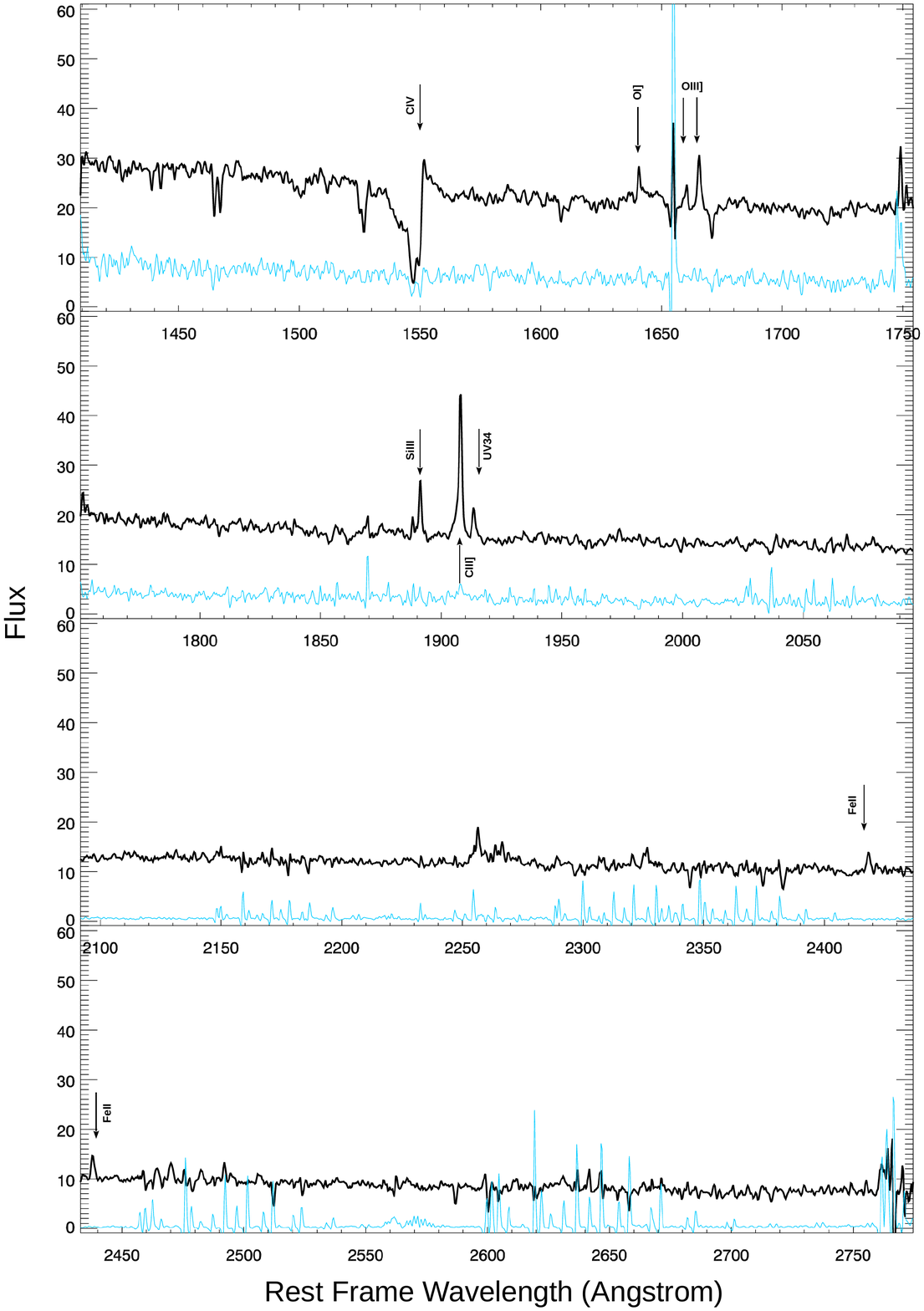}
      \caption{Spectrum of Godzilla derived from MUSE data. The blue lines show the subtracted background containing mostly sky lines. Some spectral features are marked. 
              }
         \label{Fig_Spectrum}
\end{figure}

In the rest-frame wavelength range $1400$--$2750$\,\AA\,  covered by MUSE, the most prominent absorption feature is the broad C IV 1550\,\AA\, P Cygni profile (see Fig.~\ref{Fig_Spectrum}), whose shape is curiously coincident with that of fast O-star winds. This indicates that a fast outflow at $\sim 2000\,$km/s obscures the continuum source. The outflow may be either a stellar wind or a disk wind. In the latter case, an inferred dynamic velocity of several thousand km/s of the accretion disk is consistent with Eq.~\ref{eq:vvir}.

All reported emission lines of Godzilla are associated with metal ions that can form via photoionization by OB stars~\citep{Vanzella2020}. Those include intermediate ionization species that live in a He I/H II zone, Si III, Fe III and O II, as well as high ionization species that live in the He II zone, O III, N III, C III and Ne III. Emission lines from higher ionized species that require a harder ionizing spectrum typical of an accreting BH are not seen. As for the He II 1640\,\AA\, recombination line reported in \cite{Vanzella2020}, we instead suggests O I] 1641\,\AA\, based on the line center measurement. This line is powered by Ly$\beta$ pumping in a high optical depth neutral gas, and is seen from gas condensations around $\eta$ Car~\citep{Hamann2012EtaCarInnerEjecta}. These spectral features favor a stellar source more than a hot accretion disk.

\subsection{The role of microlensing}
Either a hyperluminous star or an accretion disk is a compact continuum source which would likely involve a magnification in excess of a thousand. Therefore, microlensing effects from intracluster stars are likely to be important, whether the macro caustic results from a smooth cluster lens consisting of only galaxies and galaxy-sized DM subhalos, or is caused by perturbing effects of sub-galactic, star-free DM subhalos~\citep{Dai2018subhalo}. Intracluster stars that intervene the line of sight toward Godzilla probably account for $\kappa_{\rm ml}>10^{-3}$ of the total surface mass density. This is sufficient for creating overlapping micro caustics if the macro magnification exceeds $\sim 1000$. 

Godzilla would be traversing the corrugated micro caustic network at a typical velocity $\sim 300$--$1000$ km/s. The resultant flux variations due to a time-dependent magnification may be achromatic if Godzilla is a hyperluminous star with a wavelength-independent photosphere (ignoring small limb darkening effects), but may involve subtle color changes if it is a dazzling accretion disk with a radial temperature profile. Searches for microlensing-induced flux variations would provide a key test of the hypothesized compact source under extreme magnification, and potentially offer an opportunity to resolve the source's physical structure at a Cosmic Noon distance. We further note that if Godzilla is proven to be a compact stellar source having an extreme magnification factor of thousands or even ten thousand, it could be useful as a sensitive lensing probe to constrain minuscule DM substructures such as axion minihalos~\citep{Dai2020axionminihalo}, or primordial black holes \citep{Diego2018}.

\section{Summary and Conclusions}
\label{sec_conclusions}
We study the strong lensing effect in the galaxy cluster PSZ1 G311.65-18.48, using an improved version of our hybrid method for lens reconstruction, WSLAP+. We include new constraints consisting of the positions of critical points, useful to better constrain the lens model near the regions of maximum magnification. The addition of the new constraints help improve the lens model, in particular by positioning the predicted critical curve closer to the known position of critical points. We find that the critical curve from our lens model passes near the position of a very bright transient candidate previously identified in \cite{Vanzella2020}, but does not intersect Tr position. We identify candidate counterimages for Tr based on lens model predictions, but argue that it is possible that these candidates are not real counterimages. In the most conservative scenario where these candidates are counterimages of Tr, we constrain the magnification at Tr position to be between $\approx 600$ and $\approx 7000$. Based on these values for the magnification, we conclude that the size of the source must be less than 0.3 pc, ruling out compact and very luminous HII regions or globular clusters. Time delays from the available lens models rule out classic SN candidates, as these should have been clearly observed elsewhere in the image data. We conclude that the source is most likely an outbursting hyperluminous, and extremely magnified LBV (EMLBV) star which we dub Godzilla. Other candidates like an accretion disc around an IMBH are less likely based on the lack of spectral signatures typical of this type of object, but can not be categorically ruled out with the available information. We discuss how Godzilla is the first object of its kind found at cosmological distances, and how more objects like Godzilla should be identifiable in current data by searching for persistent unresolved knots in giant arcs in regions where the magnification can reach extreme values (that is, at a fraction of an arcsecond from a critical curve). 
We have estimated that several thousand EMLBV stars can be observed at any time between $z=1$ and $z=3$ above magnitude $M_g\approx 27$. Since these stars are most likely to be found in giant arcs, and many of these are known and studied in detail by HST, a dedicated analysis of current HST data will probably unveil additional examples of EMLBV stars, near regions of maximum magnification.

We find that in order to explain the extreme magnification of Godzilla we need to include a small mass-scale perturber ($M \sim 10^8\,M_\odot$) in the lens plane, possibly one of the dwarf galaxies in the cluster. This perturber is also able to explain a pair of images near Tr. Models of dark matter (DM) such as a warm DM or wave DM can suppress the formation of structure on the scale of dwarf galaxies and below. We therefore discuss how the existence of this perturber can be used to constrain such DM models. \\ 

Future observations of Godzilla should reveal flux fluctuations due to its relative motion with respect to the cluster caustic, but also due to the motion relative to the web of microcaustics that are expected to be pervasive near cluster critical curves. This will allow us to constrain further the nature of Godzilla, as well as giving rise to pioneering constraints on a range of models of dark matter, including compact dark matter that could play a role in microlensing.

\begin{acknowledgements}
 The authors would like to thank useful comments from Francisco Carrera. J.M.D. acknowledges the support of project PGC2018-101814-B-100 (MCIU/AEI/MINECO/FEDER, UE) Ministerio de Ciencia, Investigaci\'on y Universidades.  This project was funded by the Agencia Estatal de Investigaci\'on, Unidad de Excelencia Mar\'ia de Maeztu, ref. MDM-2017-0765. L.D. acknowledges the research grant support from the Alfred P.
Sloan Foundation (award number FG-2021-16495).
 
 This research is based on observations made with the NASA/ESA Hubble Space Telescope obtained from the Space Telescope Science Institute, which is operated by the Association of Universities for Research in Astronomy, Inc., under NASA contract NAS 5–26555. These observations are associated with program(s) 
 ID 15101 (P.I H. Dahle),
 ID 15377 (P.I M. Bayliss), 
 ID 15418 (P.I H. Dahle),
 ID 15949 (P.I M. Gladders).
 
 Some of the results and discussion presented in this work are based on observations collected at the European Organisation for Astronomical Research in the Southern Hemisphere under ESO programmes 
 0103.A-0688(C), 
 297.A-5012(A).

\end{acknowledgements}

\bibliographystyle{aa} 
\bibliography{MyBiblio} 


\begin{appendix}

\section{Free-form modelling with WSLAP+} \label{Appendix1}
Our lens model optimization is based on the code WSLAP+ \citep{Diego2005,Diego2007,Sendra2014,Diego2016}. WSLAP+ models are considered a hybrid type of model since they combine a free-form decomposition of the lens plane for the smooth large-scale component with a small-scale contribution from the member galaxies. Details can be found in these earlier references. Here we give a brief description of the method which we divide in two subsections; the first one describing the classic version of WSLAP+ (that can include weak and strong lensing constraints), and a second subsection where we describe the extension of the algorithm to include the new type of constraints at the critical points. 


\subsection{WSLAP+}
We adopt the standard definition of the lens equation
\begin{equation} 
\beta = \theta - \alpha(\theta,\Sigma), 
\label{eq_lens} 
\end{equation} 
where $\theta$ is the observed position of the source, $\alpha$ is the deflection angle, $\Sigma(\theta)$ is the unknown surface mass-density of the cluster at the position $\theta$, and $\beta$ is the unknown position of the background source. 
The optimization of the WSLAP+ solution takes advantage of the fact that the lens equation can be expressed as a linear function of the surface mass density, $\Sigma$. WSLAP+ parameterizes  $\Sigma$ as a linear superposition of functions, which translates into $\alpha(\theta,\Sigma)$ being also linear in $\Sigma$. WSLAP+ takes advantage of this linear dependency with the mass in order to quickly optimize the lens model. Since the shear components can be expressed as spatial derivatives of the deflection field, they can also be linearized in terms of the mass, thus allowing shear measurements (when available) to  be easily integrated into the same optimization scheme. An example of lensing reconstruction with WSLAP+ combining weak and strong lensing can be found in \cite{Diego2015}. \\

In WSLAP+, the surface mass density, $\Sigma$, is described by the combination of two components; 
i) a soft (or diffuse) component (usually parameterized as superposition of Gaussians) corresponding to the free-form part of the model, or large scale cluster potential; and 
ii) a compact component that accounts for the mass associated with the individual galaxies in the cluster.  \\
For the diffuse component, different bases can be used, but we find that Gaussian functions provide a good compromise between the desired compactness and smoothness  of the basis function. A Gaussian basis offers several advantages, including a fast analytical computation of the integrated mass for a given radius, a smooth and nearly constant amplitude between overlapping Gaussians (with equal amplitudes) located at the right distances, and a orthogonality between relatively distant Gaussians that help reduce unwanted correlations. 
For the compact component, we adopt directly the light distribution in the IR band (F160W in the public HST data) around the brightest member elliptical galaxies in the cluster. 
For each galaxy, we assign a mass proportional to its surface brightness. This mass is later re-adjusted as part of the optimization process. The number of parameters connected with the compact component depends on the number of layers adopted. Each layer contains a number of member galaxies. The minimum number of layers is 1, corresponding to the case where all galaxies are placed in the same layer (i.e they are all assumed to have the same light-to-mas ratio). In this case, the single layer is proportional to the light distribution of all member galaxies and is assigned a fiducial mass for the entire mass of the member galaxies. There would be only one extra parameter which accounts for the renormalization constant multiplying the map of the mass distribution, that is optimized by WSLAP+. When lensing constraints are available near the central galaxy, it is customary to consider at least two layers, with the BCG galaxy having its own layer since  typically the light-to-mass ratio of BCGs differ from those of regular galaxies. In this case there would be two extra parameters being optimized; one for the mass of the BCG galaxy, and one for the mass of the remaining galaxies which would be placed in the second layer. In other cases individual galaxies near arcs can be placed in their own layers if they need to be optimized separately. For our particular case, since most of the constraints are near the Sunburst arc, and no central constraints are available, we consider only one layer and place all member galaxies, including the BCG, in the same layer. \\

\begin{figure} 
   \includegraphics[width=9cm]{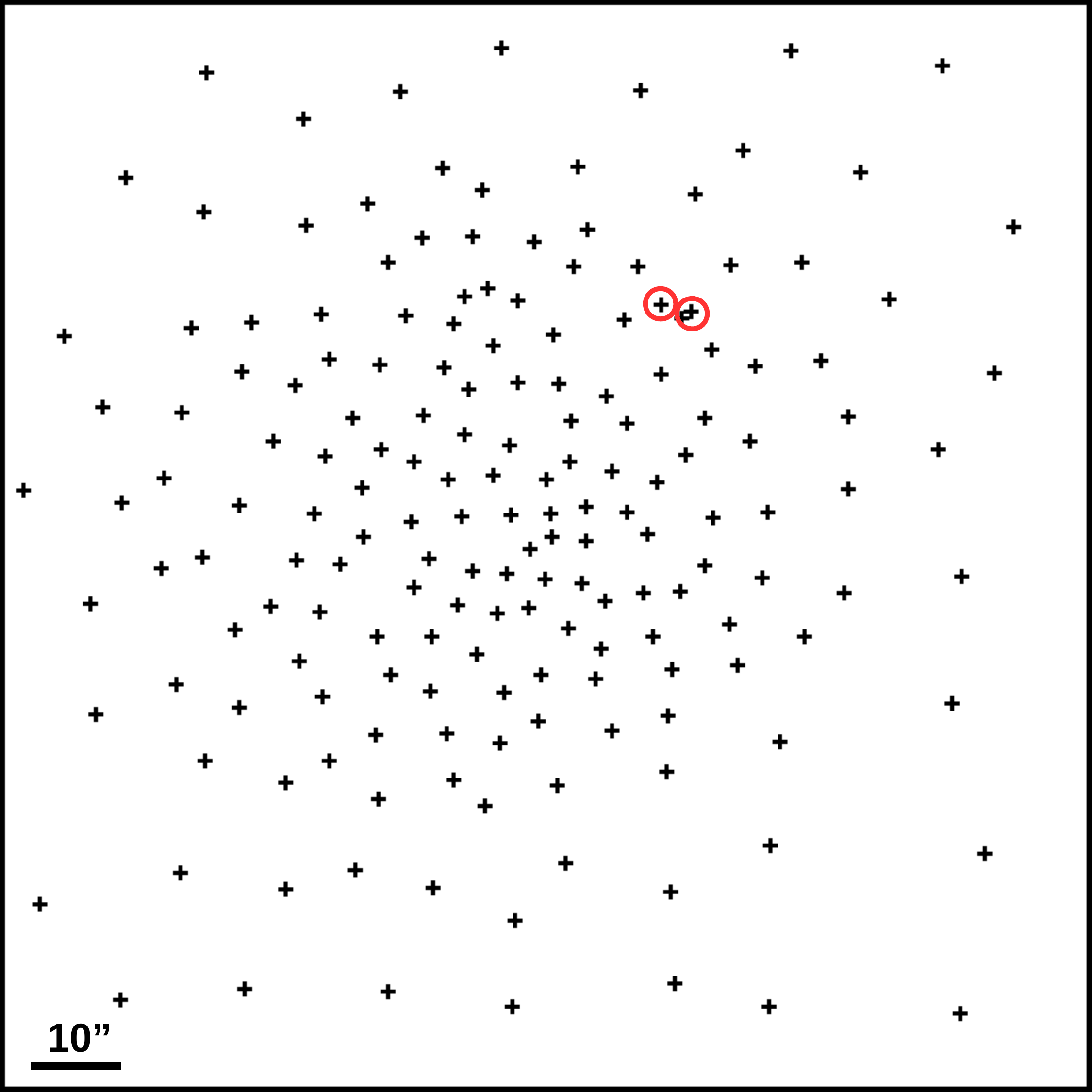}
      \caption{Distribution of grid points in the dynamical grid used to perform the reconstruction. This distribution is derived from a Monte-Carlo realization of a previous solution. Areas with a higher surface mass density have a higher concentration of grid points. The red circles mark the two additional grid points added to account for the two background galaxies near the sunburst arc at redshifts 0.5578 and 0.7346. 
              }
         \label{Fig_Grid}
\end{figure}


As shown by \cite{Diego2005,Diego2007}, the strong and weak lensing problem can be expressed as a system of linear equations that can be represented in a compact form, 
\begin{equation}
\vectTheta = \matrGamma \vectX, 
\label{eq_lens_system} 
\end{equation} 
where the measured strong lensing observables (and weak lensing if available) are contained in the array $\vectTheta$ of dimension $N_{\Theta }=2N_{\rm sl}$ (plus $2N_{\rm wl}$ if weak lensing data is available), the unknown surface mass density and source positions are in the array $\vectX$ of dimension 
\begin{equation}
N_{\rm X}=N_{\rm c} + N_{\rm l} + 2N_{\rm s}
\label{eq_Nx}
\end{equation}
and the matrix $\matrGamma$ is known (for a given grid configuration and fiducial galaxy deflection field) and has dimension $N_{\Theta }\times N_{\rm X}$.  $N_{\rm sl}$ is the number of strong lensing observables (each one contributing with two constraints, $x$, and $y$), $N_{\rm c}$ is the number of grid points (or cells) that we use to divide the field of view, $N_l$ is the number of layers ($N_l=1$ in our case as mentioned above), and $N_s$ is the number of background sources being strongly lensed (each source represent two unknowns in $X$, $\beta_x$, and $\beta_y$). Each grid point contains a Gaussian function. The width of the Gaussians are chosen in such a way that two neighbouring grid points with the same amplitude produce a plateau in between the two  overlapping Gaussians. In this work, we consider an adaptive grid configuration which is derived in an iterative manner (a first solution is derived with a regular grid and that solution is later used to derive an adaptive grid). Irregular grids are useful when there is a clear peak in the mass distribution, for instance when the cluster has a well defined centre or a single BCG.

The solution, $X$, of the system of equations \ref{eq_lens_system} is found after minimizing a quadratic function of $X$ \citep[derived from the system of equations \ref{eq_lens_system} as described in ][]{Diego2005}. The minimization of the quadratic function is done with the constraint that the solution, $\vectX$, must be positive. Since the vector $\vectX$ contains the grid masses, the re-normalisation factors for the galaxy deflection field and the background source positions, and all these quantities are always positive (the zero of the source positions is defined in the bottom left corner of the field of view).  Imposing  $\vectX>0$ helps constrain the space of meaningful solutions, and to regularise the solution, as it avoids unwanted large negative and positive contiguous fluctuations. The quadratic algorithm convergence is fast (a few minutes on a standard laptop), allowing for multiple solutions to be explored in a relatively short time. Different solutions can be obtained after modifying the first guess in the optimization and/or the redshifts of the systems without spectroscopic redshift. A detailed discussion of the quadratic algorithm can be found in \cite{Diego2005}. For a discussion of its convergence and performance (based on simulated data), see \cite{Sendra2014}.

\subsection{Adding critical points to WSLAP+}
Up to this point we have described the current version of WSLAP+. In this paper we include an additional set of constraints based on the known position of critical points, that is, positions in the lens plane where critical curves are known to be passing through. These points can be identified following symmetry arguments of pairs of lensed images near critical curves, since in this situation, the critical curve is expected to pass through (or very close to) the middle point of the image pair. For the particular case of the Sunburst arc, several critical points can be identified based on the location of knot 1 in system 5, but also other identifiable features in system 5.

At a critical point, the inverse of the magnification is zero. In the particular case of tangential critical curves (like the one near the Sunburst arc), critical points satisfy the following condition.
\begin{equation}
1-\kappa-\gamma=0    
\label{eq_muinv}
\end{equation}
We use this equality as additional constraints in each of the critical points identified in the Sunburst arc.
However, the condition in equation \ref{eq_muinv} can not be used directly in WSLAP+, since that equation is not linear in the mass. The term $\kappa$ satisfies the linearity requirement, but the term $\gamma=\sqrt{\gamma_1^2 + \gamma_2^2}$ does not (although both $\gamma_1$ and $\gamma_2$ are linear in mass). Fortunately, we can apply a simple transformation to the observables at the critical point position that will linearize equation \ref{eq_muinv}.  If the critical point is observed at a position where there is a lensed arc, one can determine the direction (given by the angle $\phi$) of the shear from the lensed arc. Then a rotation by the angle $2\phi$ can be applied to the shear components $\gamma_1$ and $\gamma_2$.  

\begin{equation}
\gamma_1^R = {\rm cos}(-2\phi)*\gamma_1 - {\rm sin}(-2\phi)*\gamma_2
\label{eq_gamma1rot}
\end{equation}
\begin{equation}
\gamma_2^R = {\rm sin}(-2\phi)*\gamma_1 + {\rm cos}(-2\phi)*\gamma_2 \\
\label{eq_gamma2rot}
\end{equation}
After this rotation, one gets and $\gamma_2^R=0$ and $\gamma_1^R=\gamma$. Consequently we also get $1-\kappa-\gamma=1-\kappa-\gamma_1^R=0$. After this transformation  of $\gamma_1$ and $\gamma_2$, we get two sets of new linear equations in the mass that can now be added to the original system of linear equations \ref{eq_lens_system}
\begin{equation}
1=\kappa+\gamma_1^R
\label{eq_New1}
\end{equation}
\begin{equation}
0=\gamma_2^R
\label{eq_New2}
\end{equation}

The vector element $\vectTheta$ in \ref{eq_lens_system} gets expanded with the left side of equations \ref{eq_New1} and \ref{eq_New2} (i.e the new observations). Similarly, the matrix $\matrGamma$ in \ref{eq_lens_system} gets expanded with the terms $\gamma_2^R$, and  $\kappa+\gamma_1^R$, where each new term $\Gamma_{i,j}$ in this expansion corresponds to the contribution to  $\gamma_2^R$ and  $\kappa+\gamma_1^R$ at position j from the cell i. Details on how the shear terms $\gamma_1$ and $\gamma_2$ are computed for each Gaussian cell are given in Appendix \ref{Appendix1}. For the compact component of the mass distribution, we similarly compute $\gamma_1$ and $\gamma_2$ from the fiducial compact mass component at the critical point positions and use that rotated shear contributions to build the corresponding column (or columns depending on how many layers are defined) in the $\matrGamma$ matrix. Once the vector  $\vectTheta$ and matrix $\matrGamma$ are expanded with the new constraints given in equations \ref{eq_New1} and \ref{eq_New2}, the optimization of the solution proceeds the same way as in the original WSLAP+ version, where the quadratic programming algorithm finds a quick solution $X$ (masses in the grid points, renormalization constants for the layers, and the source positions) of the system of linear equations.

\section{Shear components of a Gaussian cell}\label{Appendix2}
In this section we detail how the shear components are computed for each if the Gaussian cells. 
The shear components $\gamma_1$ and $\gamma_2$ can be easily computed for each one of the  Gaussian cells. The components $\alpha_x$ and $\alpha_y$ of the deflection filed for a Gaussian distribution of mass (or any circularly symmetric mass) is given by, 
\begin{equation}
\alpha_x(\theta) = \delta\,\frac{{\rm M}(\theta)\,x}{\theta^2}
\end{equation}
\begin{equation}
\alpha_y(\theta) = \delta\,\frac{{\rm M}(\theta)\,y}{\theta^2}
\end{equation}
where ${\rm M}(\theta)$ is the mass inside radius $\theta=\sqrt{x^2+y^2}$, and we define $\delta$ as:
 \begin{equation}
\delta =  \frac{4G}{c^2}\frac{D_{ls}(z)}{D_l(z)\,D_s(z)}.
\end{equation}
The shear components are obtained from the derivatives of this deflection field as:
\begin{equation}
\gamma_1 = \frac{1}{2}\left ( \frac{\partial \alpha_x}{\partial x}  -  \frac{\partial \alpha_y}{\partial y} \right ) 
\end{equation}
\begin{equation}
\gamma_2 = \frac{\partial \alpha_x}{\partial y}  = \frac{\partial \alpha_y}{\partial x}
\end{equation}

These derivatives can be easily obtained after using the chain rule for the derivatives of the mass; $\partial M/\partial x = (dM/d\theta)(x/\theta)$  and $\partial M/\partial y = (dM/d\theta)(y/\theta)$ we get

\begin{equation}
\frac{\partial \alpha_x}{\partial x}(\theta)  = \delta\left ( \frac{dM(\theta)}{d\theta}\frac{x^2}{\theta^3}  +  M(\theta)\frac{y^2-x^2}{\theta^4} \right )
\end{equation}

\begin{equation}
\frac{\partial \alpha_y}{\partial y}(\theta)  = \delta\left ( \frac{dM(\theta)}{d\theta}\frac{y^2}{\theta^3}  +  M(\theta)\frac{x^2-y^2}{\theta^4} \right )
\end{equation}

\begin{equation}
\frac{\partial \alpha_x}{\partial y}(\theta)  = \frac{\partial \alpha_y}{\partial x}(\theta) = \delta\left ( \frac{dM(\theta)}{d\theta}\frac{xy}{\theta^3} -2M(\theta)\frac{xy}{\theta^4} \right )
\end{equation}

 Although we have assumed a Gaussian distribution for the mass in each cell, these expressions are general for any circularly symmetric mass distribution \citep[see for instance equations 3.29 and 3.30 in][]{MeneghettiLectures}. The particular shape of the mass distribution determines the term $dM/d\theta$, which is easily derived for the Gaussian function.

\section{New system candidates}\label{Appendix3}
We use model $M_1$ to search for new multiply lensed galaxies. Model $M_1$ is better suited to reproduce the morphology of the arcs than model $M_2$, since the later is only more accurate in the regions near critical points. We identify 4 new sets of families that we list in table \ref{tab_2}. The redshifts listed in this table are the ones predicted by the lens model, needed to reproduce the images near the observed location.

\begin{figure} 
   \includegraphics[width=9cm]{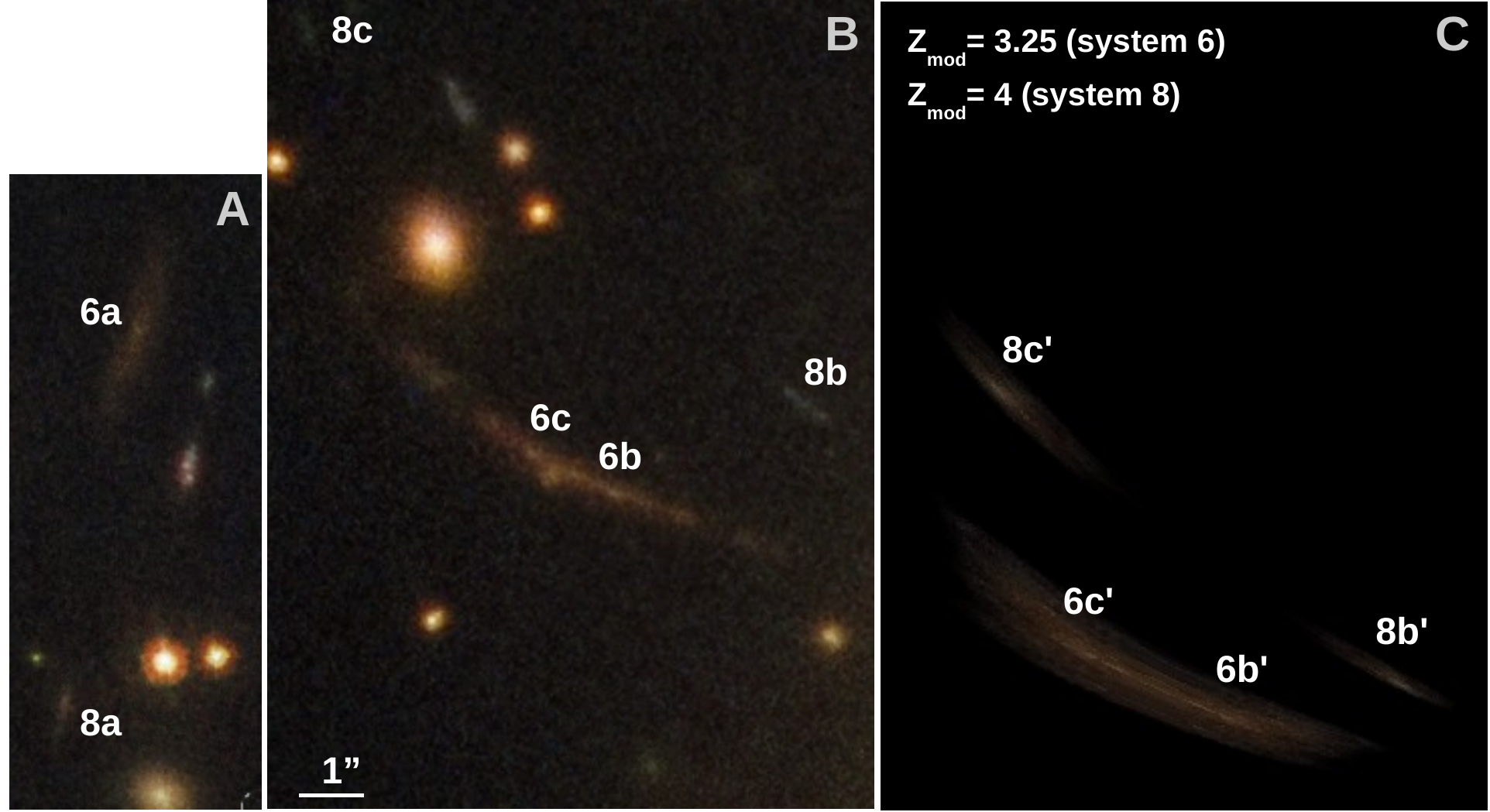}
      \caption{Prediction for new systems 6 and 8. The left panel (A) shows the images 6a and 8a used to predict the counterimages. The middle panel (B) shows the observed counterimages 6b, 6c, 8b, and 8c. The right panel (C) shows the predicted images 6b, 6c, 8b, and 8c. The   
      coordinates, orientation and dimension of panels B and C are the same.          }
         \label{Fig_Sys6and8}
\end{figure}

\begin{figure} 
   \includegraphics[width=9cm]{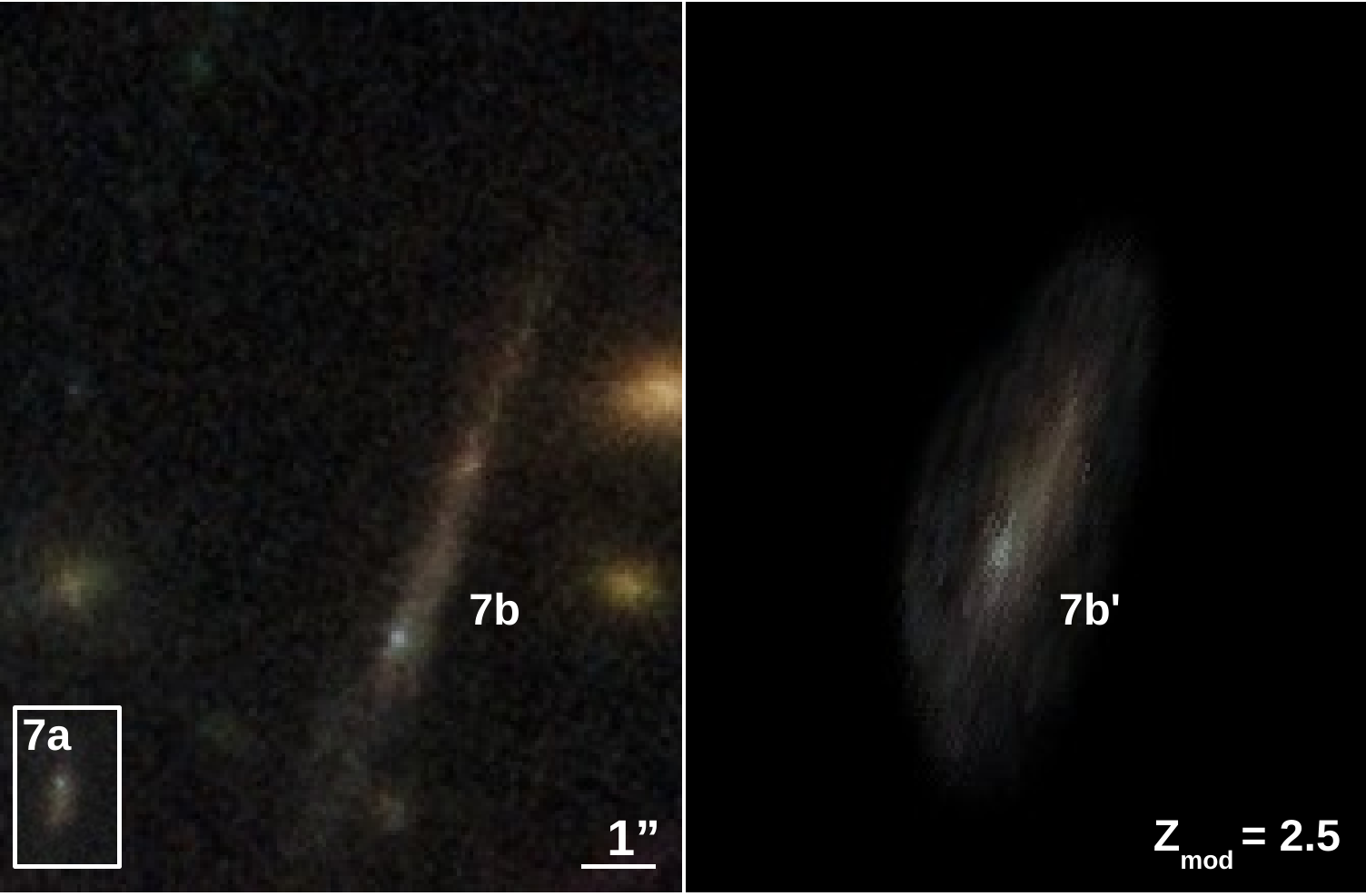}
      \caption{Prediction for new system 7. The left panel shows the observed arcs. Image 7a is used to predict the image 7b' in the right panel. Note how the tangential magnification in 7b' is smaller than in 7a, but the tangential one is larger.  
              }
         \label{Fig_Sys7}
\end{figure}

\begin{figure} 
   \includegraphics[width=9cm]{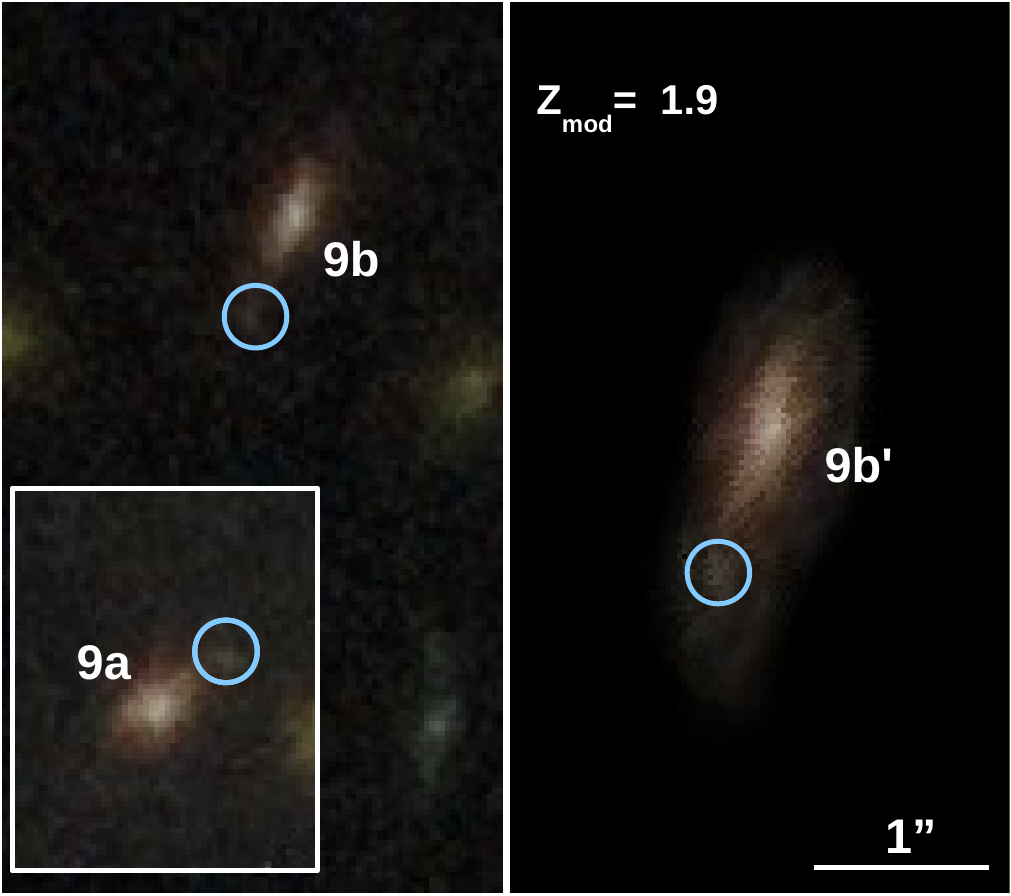}
      \caption{Prediction for new system 9. Like in figure \ref{Fig_Sys7}, image 9a is used to predict image 9'. The blue circle marks the position of a small feature that can be observed in all three images. 
              }
         \label{Fig_Sys9}
\end{figure}

The predicted images are shown in figures \ref{Fig_Sys6and8}, \ref{Fig_Sys7}, and  \ref{Fig_Sys9}. In all three cases we use counterimage "a" to predict the other counterimage(s). New system 6 is formed by a prominent red radial arc, with symmetric features. Its counterimage is a red dusty galaxy at a predicted redshift of 3.25. South of this galaxy we find a bluer lensed galaxy, 8a. This galaxy predicts a set of two radial images, 8b and 8c, north of 6b and 6c (see figure \ref{Fig_Sys6and8}). The redshift of this galaxy must be $\approx 4$ in order to reproduce these two arcs.  
System 7 forms a very prominent arc, 7b,  near image 1a. The counterimage, 7a, is much smaller but it has distinctive features that allows to match the prediction, 7b', with the observed image (see figure \ref{Fig_Sys7}). The lens model predicts a redshift of 2.5 for this system. 
Finally, system 9 corresponds to a lensed galaxy with a predicted redshift of 1.9, that has also some distinctive features. Spectroscopic confirmation of these systems, and their redshifts, will allow to improve the lens models, specially in the central region with the addition of the radial arcs, and in the North-East sector were constraints are scarce. 


\begin{table}
\begin{center}
    \caption{New candidate systems. The coordinates mark the position of the images identified in HST data. The last column shows the redshift of the system that is required for the lens model to match the predicted and observed positions (with a small error). }
 \label{tab_2}
 \begin{tabular}{|c|ccc|}   
 \hline
  Id  &   RA & DEC & $z_{model}$  \\
 \hline
6a & 15:49:55.348 & -78:11:32.66  & 3.25 \\
6b & 15:50:10.237 & -78:11:25.69  &  -- \\
6c & 15:50:10.437 & -78:11:25.44  &  -- \\
 \hline
7a & 15:50:00.263 & -78:11:39.44  & 2.5 \\
7b & 15:50:17.326 & -78:11:12.54  & --  \\
 \hline
8a & 15:49:55.696 & -78:11:38.30  & 4 \\
8b & 15:50:09.184 & -78:11:24.72  & -- \\
8c & 15:50:11.608 & -78:11:18.94  & -- \\
 \hline
9a & 15:50:01.893 & -78:11:44.57  & 1.9 \\
9b & 15:50:15.182 & -78:11:04.21  & -- \\
\hline
\end{tabular}
\end{center}
\end{table}

 \section{Lens model predictions}\label{Appendix4}
 This section presents predictions based on our lens for the counterimages of the giant arc, as well as for the geometry of the background host galaxy of Godzilla.
 
The source position for system 5 is constrained within a fraction of an arcsecond, but even within this small uncertainty, at large magnification factors small shifts in the source plane can result in large changes in the image plane.

 \begin{figure*} 
   \includegraphics[width=18cm]{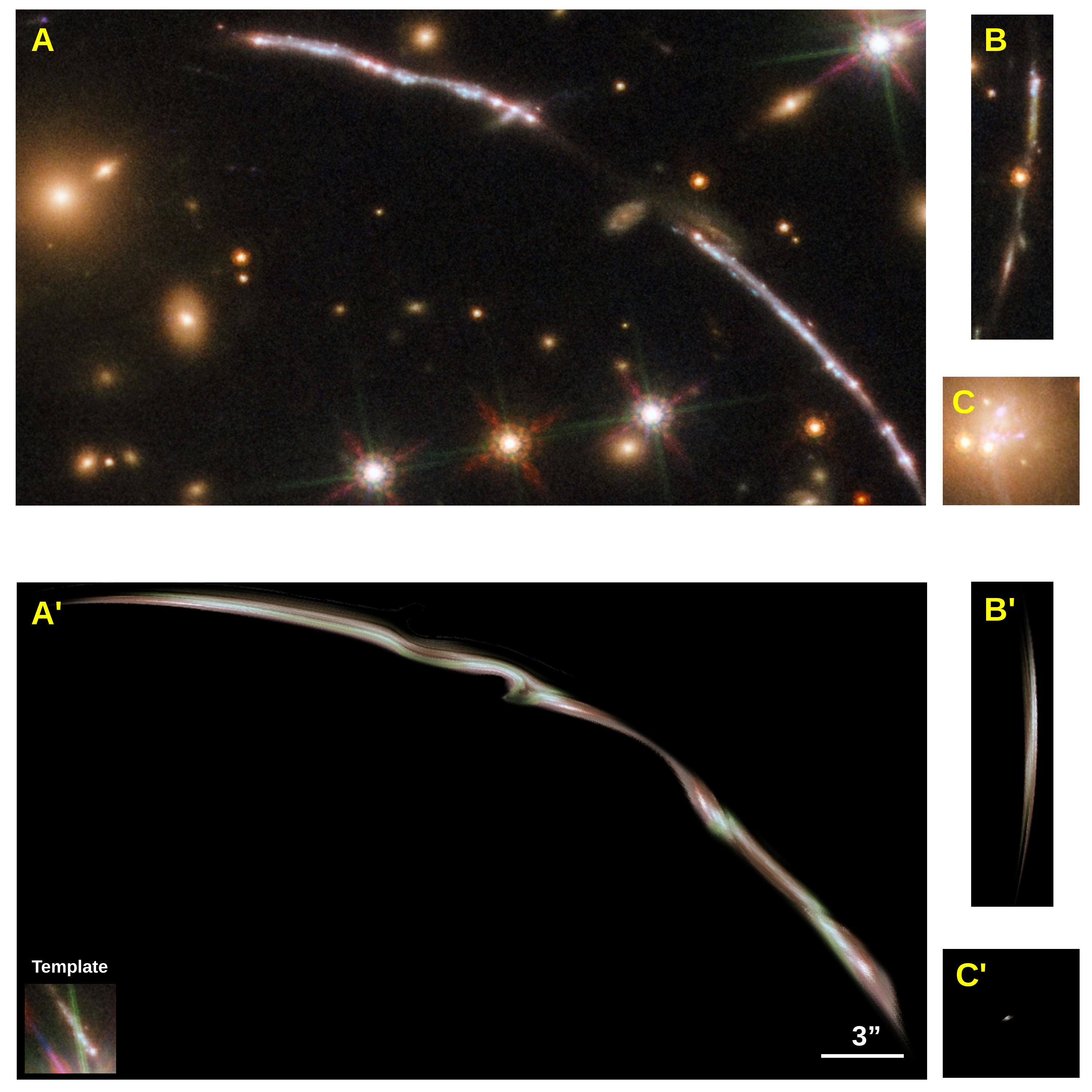}
      \caption{Predicted giant arcs from the lens model when the smallest counterimage of system 5 is used as a template. The small inset in the bottom-left part of panel A' shows the template. Panels A', B' and C' in the bottom part of the figure show the predicted counterimages from this template. Panels A, B, and C show the data version in the same portion of the sky as in panels A', B', C'. Note that the predicted central counterimage C' has not been identified spectroscopically, but small blue knots can be found near the centre of the BCG.      
              }
         \label{Fig_ReconstructionMainArc}
\end{figure*}

 Among the counterimages of system 5, we use counterimage number 12 as a template, since it contains a full image of the lensed galaxy, and the lensing distortion is more moderate. Other counterimages intersect a critical curve or are more distorted making them less than ideal to serve as templates. Using model $M_1$, we delense counterimage 12 into the source plane, and use that delensed image to relens it into the image plane. The result is shown in Fig.~\ref{Fig_ReconstructionMainArc}. Panels A, B, and C show the observed data for the counterimages, while panels A', B' and C' show the corresponding predictions. Note that the central counterimage in C has not been identified. The lens model predicts an image near the centre of the BCG, although with a small magnification factor. Also, the presence of a central mass, like a SMBH, would make this central counterimage even less magnified. In general, the predicted images shown in panels A', and B', reproduce well the position and geometry of the observed arcs. Small offsets of order 1"--2" can be appreciated but are normal in this type of reconstructions with WSLAP+, specially at large magnification factors.  \\

\begin{figure} 
   \includegraphics[width=9cm]{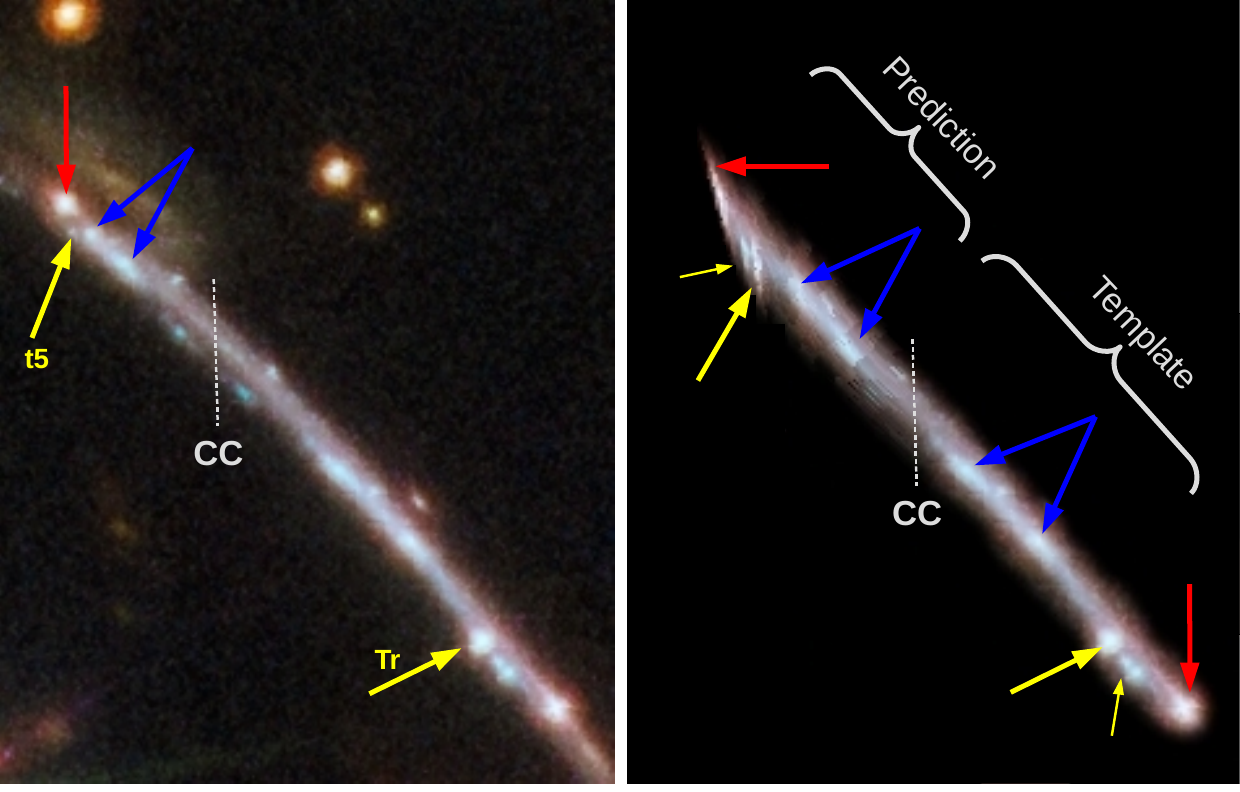}
      \caption{The left panel shows the giant double arc containing Godzilla. The approximated position of the critical curve predicted by our lens model is marked by a dashed line. The position of Godzilla is marked by a yellow line in the Southern arc. In the northern arc a small faint point source is found in the expected position of Godzilla. This is marked with a yellow arrow and a question mark. A source is also found at the expected position of Godzilla near the LyC knot 1, but is partially blended with it.  The predicted position of Godzilla based on the lens model is shown in the right panel, where we have used the southern part of the arc as a template to predict the northern part of the arc. Multiply lensed features are marked with arrows with the same color.
              }
         \label{Fig_TrPosPredicion}
\end{figure}

Next, we use counterimage 8, containing Tr, to predict the counterimage 7 on the other side of the critical curve. From simple smooth lens models,  counterimage 7 should contain also a counterimage of Tr, with similar apparent brightness. We select the portion of counterimage 8 that is to the right side if the critical curve dividing counterimages 7 and 8, and use it to delens this part of the arc and relens it into the other side of the critical curve. The result is shown in Fig.~\ref{Fig_TrPosPredicion} where in the left panel we show the observed arc, and in the right panel the lens model prediction. For convenience we mark with a vertical dotted line the approximated position of the critical curve from our lens model. In the left panel we mark Tr with a yellow arrow. A second arrow in the top-left part of this figure shows a small and faint knot that falls in the expected position of the counterimage of Tr. The right panel shows this predictions, where the yellow arrows are used again to mark Tr and its predicted counterimage. A thinner yellow arrow shows the P knots and its predicted (unobserved) counterimage. Other features are also marked with blue arrows. The small faint knot marked t5 and a yellow arrow in the left panel is the alleged counterimage t5 discussed in section \ref{sec_TrCounter}. It is possible that t5 is not a real counterimage of Tr, and that Tr remains without observable counterparts, but in this work we adopt the conservative hypothesis that t5 is a real counterimage. If this hypothesis is proven wrong (for instance through spectroscopic confirmation that t5 is not the same source as Tr), it signifies that the magnification of Tr must be even larger than the values inferred in this work, so they should be considered lower limits. 
Finally, based on the lens model prediction for the magnification of t5, and the observed flux ratios from the nearby LyC knot 1, if t5 is a counterimage of Tr, other counterimages must be also present and detectable in the image. We identify these additional counterimages as t1--t4 in section \ref{sec_TrCounter}. The same conservative hypothesis argument discussed above applies to t1--t4. That is, if they are proven to not be counterimages of Tr, then the magnification of Tr must be larger than the values inferred in this work. \\

\begin{figure} 
   \includegraphics[width=9cm]{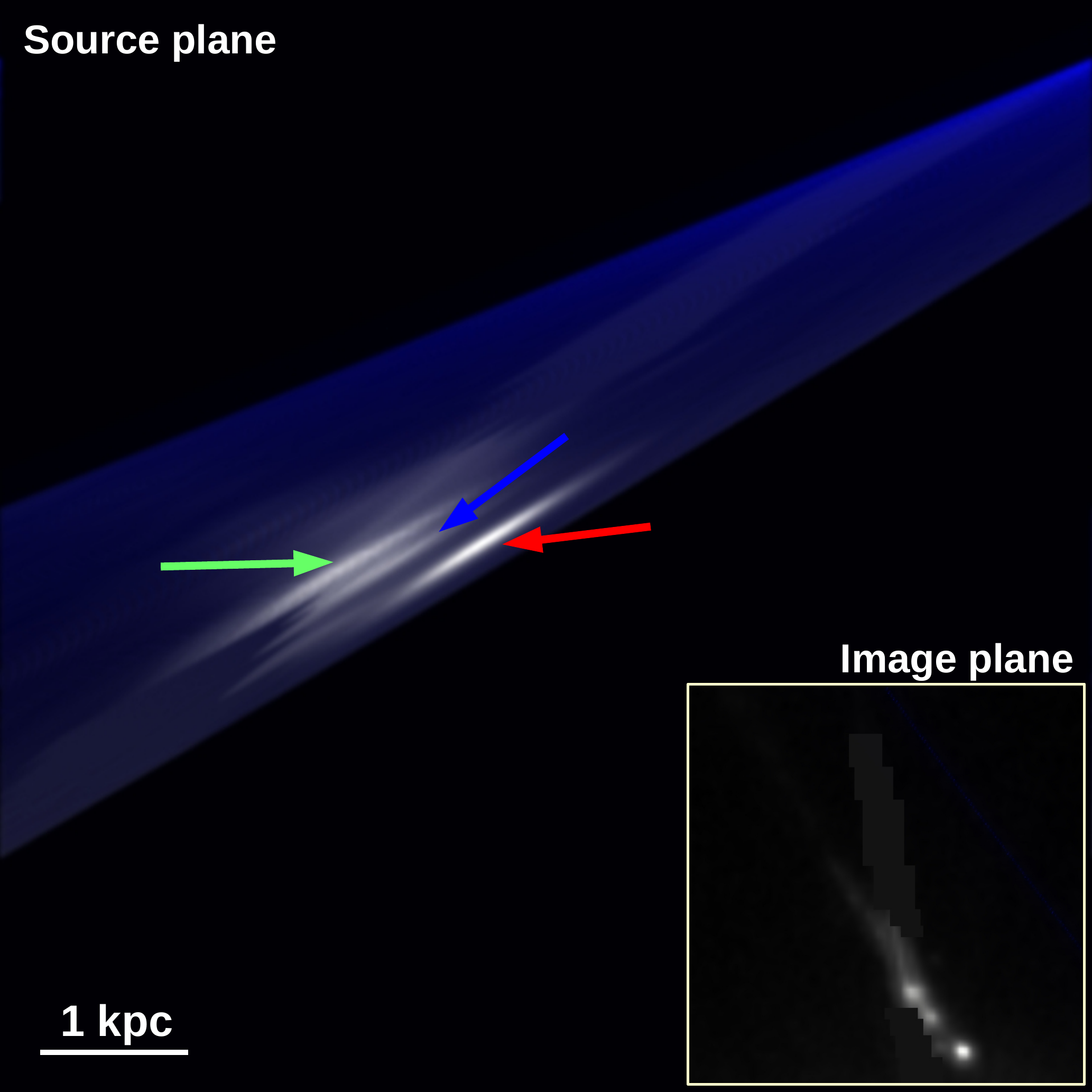}
      \caption{Reconstruction of the source based on counterimage 12 in system 5. This is the only counterimage that shows the full morphology of the arc. All other counterimages show only portions of the arc. 
      In grey we show the source surface brightness. The blue background indicates magnification. The masked region corresponds to a diffraction spike from a nearby star. The original data is shown in the bottom right part of the figure. The blue line in the image showing the original data indicates the position of critical curve at the redshift of the source.
              }
         \label{Fig_ReconstructionC12a}
\end{figure}

Next we turn out attention to the reconstruction of the source. Since there are two full images of the source, and 10 partial images, we can perform different reconstructions, depending on which image we choose to delens in the image plane. For the source reconstruction, we use model $M_1$ because it reproduces the arcs in the mage plane better than model $M_2$. 

\begin{figure*} 
   \includegraphics[width=18cm]{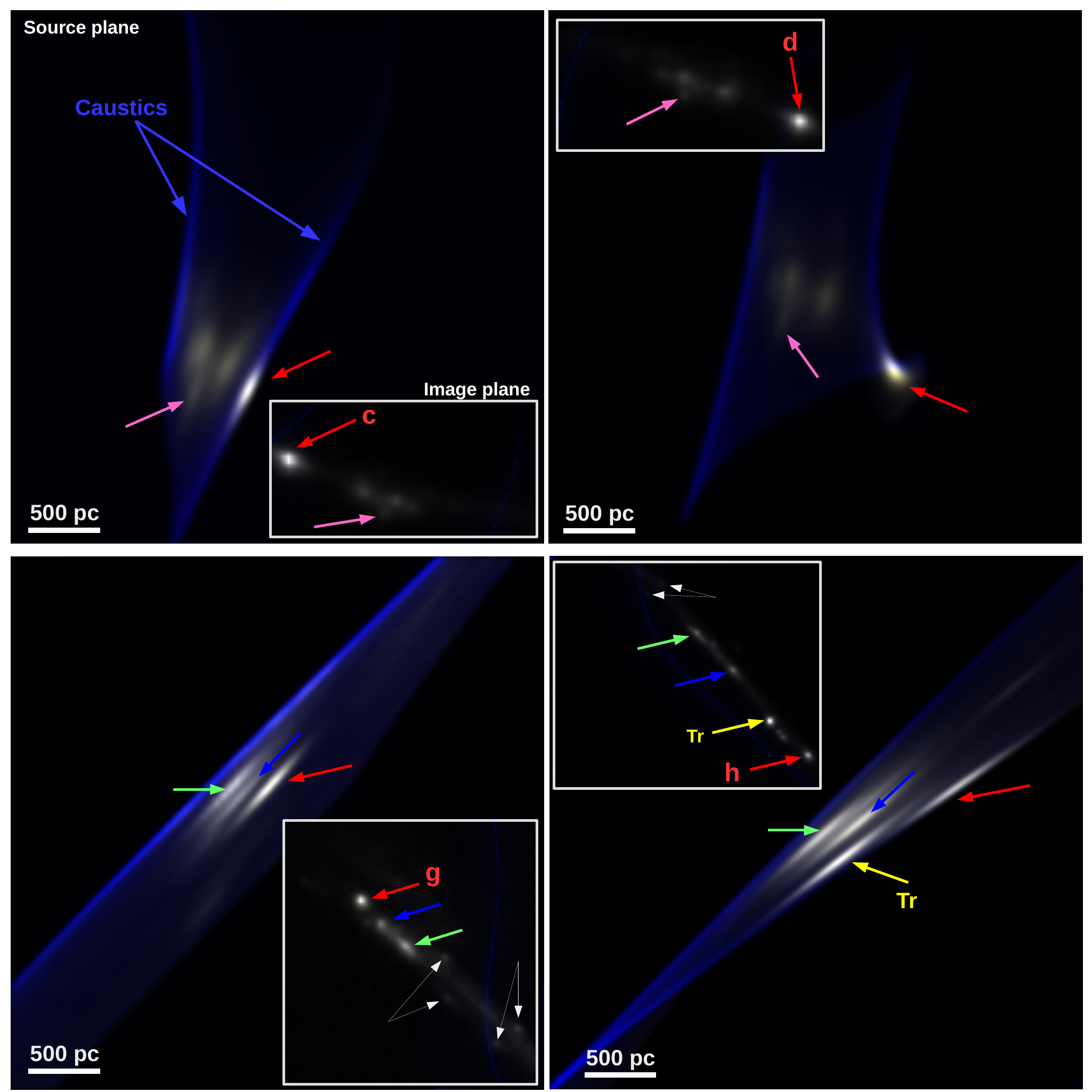}
      \caption{Partial reconstruction of the source of system 5 based on counterimages 3 (top-left), 4 (top-right), 7 (bottom-left) and 8 (bottom-right). For each panel, the original data is shown as a smaller inset in a corner of the panel. Both in the main panel, as in the inset critical curves and caustics are shown in blue. The letters denote the bright LyC knot 1, as in Fig.~\ref{Fig_Data}. 
              }
         \label{Fig_ReconstructionC12b}
\end{figure*}

First we start with counterimage 12, that as discussed earlier, offers a full view of the source, and at a moderate magnification factors.  We show the reconstructed version of the source, based on counterimage 12, in Fig.~\ref{Fig_ReconstructionC12a}. The original image being delensed is shown in the small panel in the bottom right part of the figure. The reconstructed image shows three distinctive features corresponding to the brightest LyC knot 1, and knots 2 and 3 (following the notation of \cite{Pignataro2021}). The separation between all these knots is $\approx 1$ kpc. The elongation of these three features is due to the fact that the radial magnification is much smaller than the tangential one, resulting in features more compressed in the tangential direction. This is an artifact due to the limited resolution of the telescope. This elongation would disappear if we had an instrument with much higher spatial resolution. 

In addition to counterimage 12, we use four more counterimages that provide better resolution in the source plane, thanks to the larger magnification factor. In particular we delens counterimages 3, 4, 7 and 8. Among these, counterimage 8 is the one that contains Tr, while counterimage 7 is the one where we would have expected to see a counterimage of Tr with similar brightness. However, this counterimage is clearly not seen in the other reconstructions.  To ease the identification of features, we mark them with arrows of the same color in the image and source planes. Based on the reconstructions in the top two panels, the intrinsic size of the source is larger than the size predicted from the reconstruction in Fig.~\ref{Fig_ReconstructionC12a} by a factor 2--3. This type of difference is expected since the model may not be self-consistent in terms of magnification (or flux) ratios, as these ratios are not being used as constraints.  \\

\begin{figure} 
   \includegraphics[width=9cm]{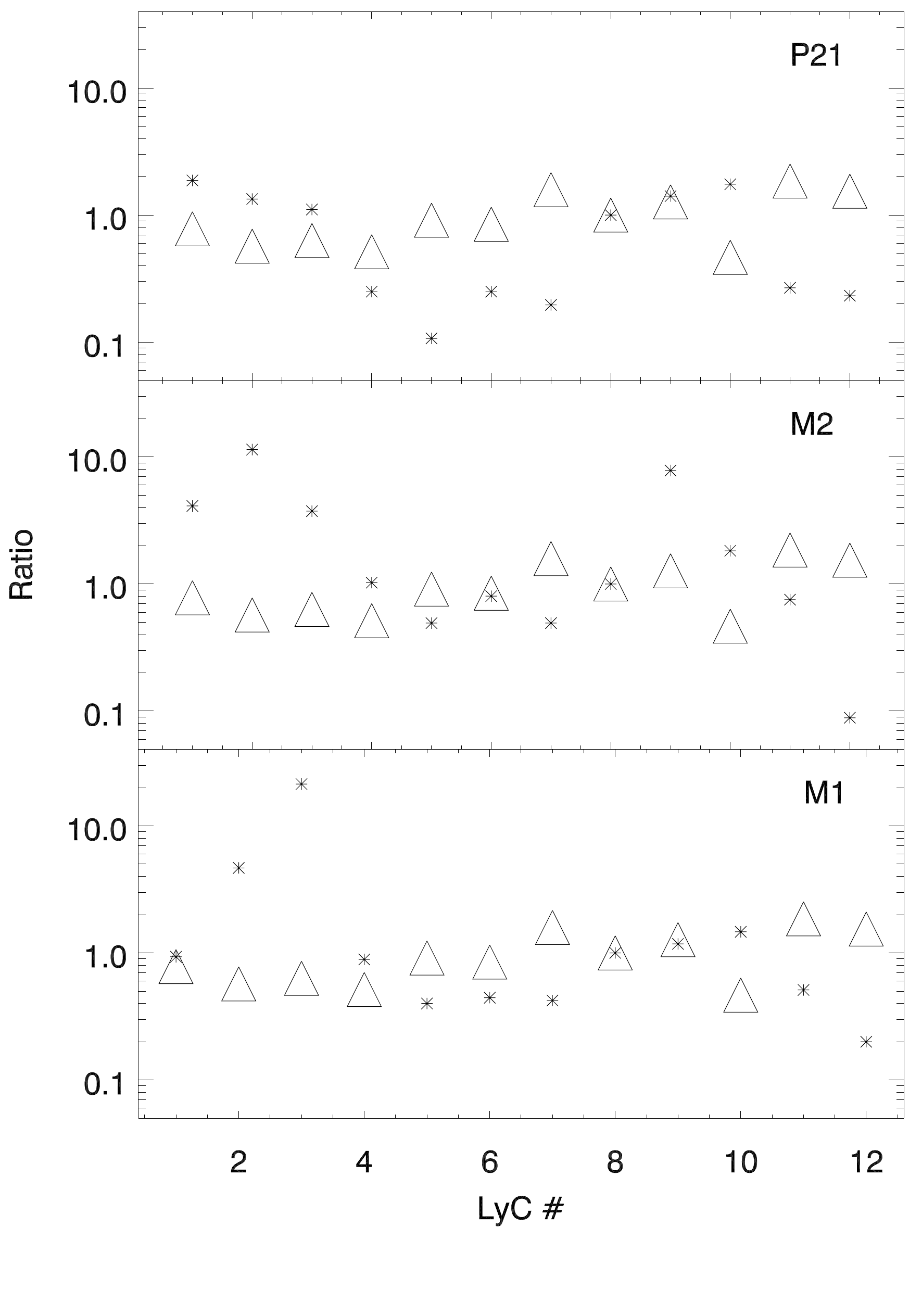}
      \caption{Flux ratio (triangles) vs magnification ratios (asterisks) predicted by different models at the 12 positions of the LyC knot 1. The bottom panel compares the flux ratios with the magnifications predicted by model 1 in this work. The middle panel is similar but for the model 2, and the top panel is for the model in \cite{Pignataro2021}. Flux measurements for the LyC knot are taken form \cite{RiveraThorsen2019}
              }
         \label{Fig_FluxRatio}
\end{figure}

Flux ratios are an interesting observable that could be exploited in future work to better constrain this lens model. In particular, we can use the observed flux at the 12 positions of the bright LyC knot 1 and compare the observed flux ratios with the observed magnification ratios. A perfect lens model should predict magnification ratios identical to the observed flux ratios. Departures between predicted and observed ratios can be used to identify regions in the lens plane where the lens model is inaccurate, although it should be emphasized that near critical curves, small changes in the lens model can result in large changes in the predicted magnification. With this caveat in mind, we compute flux ratios for knot 1 based on the photometric measurements of \cite{RiveraThorsen2019} in the F814W band. Then we compare these flux ratios with the magnification ratios predicted by the lens models, including the lens model of \cite{Pignataro2021}, that provides magnification factors at the position of knot 1. 
As a reference point, we use knot h (or LyC number 8 in the figure), which is the closest one to Tr, since we are most interested in how the models perform near this position. 
The result is sown in Fig.~\ref{Fig_FluxRatio}, where triangles indicate the observed flux ratios and asterisks the magnification ratios.  The prediction from \cite{Pignataro2021} is shown in the top panel, while the predictions from our models $M_1$ and $M_2$ are shown in the middle and bottom panel respectively. 
All three models fail at reproducing the flux ratio with accuracy in several positions. The model of \cite{Pignataro2021} predicts significantly fainter fluxes for knots 4, 5, 6, and 7. 
Models $M_1$ predicts substantially more flux in knots 2, and 3. This not surprising since knots 2 and 3 are close to each other and with a critical curve passing through. WSLAP+ models can be shallower than parametric models in certain parts of the lens plane, resulting in larger predicted magnification factors. Model $M_2$ also predicts higher fluxes in knots 2 and 3, but also in knots 1 and 9, while for knot 12 predicts a much smaller flux. 
All three lens models predict knot 7 to be fainter than knot 8, and knot 9 to be brighter. Real flux measurements agree with the fainter knot 7 but disagree with knot 9 that is fainter than knot 8, contrary to all model predictions, and indicating some bias in all lens models in this part of the lens plane. 

\section{PSF model and maximum separation for unresolved pairs}\label{Appendix5}
\begin{figure*} 
   \includegraphics[width=18cm]{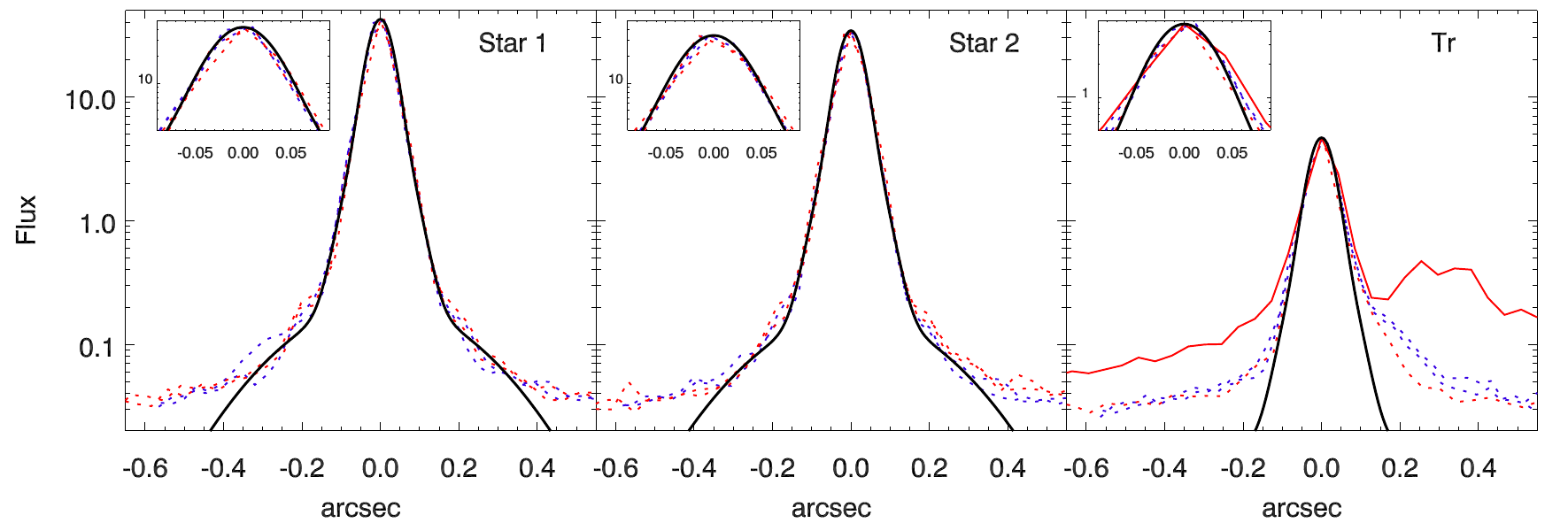}
      \caption{PSF fit model. The left and middle panel shows the profiles (dotted lines) and PSF model (solid line) for two bright stars found near Tr in the F606W band image. For each star, we derive 4 profiles. The two blue dotted lines correspond to the horizontal and vertical profiles, while the red dotted lines correspond to two profiles at 45$^{\circ}$ and -45$^{\circ}$.  
      The right panel shows the profiles in the same 4 directions at the position of Tr. The red dotted line is in a direction 45 degrees North to East, that is nearly perpendicular to the arc.  The red solid line is in a direction -45 degrees North to East, that is close to the direction of the arc. At $\approx 0.3"$, this profile intersects the P knots. The red solid curve is where we best expect to see a possible resolved image since it follows the direction of the shear. In all three panels, the inset shows a zoomed version near the peak, and covering approximately between the maximum of the peak to 1/10 the maximum.   
              }
         \label{Fig_PSFfit}
\end{figure*}

Under the hypothesis that Tr is is forming an unresolved double image, the separation between the images forming the pair must be small enough so the double image appears resolved.   In this section we constrain the maximum allowed separation between the alleged pair of lensed images in HST data, so they still appear unresolved. 
 We use the image in the F606W band, which offers a good compromise between spatial resolution and signal-to-noise.  We identify two stars near Tr, which we use to model the PSF in this band. Since we are interested in the resolving power of HST in the direction where the magnification is maximum (i.e. in the direction of the giant arc), we restrict our analysis to one-dimensional profiles that intersect the stars and/or Tr. For each star, and for Tr we derive 4 one-dimensional profiles. Each profile intersects the maximum of the source and follows a different direction. Two of the profiles go in the horizontal and vertical direction, while the other two go at 45 degrees and -45 degrees. The direction at -45 degrees is very close to the direction of the giant arc at the position of Tr. Hence the profile at -45 degrees is where we can impose the tightest constraints on the separation of the pair of lensed images. 
 The profiles for all three sources are shown in Fig.~\ref{Fig_PSFfit}. 
 
 In order to get a sense of the error in the PSF model, we perform a fit to six nearby and unsaturated stars. We fit a model for the PSF of the following form,
 \begin{equation} 
     PSF = exp\left( \frac{d^2}{2\sigma_1^2} \right ) + B\times exp\left( \frac{d^2}{2\sigma_2^2} \right) + C\times exp\left( \frac{d^2}{2\sigma_3^2} \right)
     \label{Eq_PSF}
 \end{equation}
We find that a model with values $\sigma_1=0.027" \pm 0.002"$, $\sigma_2=0.053" \pm 0.003$, $\sigma_3=0.39" \pm 0.008$, $B=0.25 \pm 0.04$, and$C=0.0019 \pm 0.0005$ reproduces well the observed profiles for the six stars (see black solid curve in Fig.\ref{Fig_PSFfit}, where for clarity we only show the first two stars). 
The right panel of Fig~\ref{Fig_PSFfit} shows the 4 derived profiles at the position of Tr, compared with the PSF model from Eq.~\ref{Eq_PSF}. The red solid curve shows the direction at -45 degrees (measured from North to East, assuming the pixels are oriented this way) which is a direction very close to the arc. The P knots can be easily appreciated in this profile at $\approx 0.3"$. The inset plot show zoomed versions of the profiles, and span approximately one order of magnitude in flux from the maximum flux.   
Comparing the zoomed versions of the plot, we appreciate that the red solid curved intersecting Tr departs slightly from the PSF model at 1/10 the maximum of the peak. 

It is unclear where this deviation is due to a partially resolved source underneath, or the contribution from the rest of the arc, but we can use this deviation to set an upper limit on the separation of two unresolved counterimages. 

\begin{figure} 
   \includegraphics[width=9cm]{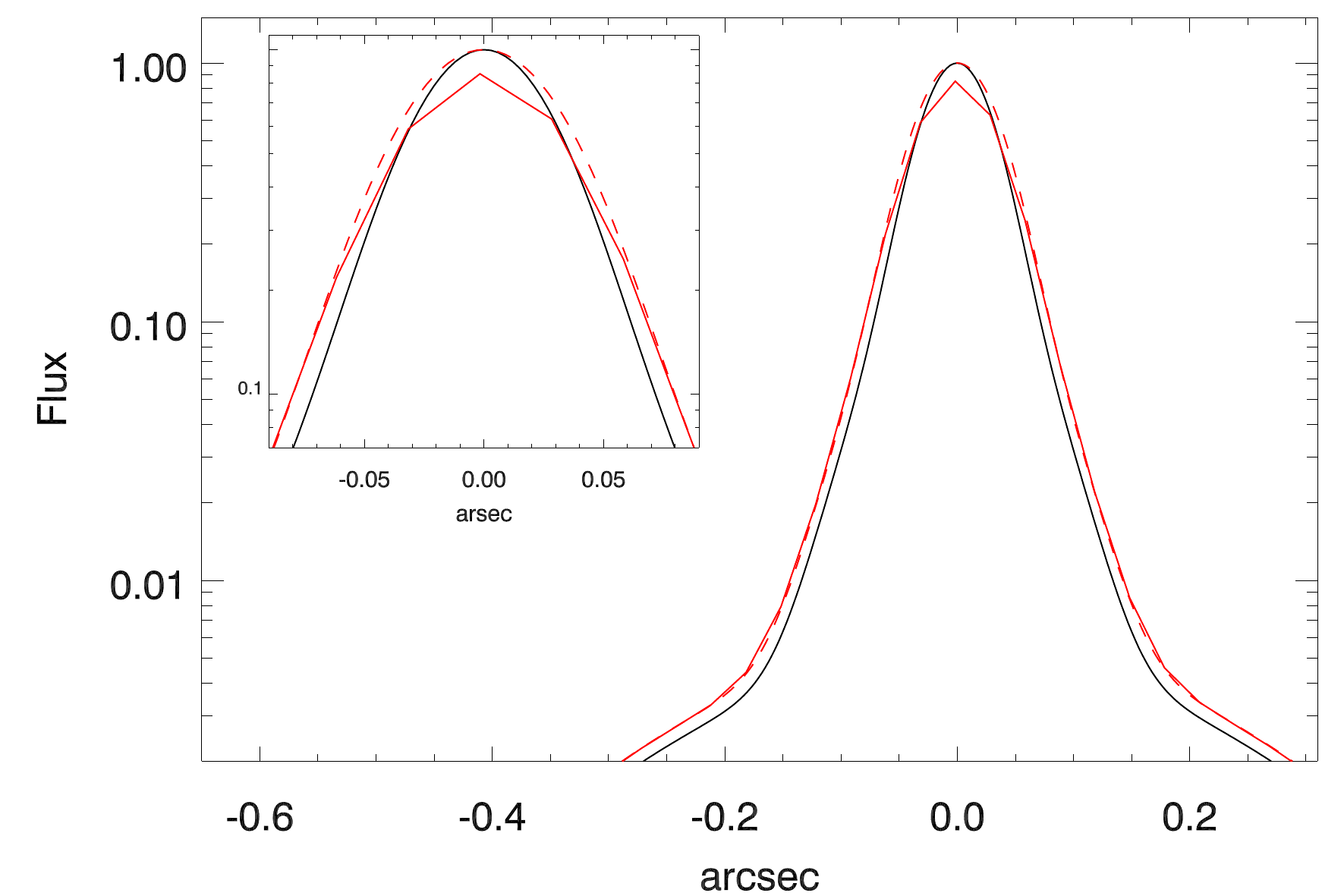}
      \caption{Simulated profile of a pair of images separated by 30 mas. The solid black curve shows the PSF model. The solid red curve is the resulting profile after convolving the two images by the PSF model. The red dashed line is the corresponding profile after re-pixelizing the simulated data to the 30 mas pixel in HST. Like in Fig.~\ref{Fig_PSFfit}, the inset covers a zoomed version near the peak. 
              }
         \label{Fig_PSFfit_30mas}
\end{figure}

\begin{figure} 
   \includegraphics[width=9cm]{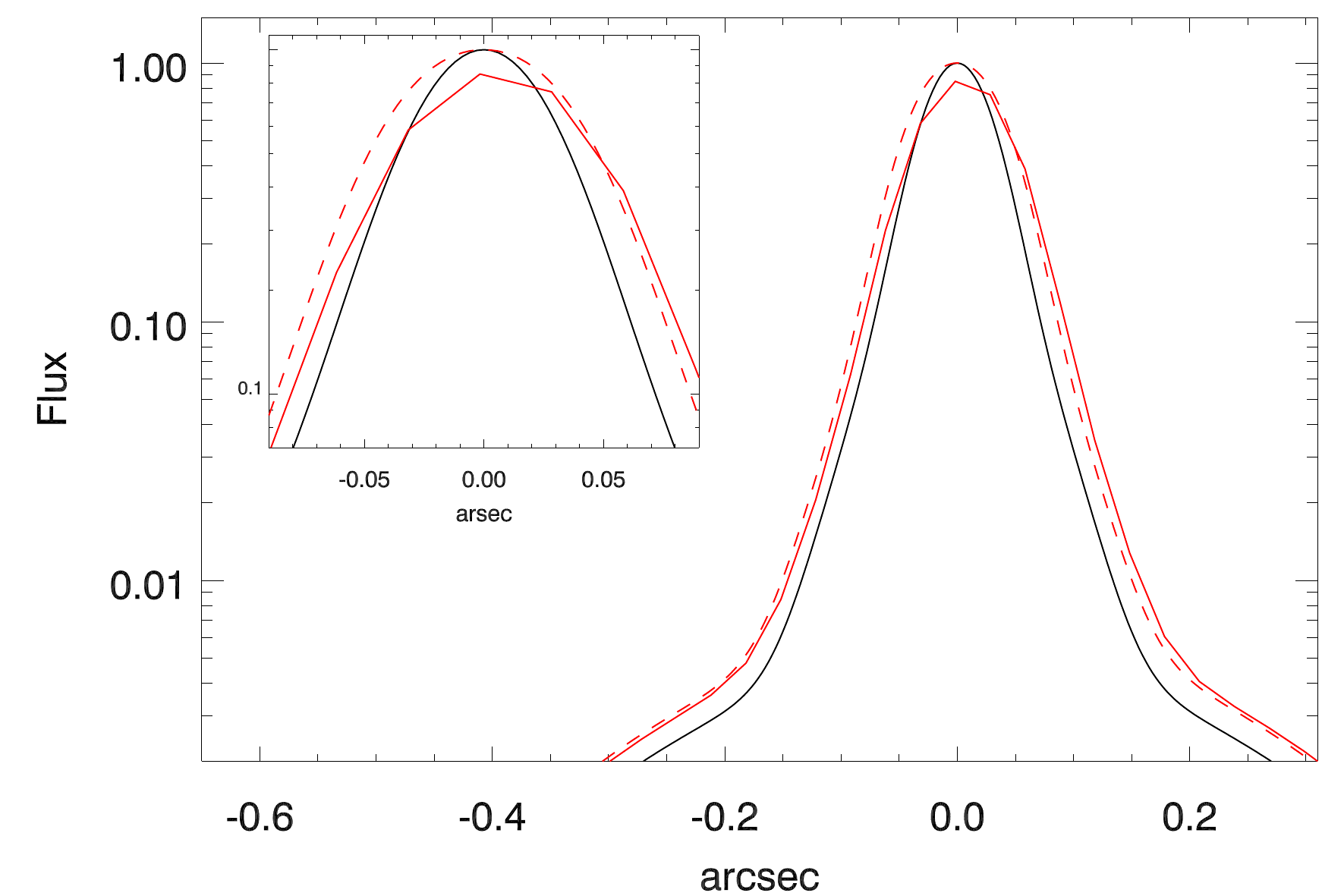}
      \caption{Like in Fig.~\ref{Fig_PSFfit_30mas} but for a separation of $30\times\sqrt{2}$ mas, which is the distance between the extreme of diagonal points in the 30 mas pixel. 
              }
         \label{Fig_PSFfit_42mas}
\end{figure}

Using the PSF model derived previously, we can simulate two point sources at a given separation and convolve them with the lens model. The simulation is done with a pixel scale of 3 mas, i.e., ten times better than the native 30 mas pixel in the HST image.
We test two separations, 30 mas and $30\times\sqrt{2}$, the first separation is similar to the size of the HST pixel, while the second is the size of its diagonal. 
The simulated convolved image is finally repixelized to match the 30 mas HST pixel size. 
The two point sources are placed in the horizontal axis, and the profile is computed in the same axis, in order to match the direction of maximum elongation along the arc. The resulting profiles are shown in Fig.~\ref{Fig_PSFfit_30mas} and   Fig.~\ref{Fig_PSFfit_42mas}. Clearly the larger separation of $30\times\sqrt{2}$ exceeds the observed profile shown in the right panel of Fig.~\ref{Fig_PSFfit}. On the contrary, the profile corresponding to the 30 mas separation is still consistent with the observed profile. Hence we can conclude that the separation must be $\approx 30$ mas at most. As mentioned earlier, it is possible that the deviation from the PSF model observed in the profile of Tr (red solid curve in the right panel of  Fig.~\ref{Fig_PSFfit}) is due to the contribution from a nearby source in the underlying arc. This would result in an even smaller separation between the pair of images, but we adopt the separation of 30 mas as a conservative upper limit.

An upper limit of 30 mas in the separation of the double image can be directly applied to infer the maximum size, or separation of the source, to caustic of the perturber. The distance from each image in the pair to the critical curve must be $d< 15$ mas, or $d<85.5$ pc. Adopting the most conservative estimate of the magnification, derived from the model of \cite{Pignataro2021} of $\mu\approx600$ (see section \ref{Sect_MuAtTr}), and assuming the radial component of the magnification is 2, the distance in the source plane, or radius of the source must be $r<0.3$ pc. Such a small radius rules out bright compact sources like globular clusters, which are typically about an order of magnitude larger. 
Adopting instead the upper limit of the magnification, $\mu \approx 7000$, and the same value for the radial magnification $\mu_r = 2$, we derive an upper limit for the size of $r<0.024$ pc or $r<5000$ AU.

\begin{figure} 
   \includegraphics[width=9cm]{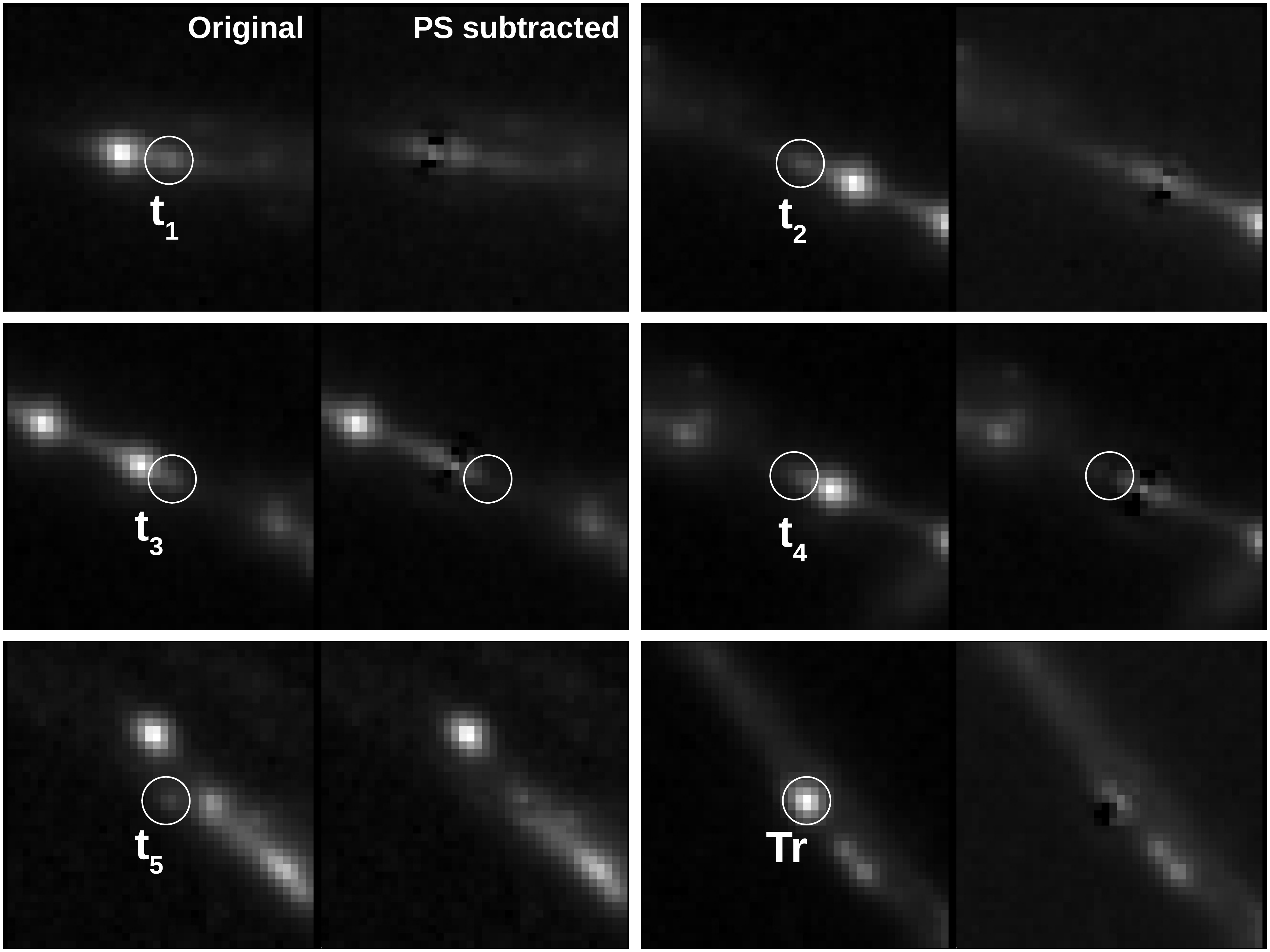}
      \caption{PSF model fitting to the sources in t1--t5 and Tr. For each pair of images, the left panel shows the original F606W band, with the source being fitted marked by a circle. In all panels, except in the bottom right, a brighter nearby source is subtracted before.  
      In each case, the right panel shows the image after subtracting the point sources. For t3 and t4, the right panel also marks with a circle the original position if the source being subtracted. 
      }
         \label{Fig_PSFfit_T1T5}
\end{figure}

The PSF model above is also used to derive fluxes at the positions of t1--t5, as well as in Tr. In order to account for the uncertainty in the PSF model, we model six nearby and unsaturated stars with a model of the form given in Eq.~\ref{Eq_PSF}. 
For each of the six star models, we determine the flux by minimizing the variance of the residual, $R$, given by $R=data-model$ in an aperture of radius 0.1". 
Since t1--t5 are partially blended with nearby brighter unresolved sources, we fit and subtract the flux from these sources before estimating the flux at t1--t5. For some counterimages, accurate photometry is harder to obtain since the position of the source can not be established with clarity. Clear examples are t3 and t4 which are the ones that are closest to knot 1.  To account for this uncertainty, we also modify slightly the position of the source being subtracted several ties and obtain different measurements. At the end of the process we have several photometric measurements where both the PSF model and the source position are varied. With these measurements we obtain the mean and dispersion of the flux and transform them into magnitudes.  The magnitudes derived this way are listed in the fourth column of table \ref{tab_3}. 
The resulting PSF subtracted images are shown in Figure~\ref{Fig_PSFfit_T1T5}, together with the original data before subtraction. Imperfections from the PSF model can still be appreciated, specially near bright sources. These imperfections are both positive and negative, partially cancelling each other, and are mostly due to a non-symmetric PSF, but also photon noise.  

\begin{figure} 
   \includegraphics[width=9cm]{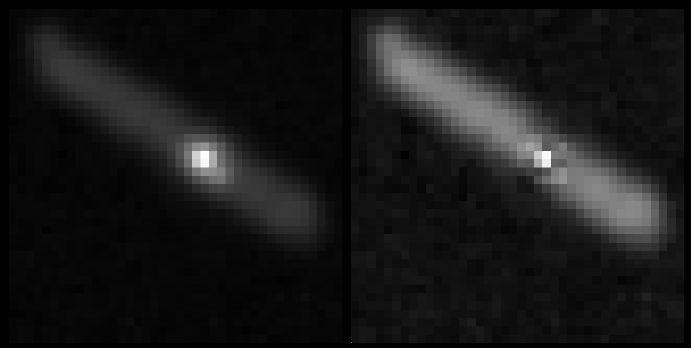}
      \caption{Performance of the flux estimation in the case of simulated data mimicking the blended sources in the Sunburst arc. The left panel shows the original simulated data and the right panel the residual after the two point source subtractions. The fainter source is placed south-west from the brighter source, and partially blended with it. The flux of the two sources are 30 and 5 in the same units as the native F606W image. These fluxes are recovered with typical errors $\sim 5\%$ and $\sim 10\%$ for the brighter and fainter source respectively. 
      }
         \label{Fig_PSFfit_Sims}
\end{figure}

Finally, we test the performance of the PSF model when extracting fluxes by comparing the input and recovered fluxes of simulated sources. In order to test possible errors emerging from the fact that t1-t5 appear blended with the bright LyC cluster, we simulate two point sources that are close enough so their profiles are partially blended. The simulation corresponds to a region of $40\times40 pixels$ in the F606W filter, and includes also an elongated feature mimicking the underlying arc. This arc includes small random features and photon noise in order to  make it more realistic (added after smoothing as the square root of the counts). The arc is smoothed assuming a Gaussian kernel with a FWHM larger than the PSF model in order to account for the fact that the arc is resolved in the radial direction. The surface brightness from the arc is comparable to the observed flux. 
Instrumental noise is added directly from the data by selecting a region of the same dimension near the arc but with no visible background objects, and which is added to the simulated arc after including the photon noise. Finally, to simulate the two point sources, and in order to mimic the asymmetric shape of the PSF, we simply consider a region of $40\times40$ pixels around an unsaturated and isolated star from the image, and re-scale it to the desired fluxes. The two re-scaled stars are then added to the arc. The fluxes of the two point sources are chosen to mimic the fluxes of the LyC knot and t1-t5. The final image is shown in the left panel of Fig.~\ref{Fig_PSFfit_Sims}. We simulate the same data set multiple times varying the random features in the underlying arc. We recover the fluxes with typical errors of $\approx 5\%$ and $\approx 10\%$ respectively for the brighter and fainter point sources.

Note that the configuration shown in  Fig.~\ref{Fig_PSFfit_Sims} corresponds to the worst case scenario found in t3 and t4, where the overlap with the LyC knot is largest. Images t1, t2, and t5 have less overlap, and the errors are typically less than 10\% in those cases. If we simply simulate one point source, with a surface brightness comparable to the one found at Tr (i.e, like in the bottom-right panel of Fig.~\ref{Fig_PSFfit_T1T5}), the typical error in the flux is $\approx 5\%$. 

\end{appendix}

\end{document}